\documentclass[a4paper, 11pt]{article}
\usepackage[T1]{fontenc} 
\usepackage[utf8]{inputenc}
\usepackage[english]{babel}
\usepackage{booktabs}
\usepackage{adjustbox}
\usepackage{multirow}
\usepackage{siunitx}
\usepackage{graphicx}
\usepackage{amssymb}
\usepackage{amsmath}
\usepackage{rotating}
\usepackage{textcomp} 
\usepackage{subfig} 
\usepackage[hidelinks]{hyperref}
\usepackage{xcolor}
\usepackage{arydshln}
\usepackage{rotating}
\usepackage{authblk}
\usepackage{csquotes}
\usepackage{makecell}
\usepackage{colortbl}
\usepackage{array} 
\usepackage{bm}

\usepackage{pifont}

\newcommand{\mko}{\ding{55}}

\newcommand{\lk}{\bfseries}
\newcommand\qm{\boldmath}

\newcommand{\uPnue}{\nu_{\mbox{\tiny e}}}

\newcommand{\uPnux}{\nu_{\mbox{\tiny $x$}}}
\newcommand{\Pnue}{\nu_{\mathrm{e}}}
\newcommand{\APnue}{\overline{\nu}_{\mathrm{e}}}
\newcommand{\Pnux}{\nu_{x}}
\newcommand{\Tnue}{$\nu_{\mathrm{e}}$}
\newcommand{\ATnue}{$\overline{\nu}_{\mathrm{e}}$}
\newcommand{\Tnux}{$\nu_{x}$}
\newcommand{\ETOT}{\mathcal{E}_{\text{tot}}}
\newcolumntype{R}{>{$}r<{$}}
\newcolumntype{L}{>{$}l<{$}}
\newcolumntype{C}{>{$}c<{$}}

\newcolumntype{A}{r@{${}\pm{}$}l}
\newcolumntype{M}{R@{${}\,{}$}L}

\newcommand{\tsup}[1]{\textsuperscript{#1}}

\newcommand{\df}[1]{\ensuremath{\operatorname{d}\!
{#1}}}

\usepackage[top=1.25in, bottom=1.75in,
left=1.0in, right=1.0in]{geometry}

\makeatletter
\def\adl@drawiv#1#2#3{%
        \hskip.5\tabcolsep
        \xleaders#3{#2.5\@tempdimb #1{1}%
        #2.5\@tempdimb}%
                #2\z@ plus1fil minus1fil\relax
        \hskip.5\tabcolsep}
\newcommand{\cdashlinelr}[1]{%
  \noalign{\vskip\aboverulesep
           \global\let\@dashdrawstore\adl@draw
           \global\let\adl@draw\adl@drawiv}
  \cdashline{#1}
  \noalign{\global\let\adl@draw\@dashdrawstore
           \vskip\belowrulesep}}
\makeatother

\usepackage[backend=bibtex, 
sorting=none,citestyle=numeric-comp]{biblatex}
\bibliography{biblio}

\DeclareSIUnit{\nn}{\relax}
\DeclareSIUnit{\erg}{erg}
\DeclareSIUnit{\ton}{ton}
\DeclareSIUnit{\tonne}{tonne}
\DeclareSIUnit{\parsec}{pc}

\author[1]{A.\ Gallo Rosso\thanks{Email: 
\href{mailto:agallorosso@laurentian.ca}
{agallorosso@laurentian.ca}.}}
\affil[1]{Department of Physics, Laurentian
University, Sudbury ON P3E 2C6, Canada}
\title{Supernova neutrino fluxes in HALO-1kT,
Super-Kamiokande, and JUNO}
\date{}

\begin{document}

\maketitle

\begin{abstract}
When the next galactic core-collapse supernova
occurs, we must be ready to obtain as much
information as possible. Although many present and
future detectors are well equipped to detect
\ATnue\ and \Tnux\ neutrinos, the detection of the
\Tnue\ species presents the biggest challenges. We
assess the impact that a 1 ktonne
lead-based detector, such as HALO-1kT, can have in
constraining electron neutrino time-integrated
fluxes. The study involves the detector taken alone 
as well as when combined with massive
\ATnue-sensitive detectors such as Super-Kamiokande
and JUNO. We find that HALO-1kT alone is not
able to strongly constrain the emission parameters.
When combined with other detectors, however, the
orthogonal information might be helpful in
improving the \Tnue\ total emitted energy and mean
energy accuracy, up to about $50\%$, if no other
\Tnue-sensitive channel is implemented. A
discussion on the reconstruction of \ATnue\ and
\Tnux\ species, as well as the total emitted
energy, is also presented.
\end{abstract}

\section{Introduction}

The direct detection of neutrinos coming from a
galactic core-collapse supernova\footnote{In the
following, supernova for short.} will have
invaluable physical importance. Under an
astronomical point of view, supernova neutrinos can
provide astronomers with an early warning for
optical detection \cite{Antonioli:2004zb,
Kharusi:2020ovw}. Their sensitivity to the
properties of the compact object, such as the
binding energy or the equation of state (see e.g.\ 
\cite{Lattimer:2006xb,Mirizzi:2015eza} for a
review) could help constraining such quantities
\cite{GalloRosso:2017hbp,GalloRosso:2018ugl}. A
neutrino signal is also a useful trigger for
gravitational-wave detectors \cite{Arnaud:2003zr,
Abbott:2008mr,Pagliaroli:2009qy,Casentini:2016vyp,
Nakamura:2016kkl} opening the door to the
possibility of combined, multi-messenger analyses
\cite{Mueller:2012sv,Ott:2012jq,Nishizawa:2014zna,
Kuroda:2017trn,Shibagaki:2020ksk}.

The handful of electron antineutrinos detected back
in 1987 \cite{Hirata:1987hu,Hirata:1988ad,
Bionta:1987qt,Bratton:1988ww,Alekseev:1988gp} have
already proved to be a unique source of information
on the explosion mechanism of massive stars
\cite{Loredo:2001rx,Pagliaroli:2008ur,
Vissani:2014doa}. This helped affirm the current
paradigm of neutrino-driven explosions and
delayed-accretion shock mechanism
\cite{Colgate:1966ax,Wilson:1971,Nadyozhin:1978zz,
Bethe:1984ux} in which neutrinos play a crucial
role in carrying out the energy and triggering the
explosion.

Since the last detected event, numerical
simulations have undeniably pushed forward our
understanding of the theoretical framework
underlying the explosion \cite{Janka:2006fh,
Janka:2012wk,Burrows:2012ew,Foglizzo:2015dma} and
one-dimensional models have come to a general
agreement \cite{OConnor:2018sti}. However, despite
this outstanding improvement, the global picture is
far from being understood completely. The huge
computational resources required by state-of-the-%
art three-dimensional simulations require some
degree of approximation in the treatment of
neutrino transport which could lead to
misrepresenting results \cite{Skinner:2015uhw,
Just:2018djz,Cabezon:2018lpr,Pan:2018vkx}.
Moreover, a first look to the outcomes of three-%
dimensional simulations suggest that a supernova
explosion is a multi-dimensional phenomenon, with
peculiar features that could emerge in the detected
fluxes, as e.g.\ discussed in Refs.\ 
\cite{Mirizzi:2015eza,Janka:2016fox,
Horiuchi:2017sku}.

A direct detection will also be an invaluable
opportunity to cast a light onto the neutrino
behavior in dense environments. The idea that
neutrino self-interaction could lead to flavor
conversion phenomena other than the MSW effect
\cite{Wolfenstein:1977ue,Mikheev:1986gs} dates back
to the early 90s \cite{Pantaleone:1992eq,
Samuel:1993uw,Sigl:1992fn} and is still developing
(for a review, see e.g.\ \cite{Duan:2009cd,
Duan:2010bg,Mirizzi:2015eza,Tamborra:2020cul}).
Recently, the fast
pair-wise flavor conversion of neutrinos on very
small scales has been shown to be possible
\cite{Sawyer:2015dsa,Izaguirre:2016gsx} remaining
a matter of discussion \cite{Capozzi:2017gqd,
Dasgupta:2018ulw,Capozzi:2018clo,Abbar:2018beu,
Abbar:2018shq,Nagakura:2019sig,Morinaga:2019wsv,
DelfanAzari:2019tez,Glas:2019ijo,Abbar:2019zoq,
Shalgar:2019qwg}. Finally, the possibility of
inferring new neutrino physics from a supernova
explosion, such as the existence of sterile
neutrinos \cite{Peres:2000ic,Sorel:2001jn,
Choubey:2006aq,Choubey:2007ga,Raffelt:2011nc,
Tamborra:2011is,Wu:2013gxa,Esmaili:2014gya,
Warren:2014qza,Franarin:2017jnd,
Mastrototaro:2019vug,Suliga:2019bsq,Tang:2020pkp}
or non-standard interactions
\cite{EstebanPretel:2007yu,Wu:2013gxa,
Stapleford:2016jgz,Yang:2018yvk,Sung:2019xie,
Suliga:2019bsq,deGouvea:2019goq,Shalgar:2019rqe},
have also been investigated.

When the next supernova happens, we will have to be
ready to get the most of the signal. That is, a
good time resolution and a good reconstruction of
the flux for each neutrino flavor. Although the
detectors built with the primary purpose of
detecting supernovae are quite rare, at present
there are many able to detect a signal employing
different technologies, such as water/ice (e.g.\ 
Super-Kamiokande \cite{Ikeda:2007sa}, IceCube
\cite{Abbasi:2011ss}), liquid scintillator (e.g.\ 
KamLAND \cite{Tolich:2011zz}, SNO+
\cite{Rumleskie:2020iip}), lead (e.g.\ HALO
\cite{Zuber:2015ita}) or noble gases (e.g.\ 
XENON \cite{Reichard:2017guz}). Moreover, the next
decade will see the commissioning of new, massive
experiments able to see a supernova event with
impressive statistics. The Jiangmen Underground
Neutrino Observatory (JUNO) \cite{An:2015jdp,
Miramonti:2020bwk}, the Deep Underground Neutrino
Experiment (DUNE) \cite{Acciarri:2015uup,
Abi:2020oxb}, Hyper-Kamiokande (HK)
\cite{Abe:2018uyc,Itow:2020pgf}, and IceCube-Gen2
\cite{Aartsen:2020fgd} will possibly open a golden
era for supernova neutrino detection.

Dealing with supernova neutrinos, this kind of
differentiation and overlapping is crucial. On one
hand, the time-scale of a supernova is greater than
that of a standard experiment. For instance,
assuming a reasonable rate of $1.63\pm 0.46$
galactic supernovae per century
\cite{Rozwadowska:2021lll}
the probability for a detector to see one event is
about 7.8\%, 15.0\%, 27.8\%, and 55.7\%, assuming 
it stays constantly operational for 5, 10, 20, and
50 years respectively. It is not unlikely that,
once very massive (and expensive) experiments
finish their life cycle, detection of supernova
neutrinos will be left to stable, small-scale ones.

On the other hand, it seems now well-established
that the reconstruction of the next supernova
signal --- and thus the breaking of the
degeneracies introduced by the uncertainties on
the spectral shape \cite{Minakata:2008nc} --- will
require the combination of many sources of
information, coming from different detection
channels \cite{Laha:2013hva,Laha:2014yua,
Lu:2016ipr,GalloRosso:2017hbp,Nikrant:2017nya,
GalloRosso:2017mdz,Li:2017dbg,Li:2019qxi,
Abi:2020lpk,Nagakura:2020bbw}. However, the vast
majority of the supernova detectors
are mostly sensitive to electron antineutrinos
through the capture on protons or Inverse Beta
Decay (IBD), while the nuclear reactions used for
detecting the \Tnue\ species usually come with
low statistics and difficult tagging (see e.g.\ 
\cite{Scholberg:2012id}).

For instance, recent calculations on the
sensitivity of massive liquid scintillator detector
such as JUNO \cite{Li:2017dbg,Li:2019qxi} have
shown that a complete reconstruction of the three
neutrino fluxes is indeed possible thanks to its
major three channels (IBD and elastic scattering on
protons and electrons). However, an accurate
reconstruction of the \Tnue\ species is difficult
to reach if the supernova is not extremely close.
As a consequence, a future detection by DUNE seems
our only chance to constrain the \Tnue\ flux (see
e.g.\ \cite{Nagakura:2020bbw}).

Among the possible alternatives, lead is
particularly promising. It is mostly sensitive to 
\Tnue\ Charged-Current (CC) interactions, while the 
ones involving \ATnue\ are suppressed thanks to the
high number of neutrons (Pauli-blocking). Ref.\ 
\cite{Cline:1993rx} was the first one to suggest
lead as an active material for supernova neutrinos.
In principle, lead can be use to retrieve useful
information on neutrino temperatures
\cite{Fuller:1998kb} as well as to assess the
deviation of the fluxes from a pure thermal
behavior \cite{Engel:2002hg}. Neutrino interactions
on lead have been studied in many theoretical
calculations \cite{Kolbe:2000np,Engel:2002hg,
Lazauskas:2007bs,Almosly:2016mse,Almosly:2019han}
and a direct measurement is planned at the Oak
Ridge National Laboratory (ORNL) facility
\cite{Akimov:2017ade}.

This technology has been concretely implemented by
the Helium and Lead Observatory (HALO). A \SI{79}
{\ton} version\footnote{That is, metric tons, as it
will always be understood in the following.} is
currently running at SNOLAB \cite{Zuber:2015ita},
while an upgraded \SI{1}{\kilo\ton} version
(HALO-1kT) will possibly be built at Laboratori 
Nazionali del Gran Sasso (LNGS). Under an
experimental point of view, the whole configuration
has been proven to be stable, reliable and
characterized by a high livetime. On the
theoretical side, the impact of HALO detectors has
been investigated in Ref.\  \cite{Vaananen:2011bf}.
In the paper, it is shown how the ratio between 1-
and 2- neutron events detected in the actual
configuration is sensitive to different neutrino
temperatures and patterns of flavor transformation.

The aim of this paper is to deepen and broaden the 
discussion on the impact that a detector like
HALO-1kT could have in the reconstruction of the
detected time-integrated fluxes (fluences), with
a special emphasis on the reconstruction of the
\Tnue\ species as detected \emph{on Earth}.%
\footnote{In fact, this is the first step of any
further analysis.} HALO-1kT is considered both
alone and in combination with Water-Cherenkov
Super-Kamiokande (SK) and liquid scintillator-based
JUNO. This is in order to asses the significance of
the orthogonal information brought by HALO-1kT to
massive supernova \ATnue\ detectors. In order 
to compare the concept of the different detectors,
we consider a perfect version of them:\footnote{%
This choice is especially well-suited for HALO-1kT
and JUNO; since not yet operational, it is hard to
predict the exact response and general 
characteristics.} namely, 100\% efficiency, perfect
response and no uncertainties on the nuclear cross
sections.

In this work, the supernova explosion is assumed
to be described by the models provided by Ref.\ 
\cite{Mirizzi:2015eza} and the neutrino
time-integrated fluxes are parameterized by
standard quasi-thermal Garching distributions
\cite{Keil:2002in,Tamborra:2012ac}. This results
in a problem with 9 degrees of freedom, tackled
by a Monte Carlo based likelihood analysis. This
is the first analysis of this kind for a HALO-like
detector.

The present paper is structured as follows: in
Section \ref{sec:Sec2} we discuss in detail the
hypotheses underlying our analyses, such as the
assumptions on supernova models and detection
channels. A description of the numerical approach
implemented is presented in Section \ref{sec:lik}.
The outcomes of the analyses are presented in
Section \ref{sec:res}. Section \ref{sec:anue} is a
parenthesis on \ATnue\ fluences and Section
\ref{sec:etot} on the total emitted energy
$\ETOT$. Finally, section \ref{sec:conclu} is a
conclusion.

\section{Hypotheses and method}
\label{sec:Sec2}

In our analyses, we assume a supernova explosion
happening at a distance $D$ of exactly \SI{10}
{\kilo\parsec}. We take into account two different
progenitor masses, as described by the
one-dimensional models reported in Ref.\ 
\cite{Mirizzi:2015eza}: 27 $M_{\odot}$ in the first
case (LS220-s27.0co) and 9.6 $M_{\odot}$ in the
second one (LS220-z9.6co). In both models, the
equation of state (EoS) implemented comes from
Ref.\ \cite{Lattimer:1991nc}, convection has a
mixing-length treatment, and beta reactions include
self-energy shifts of nucleons. Since our goal is
to assess the reconstruction of detected fluxes
on Earth, as a first approximation no flavor
transformation is implemented.

In the following, we will consider the standard
effective description that deals with just three
neutrino species: \Tnue,\ \ATnue, and \Tnux, where
the latter indicates one of the species $\nu_\mu$,
$\overline{\nu}_{\mu}$, $\nu_\tau$, $\overline{\nu}
_{\tau}$. The fluences are described by standard
quasi-thermal  Garching distributions
\cite{Keil:2002in,Tamborra:2012ac}, differential in
the neutrino energy $E$:
\begin{equation}
	\label{eq:flu}
	\frac{\df{F}(\nu_i)}{\df{E}} =
	\frac{\mathcal{E}(\nu_i)}{4\pi D^2}
	\frac{[\alpha(\nu_i)+1]^{[\alpha(\nu_i)+1]}}{
	\Gamma[\alpha(\nu_i)+1]}\frac{E^{\alpha(\nu_i)}}
	{\langle E(\nu_i)\rangle^{\alpha(\nu_i)+2}}
	\exp\left[-[\alpha(\nu_i)+1]\frac{E}
	{\langle E(\nu_i)\rangle}\right].
\end{equation}
Here, $\mathcal{E}(\nu_i)$ is the total emitted
energy for a given neutrino species, $\langle E(\nu
_i)\rangle$ is the mean energy and $\alpha(\nu_i)$
is the pinching parameter, describing the width of
the distribution and thus its shape. Here, $\nu_i$
is the neutrino species. A fit of the fluxes
integrated over time gives the sets of parameters
assumed for the two models. They are reported in
Table \ref{tab:param}.

\begin{table}[t]
	\centering
	\begin{tabular}{rccccccc}
		\toprule
		&\multicolumn{3}{c}{LS220-s27.0co}
		&&\multicolumn{3}{c}{LS220-z9.6co}\\
		\cmidrule{2-4}
		\cmidrule{6-8}
		& $\mathcal{E}^*(\nu_i)$
		& $\langle E(\nu_i)\rangle^*$
		& $\alpha^*(\nu_i)$
		&& $\mathcal{E}^*(\nu_i)$
		& $\langle E(\nu_i)\rangle^*$
		& $\alpha^*(\nu_i)$\\
		&[\SI{e53}{\erg}]
		&[\si{\mega\electronvolt}] &
		&&[\SI{e53}{\erg}]
		&[\si{\mega\electronvolt}] &\\
		\midrule
		\Tnue
		& 0.571	& 10.8	& 2.42
		&& 0.316 & 9.9 & 2.75\\
		\ATnue
		& 0.568	& 13.6	& 2.26
		&& 0.338 & 12.3 & 2.19\\
		\Tnux
		& 0.526	& 12.9	& 1.85
		&& 0.295 & 12.5 & 2.46\\
		\bottomrule
	\end{tabular}
	\caption{True parameters describing the
	time-integrated fluxes (fluences) from the two
	models LS220-s27.0co and LS220-z9.6co
	\protect\cite{Mirizzi:2015eza}
	fitted with distribution 
	\protect\eqref{eq:flu}.}
 	\label{tab:param}
\end{table}

Since all three parameters for each neutrino
species are left free to vary, this results in a
problem with 9 degrees of freedom. The signal is
considered to be given by three different
detectors: HALO-1kT (HALO),\footnote{In the
following, HALO-1kT will simply be denoted as HALO
for short, since there cannot be room for confusion
with the SNOLAB experiment.} Super-Kamiokande (SK)
and JUNO. In the following, we discuss the
hypotheses underlying the description of each 
detector.

\subsection{HALO}

In HALO detector, supernova neutrinos interacting
in \SI{1}{\kilo\ton} of lead produce neutrons
captured by \tsup{3}He proportional counters. The 
signal is given by two classes of events: 1-neutron
events (1n) and 2-neutron events (2n), given by a
combination of both CC and NC interactions:
\begin{equation}
	\text{CC}
	\begin{cases}
		\Pnue+\text{\tsup{$x$}Pb}\to\text{e}^-
		+\text{\tsup{$(x-1)$}Bi}+\text{n};\\
		\Pnue+\text{\tsup{$x$}Pb} \to\text{e}^-
		+\text{\tsup{$(x-2)$}Bi} + 2\text{n}.\\
	\end{cases}\quad
	\text{NC}
	\begin{cases}
		\nu+\text{\tsup{$x$}Pb}\to\nu
		+\text{\tsup{$(x-1)$}Pb} + \text{n};\\
		\nu + \text{\tsup{$x$}Pb} \to\nu
		+\text{\tsup{$(x-2)$}Pb} + 2\text{n}.\\
	\end{cases}
\end{equation}

The kinematical thresholds, weighted over the
isotopic abundances, are \SI{7.61}{\mega%
\electronvolt} (10.47 \si{\mega\electronvolt})
for NC (CC) 1n events and \SI{14.64}{\mega%
\electronvolt} (\SI{18.53}{\mega\electronvolt})
for NC (CC) 2n events. The 1n and 2n neutrino-%
nucleus cross-sections assumed in this work are the
ones provided by Engel et.\ al \cite{Engel:2002hg}.
In principle, the theoretical calculation describes
just the isotope \tsup{208}Pb but we apply it to
lead in general. In fact, this is the only
reference in the literature providing the 1n and
2n partial cross sections. Hopefully, the
uncertainties affecting these quantities will be
reduced to some 10\%, thanks to direct measurements 
on mini-HALO detector at ORNL. The values are
interpolated outside the tabulated ranges: $(10-95
)$ \si{\mega\electronvolt} for 1n events and $(25
-95)$  \si{\mega\electronvolt} for 2n events. The
results show no significant dependency over the
implemented interpolation method.

We assume 100\% detection efficiency, a perfect
reconstruction of the two classes of events, as
well as an error-free cross-section. This is done
as a proof of principle, to assess the impact HALO
can theoretically have among other perfect
detectors. In fact, it is worth mentioning that,
dealing with supernova neutrinos, the limiting
factor is rather the statistics than the detector
response (see e.g.\ Ref.\ \cite{Abi:2020lpk}). It
is worth mentioning that the current
\SI{79}{\ton} version of HALO has a measured
single-neutron detection efficiency of about 28\%
on average. This value is expected to increase up
to about 50\% in the \SI{1}{\kilo\ton}
version.\footnote{C.J.\ Virtue, private
communication.}
The number of expected 1n and 2n events for the two
models are reported in Table \ref{tab:eve}.

\begin{table}[t]
\centering
\resizebox{\textwidth}{!}{
	\begin{tabular}{lccccccccccc}
		\toprule
		&\multicolumn{2}{c}{HALO-1kT}
		&&\multicolumn{2}{c}{Super-K}
		&&\multicolumn{5}{c}{JUNO}\\
		\cmidrule{2-3}\cmidrule{5-6}\cmidrule{8-12}
		& 1n & 2n && IBD & eES && pES & eES & IBD
		& \Tnue-\tsup{12}C & \ATnue-\tsup{12}C\\
		\midrule
		\multirow{2}{*}{\Gape[2.52ex]%
		{LS220-s27.0co}}
		&\multirow{2}{*}{\Gape[2.52ex]{152}}
		&\multirow{2}{*}{\Gape[2.52ex]{17.3}}
		&&\multirow{2}{*}{\Gape[2.52ex]{5515}}
		&\multirow{2}{*}{\Gape[2.52ex]{228}}
		&& --- & 362 & 5376
		&\multirow{2}{*}{\Gape[2.52ex]{40.2}}
		&\multirow{2}{*}{\Gape[2.52ex]{111}}\\
		\cmidrule{8-10}
		&&&&&&& 1319 & 204 & 5324 & & \\
		\hline
		\multirow{2}{*}{\Gape[2.52ex]%
		{LS220-z9.6co}}
		&\multirow{2}{*}{\Gape[2.52ex]{58.7}}
		&\multirow{2}{*}{\Gape[2.52ex]{4.8}}
		&&\multirow{2}{*}{\Gape[2.52ex]{2943}}
		&\multirow{2}{*}{\Gape[2.52ex]{120}}
		&& --- & 2883 & 203
		&\multirow{2}{*}{\Gape[2.52ex]{11.2}}
		&\multirow{2}{*}{\Gape[2.52ex]{48.2}}\\
		\cmidrule{8-10}
		&&&&&&& 491 & 2840 & 120 & & \\
		\bottomrule
	\end{tabular}
	}
	\caption{Number of expected events for each
	detection channel: 1-neutron events on lead
	(1n), 2-neutron events on lead (2n), inverse
	beta decay (IBD), neutrino-electron elastic
	scattering (eES), neutrino-proton elastic
	scattering (pES), and CC scatterings on
	\tsup{12}C. The fluences are described by the
	parameters reported in table
	\protect\ref{tab:param}, assuming the supernova
	models LS220-s27.0co and
	LS220-z9.6co \protect\cite{Mirizzi:2015eza}.}
	\label{tab:eve}
\end{table}

\subsection{Super-Kamiokande}

Super-Kamiokande is a water-based Cherenkov
detector with a fiducial mass of \SI{22.5}
{\kilo\ton} of  water and a threshold of
\SI{5}{\mega\electronvolt} \cite{Beacom:1998ya,
Smy:2010zza}. The main detection channels are the
CC inverse beta decay (IBD)
\begin{equation}
	\APnue+\mathrm{p}\to\mathrm{n}+\mathrm{e}^+
\end{equation}
and the neutrino-electron elastic scattering (eES)
\begin{equation}
	\nu+\mathrm{e}^-\to\nu + \mathrm{e}^-.
\end{equation}
Those channels are also the ones considered in the
present work, and the way they are treated is the
same as Refs.\ \cite{GalloRosso:2017hbp,
GalloRosso:2017mdz}. As for HALO, the number of
expected events for the two models is reported in
Table \ref{tab:eve}.

In the following, we assume 100\% detection and
tagging efficiency. This is not far from the 90\% 
expected from gadolinium doping
\cite{Beacom:2003nk,Laha:2013hva}.

\subsection{JUNO}
\label{sec:JUNO}

JUNO is a liquid scintillator-based detector,
assumed to be composed by \SI{20}{\kilo\ton}
\cite{Lu:2016ipr,Li:2017dbg} of Linear Alkyl 
Benzene (LAB) with chemical formula $\mathrm{H}
_{28.360}\mathrm{C}_{17.195}\mathrm{N}
_{0.002}\mathrm{O}_{0.002}$. The considered
channels are: IBD, eES, neutrino-proton elastic
scattering (pES) and Charged-Current reactions on
\tsup{12}C (\tsup{12}C-CC):%
\footnote{We neglect the subdominant contribution
from the Neutral-Current reactions on \tsup{12}C,
since their statistics are outnumbered by other NC
reactions.}
\begin{align}
\label{eq:4:CCsuC1}
\Pnue+\text{\tsup{12}C} &\to\mathrm{e}^-+
\text{\tsup{12}N}^{(*)};\\
\label{eq:4:CCsuC2}
\APnue+\text{\tsup{12}C} &\to \mathrm{e}^++
\text{\tsup{12}B}^{(*)}.
\end{align}

Charged-current reaction \eqref{eq:4:CCsuC1} has
a threshold of \SI{17.4}{\mega\electronvolt} and
reaction \eqref{eq:4:CCsuC2} of \SI{14.4}
{\mega\electronvolt}. The emitted
$\mathrm{e}^{\pm}$ is observable, in principle,
in coincidence with the $\beta^{+}$ ($\beta^-$)
decay of \tsup{12}B (\tsup{12}N), happening with a
half-life of \SI{20.2}{\milli\second} (\SI{11.0}
{\milli\second}). On one hand, the signature is
similar to the IBD one; on the other, it is much
more difficult to tag, since the (anti)electrons
from the $\beta^{\pm}$ decay distribute broadly
in energy and time \cite{Scholberg:2012id}.

In the likelihood analysis, the \tsup{12}C-CC
events are analyzed separately. As for HALO,
100\% efficiency and a perfect knowledge of the
cross-section are assumed. In order to assess the
impact of \tsup{12}C-CC interactions in the
analysis, three possible scenarios are considered:
\begin{enumerate}
	\item The \tsup{12}C-CC channel is not
	implemented. That is, \tsup{12}C-CC \Tnue\
	and \ATnue\ events are assumed undetectable
	and thus not considered in the analysis.
	\item The \Tnue\ and \ATnue\ \tsup{12}C-CC
	events can be extracted from other channels
	(IBD, eES) but cannot be distinguished.
	That means that, although detectable and
	separable from IBD and eES events,
	reactions \eqref{eq:4:CCsuC1} and
	\eqref{eq:4:CCsuC2} cannot be told apart
	between each other. This scenario is indicated
	by the letter ``C'' in the label.
	\item The \Tnue\ and \ATnue\ \tsup{12}C-CC 
	events can be extracted from other channels and
	can also be distinguished. It is
	indicated by the letters ``Cd'' in the label.
\end{enumerate}

The pES events are treated in the same way as in
Ref.\ \cite{Li:2017dbg}. The energy dependance of
the light quenching has been described from the
study of Von Krosigk \cite{VonKrosigk:2015yio} and
it has subsequently been rescaled in order to match
with the number of expected events given by Ref.\ 
\cite{Lu:2016ipr}. Also in this case, and to assess
the impact low-energy events have on the analysis,
two possible scenarios are taken into account in
the following:
\begin{enumerate}
	\item Assuming the pES is detectable (``p'' in
	the label), it is analyzed in the window
	that goes from 0.2 to
	\SI{4}{\mega\electronvolt}. In this window,
	the contribution from other channels is
	neglected. The IBD and eES are then analyzed
	from \SI{5}{\mega\electronvolt}, assuming no
	contamination from pES and 100\% tagging
	efficiency for the IBD.
	\item Assuming the pES is not detectable (``n''
	in the label), the
	eES and IBD reactions are analyzed in the
	whole range $(0.2 - 100)$
	\si{\mega\electronvolt}, assuming 100\% tagging
	efficiency for the IBD.
\end{enumerate}

Summarizing, each one of the analyses involving
JUNO is performed six times, depending on the
combination of the two assumptions about the pES
channel (not implemented/included) and three
assumptions about the \tsup{12}C-CC channel (not
implemented, \Tnue-\ATnue\ indistinguishable or
\Tnue-\ATnue\ tagged). As for the aforementioned
detectors, 100\% efficiency is assumed for all
channels. The six scenarios are summarized in Table
\ref{tab:jcon}, together with the label to
designate them, while the number of expected events
given the two supernova models are reported in
Table \ref{tab:eve}.

\begin{table}
	\centering
	\begin{tabular}{ccccll}
	\toprule
	pES & eES & IBD & \multicolumn{1}{c}{
	\tsup{12}C-CC}
	& & Label\\
	\midrule
	&&\multirow{3}{*}{\Gape[3ex]{\makecell{100\%\\
	[-0.3em]tagging}}}
	&\mko & & n\\
	\cmidrule{4-6}
	\mko &$>\SI{0.2}{\mega\electronvolt}$
	& &\multirow{2}{*}{\Gape[0.7ex]{\makecell{indep.
	\\[-0.3em]signal}}}
	&\Tnue/\ATnue\ indistinguishable & nC\\
	\cmidrule{5-6}
	& & & &\Tnue/\ATnue\ distinguishable & nCd\\
	\midrule
	$(0.2-4)\si{\mega\electronvolt}$
	&$>\SI{5}{\mega\electronvolt}$
	&$>\SI{5}{\mega\electronvolt}$
	&\mko & & p\\
	\cmidrule{4-6}
	\multirow{2}{*}{\Gape[0.7ex]{\makecell{no eES\\
	[-0.3em]contamination}}}
	&\multirow{2}{*}{\Gape[0.7ex]{\makecell{no pES\\
	[-0.3em]contamination}}}
	&\multirow{2}{*}{\Gape[0.7ex]{\makecell{100\%\\
	[-0.3em]tagging}}}
	&\multirow{2}{*}{\Gape[0.7ex]{\makecell{indep.\\
	[-0.3em]signal}}}
	&\Tnue/\ATnue\ indistinguishable & pC\\
	\cmidrule{5-6}
	& & & &\Tnue/\ATnue\ distinguishable & pCd \\
	\bottomrule
	\end{tabular}
	\caption{Different assumptions for JUNO detector
	and its channels, as well as the label for each
	configuration, as explained in Section
	\protect\ref{sec:JUNO}. Namely: pES un/%
	implemented (n/p), $\Pnue/\APnue$ \tsup{12}C-CC
	reactions included (C), $\Pnue/\APnue$
	\tsup{12}C-CC reactions
	included and distinguishable (d).}
	\label{tab:jcon}
\end{table}

\subsection{Likelihoods}
\label{sec:lik}

With the exception of the HALO events, all the
detected channels are analyzed from their
differential spectra by means of a binned
likelihood:
\begin{equation}
\label{eq:5:LikBin}
	\mathcal{L}_{j}\propto
	\prod_{i=1}^{N_{\text{bin}}}
	\frac{\nu_i^{n_i}}{n_i!} e^{-\nu_i}
	\quad\text{with}\quad j=\text{IBD, eES, pES,
	\Tnue-\tsup{12}C, \ATnue-\tsup{12}C};
\end{equation}
where $\nu_i$ is the number of expected events for
the process $j$ (as a function of parameters) in
the $i$-th bin and $n_i$ is the number observed in
the simulation, in the same bin. Bin widths are not
uniform but the coverage is denser at low energies;
the partition depends on the detectors, the
channels and the analyses. However, we found no
significant dependency upon the different binnings
that can be assumed.

Concerning the lead events in HALO, there is
no spectrum to be detected. The only kind of
available information is the two numbers of 1n and
2n events --- see e.g.\ Ref.\ 
\cite{Vaananen:2011bf}. Therefore, the likelihood
simply confronts the Poissonian deviation between
the detected number $n_i$ and the expected number
$N_i$ as a function of the parameters:
\begin{equation}
	\mathcal{L}_{\text{HALO}}\propto
	\frac{1}{\sqrt{N_1 N_2}}\:\text{exp}\!\left[
	-\frac{\left(n_1-N_1\right)^2}{2N_1}
	-\frac{\left(n_2-N_2\right)^2}{2N_2}
	\right].
\end{equation}

The procedure of analysis follows a Monte Carlo
approach. The $9$ analyzed parameters define a 9-%
dimensional phase-space, explored by throwing in it 
random points $P$ extracted uniformly within some
priors. Each point describes a particular set of
neutrino fluences, which can be accepted or
rejected according to the value of its likelihood.
A point is accepted within a certain confidence
level (CL) if its likelihood satisfies the relation
\begin{equation}\label{eq:5:likfun}
	\log\mathcal{L}\left(P\right)\ge
	\log\mathcal{L}_{\text{max}} - \frac{A}{2},
\end{equation}
where $\mathcal{L}_{max}$ is the likelihood maximum
inside the prior and $A$ is defined with an
integral of a chi-square distribution with
$N_{\text{dof}}$ degrees of freedom;
\begin{equation}
\int_0^{A} \chi^2(N_{\text{dof}}; z)
	\,\df{z} = \text{CL}.
\end{equation}
In the following, we consider
\begin{equation}
	\text{CL} = 3\sigma = 0.9973.
\end{equation}

The priors are broad enough to cover any plausible
scenario. The total energies are constrained to
vary  in the interval
\begin{equation}\label{eq:pris}
\SI{1e52}{\erg}\le\mathcal{E}(\nu_i)\le\SI{2e53}
{\erg},
\end{equation}
while the mean energies have the prior
\begin{equation}\label{eq:prie}
	\SI{2}{\mega\electronvolt} \le \langle E(\nu_i)
	\rangle \le \SI{70}{\mega\electronvolt}.
\end{equation}
Concerning the pinching parameters, we recall that
we are dealing with time-integrated fluxes.
They are observed to be closer to a thermal
distribution than the time-dependent ones, as
discussed in Refs.\ \cite{Hudepohl:2013zsj,
Vissani:2014doa}. According to this latter
reference, a reasonable conservative interval is
$\alpha\in[1.5,\,3.5]$. In the following, we 
consider as a prior
\begin{equation}
	\label{eq:pria}
	1.0 \le \alpha(\nu_i)\le 4.0.
\end{equation}

\begin{figure}[t!]
	\centering
	\subfloat[LS220-s27.0co, \Tnue\ species.]
	{\label{fig:1h14}
	\includegraphics[width=.45\textwidth]{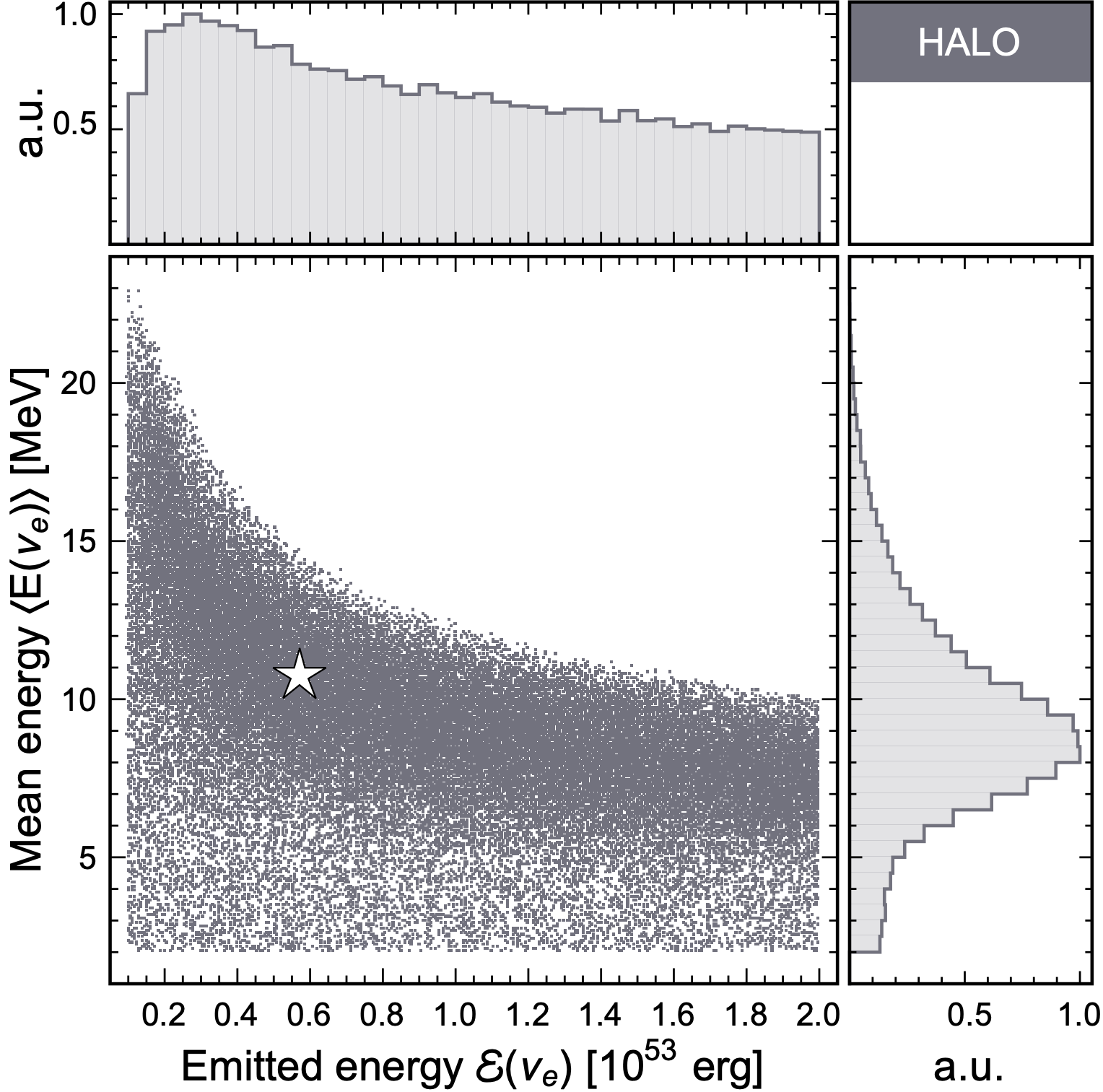}}
	\hfill
	\subfloat[LS220-z9.6co, \Tnue\ species.]
	{\label{fig:2h14}
	\includegraphics[width=.45\textwidth]{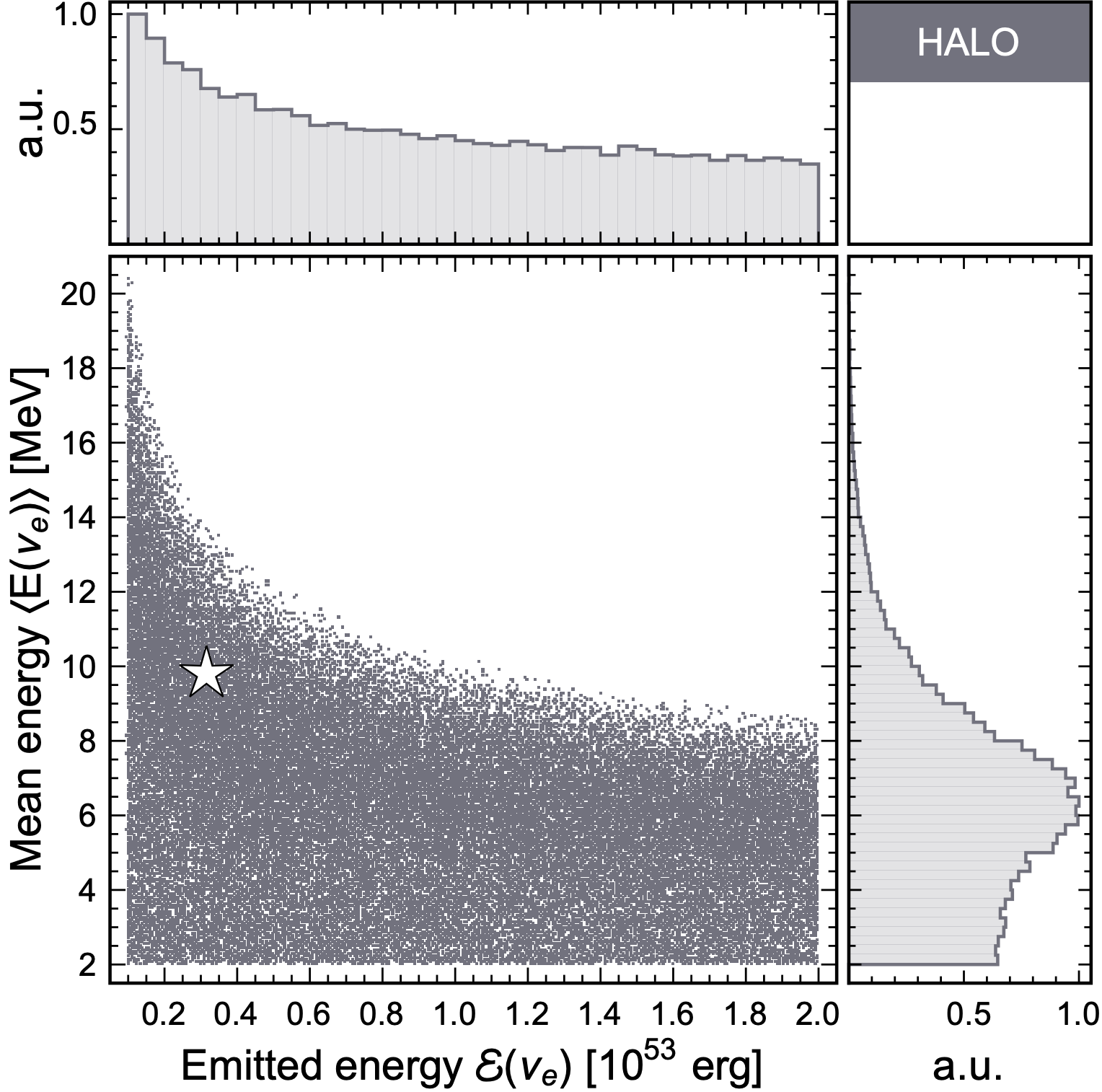}}%
	\\ \subfloat[LS220-s27.0co, \Tnux\ species.]
	{\label{fig:1h36}
	\includegraphics[width=.45\textwidth]{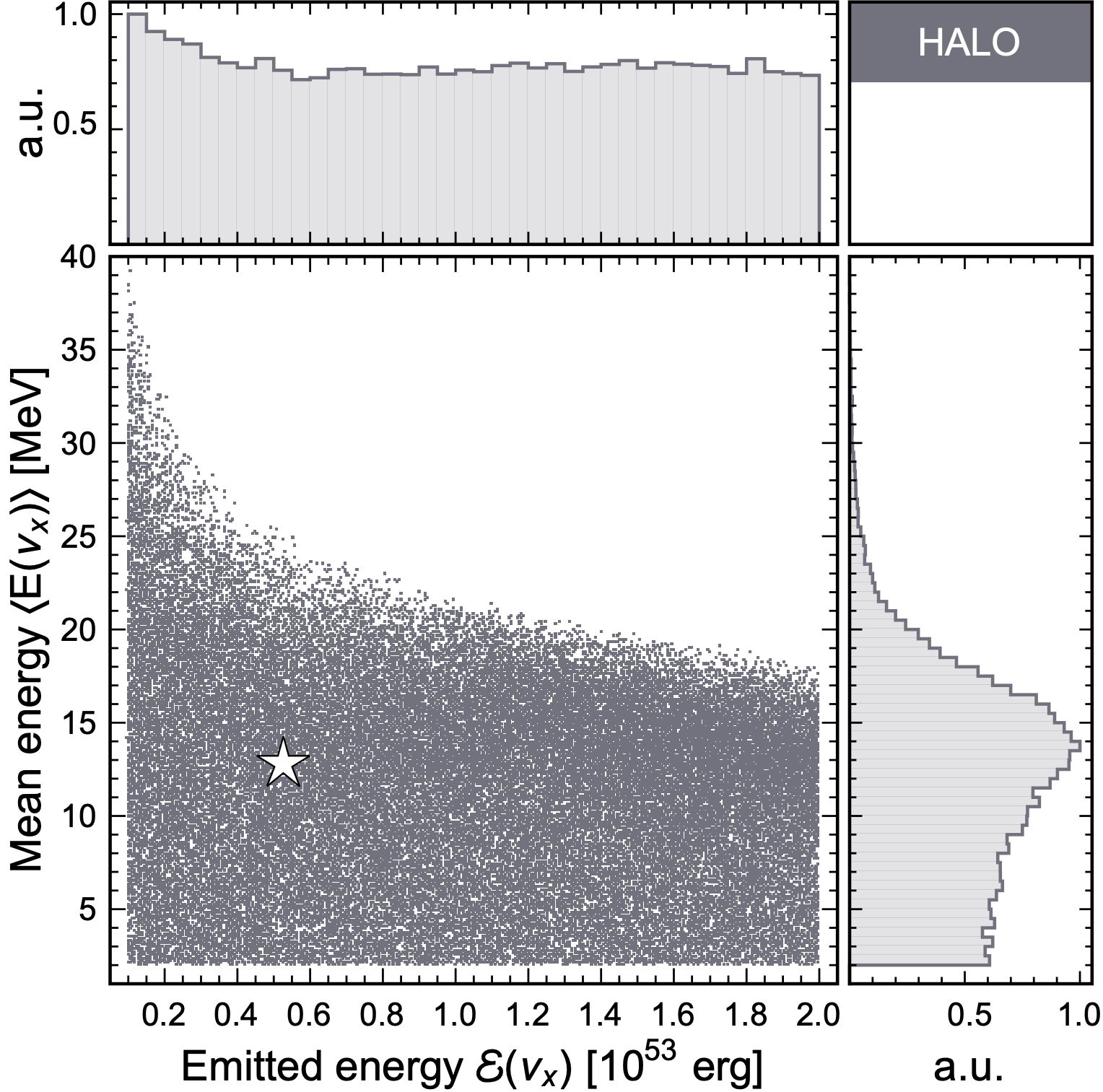}}
	\hfill
	\subfloat[LS220-z9.6co, \Tnux\ species.]
	{\label{fig:2h36}
	\includegraphics[width=.45\textwidth]{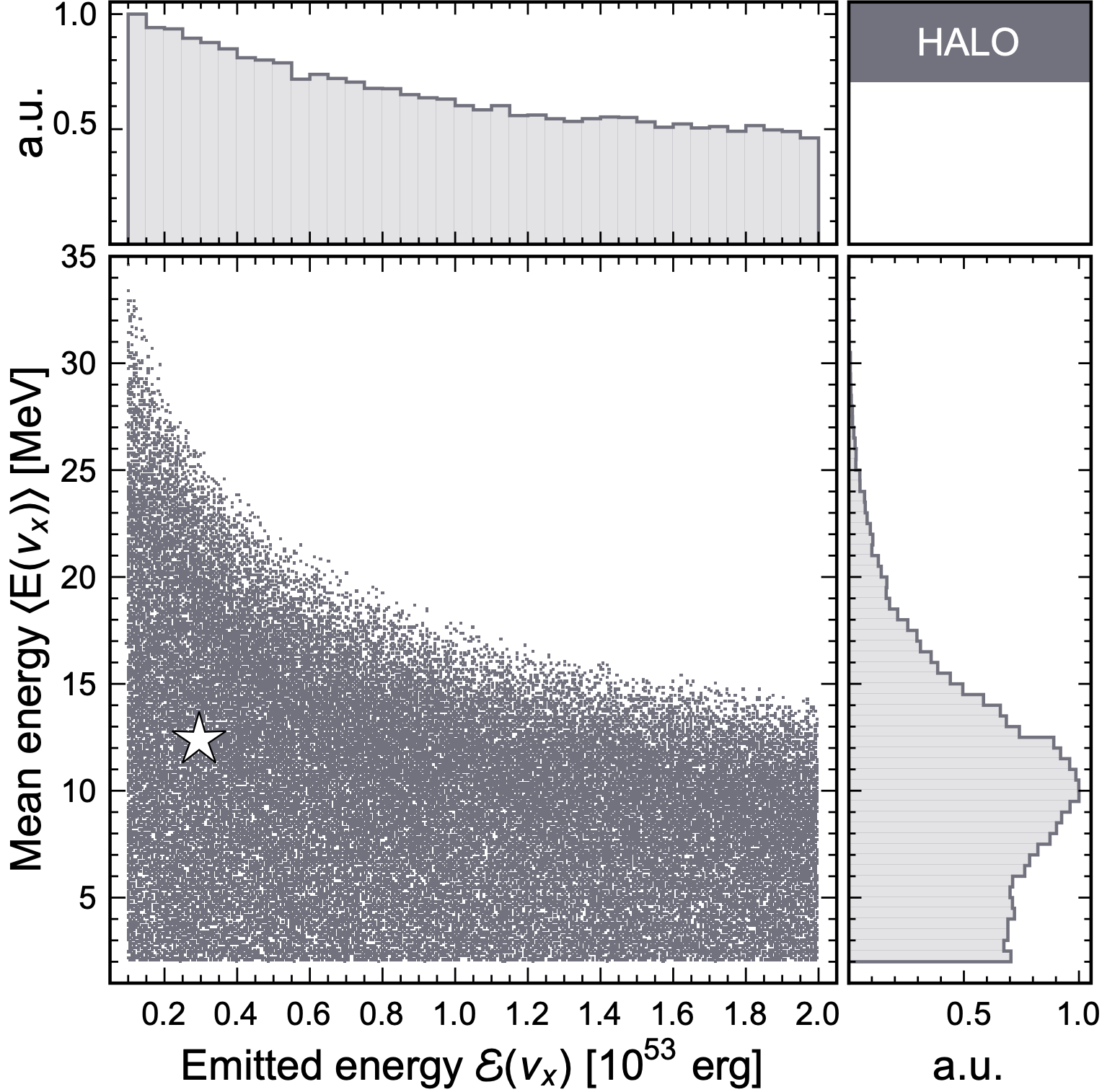}}%
	\\
	\caption{Projections onto $\mathcal{E}(i)$--$
	\langle E(i)\rangle$ planes and probability
	distributions of the reconstructed parameters,
	from the HALO signal alone, given the supernova
	models LS220-s27.0co (left column) and
	LS220-z9.6co (right column). The top (bottom)
	row refers to the \Tnue\ (\Tnux) species. The
	true values are marked by a star.}
	\label{fig:Hsol}
\end{figure}

\section{Results}
\label{sec:res}

Each analysis involves a different combination of
detectors, namely: HALO alone, SK alone, JUNO
alone, the combination HALO+SK, the combination
HALO+JUNO, the combination SK+JUNO, and the
combination HALO+SK+JUNO. We recall that each
analysis involving JUNO has been performed six
times, depending on the different assumptions on
the detection channels (see Table \ref{tab:jcon}).

Each analysis is performed collecting \num{7e4} 
accepted points at $3\sigma$ CL. The results are
quantified by projecting the extracted points onto
selected 2-dimensional scatter plots and
1-dimensional histograms. The parameters defining 
the neutrino spectra are extracted from the
histograms. Their means and standard deviations are
taken as reference values and associated errors. In
the following, we focus our attention exclusively
on \Tnue\ and \Tnux; for a discussion on the
\ATnue\ species see Section \ref{sec:anue}.

\subsection{HALO alone}
\label{sec:HALOalone}

The results of the analyses involving HALO alone
are graphically shown in Figure \ref{fig:Hsol}, and
quantitatively in Table \ref{tab:Hsol}. From those
results, one can see that the only parameters
somewhat constrained by HALO alone are \Tnue\ and
\Tnux\ mean energies, although presenting
huge low-energy tails. Indeed, the percentage
precision for the reconstructed values is about $30
\%-40\%$ for the \Tnue\ species and $\approx 50\%$
for the \Tnux\ one. Even if the accuracies for the
total energies $\mathcal{E}(\Pnue)$ and $\mathcal{
E}(\Pnux)$ are just slightly worse $(50\%-60\%)$,
in that case there is no real improvement over the 
prior. In fact, the standard deviations are
identical to the ones of the flat prior, showing
that no real information is brought up by the data
--- this is also clear from the $\mathcal{E}(i)$ 
distributions in Figure \ref{fig:Hsol}. As we show
in the following, the reconstructing precision of
the $\mathcal{E}(i)$ parameters improves together
with the inclusion of other detection channels.
However, this is not true for the pinching
parameters $\alpha(\Pnue)$ and $\alpha(\Pnux)$. In
fact, their standard deviations hardly improve from
their prior values, even given that all the
detection channels considered in this study; that
is, their distributions are almost flat. For this
reason, the pinching parameters of the two species
\Tnue\ and \Tnux will no be further discussed.

\begin{table}[t!]
	\centering
		\begin{tabular}{r@{\hskip0.15em}l
		r@{\hskip0.15em}c@{\hskip0.15em}l%
		r@{\hskip0.15em}c@{\hskip0.15em}%
		lcc%
		r@{\hskip0.15em}c@{\hskip0.15em}l%
		r@{\hskip0.15em}c@{\hskip0.15em}%
		lc}
			\toprule
			&&\multicolumn{7}{c}{LS220-s27.0co}
			&&\multicolumn{7}{c}{LS220-z9.6co}\\
			\cmidrule{3-9}\cmidrule{11-17}
			&& $(\mu^*$ & $\pm$ &$\sigma^*)$  
			& $\mu$   & $\pm$ &$\sigma$ &\%
			&
			& $(\mu^*$ & $\pm$ &$\sigma^*)$  
			& $\mu$   & $\pm$ &$\sigma$ &\% \\
			\midrule
			$\mathcal{E}(\Pnue)\hphantom{\rangle}$
			& {[\SI{e53}{\erg}]}
			& 0.57 &$\pm$ & 0.55
			& 0.93 &$\pm$ & 0.54 & 58
			&& 0.32 &$\pm$ & 0.55
			& 0.91 &$\pm$ & 0.56 & 62\\
			$\langle E(\Pnue)\rangle$
			& {[\si{\mega\electronvolt}]}
			& 10.8 &$\pm$ & 19.6
			& 9.3 &$\pm$ & 3.2 & 34
			&& 9.9 &$\pm$ & 19.6
			& 6.4 &$\pm$ & 2.7 & 43\\
			$\alpha(\Pnue)\hphantom{\rangle}$ &
			& 2.42 &$\pm$ & 0.87
			& 2.59 &$\pm$ & 0.86 & 33
			&& 2.75 &$\pm$ & 0.87
			& 2.59 &$\pm$ & 0.85 & 33\\
			\cmidrule{3-9}\cmidrule{11-17}
			$\mathcal{E}(\APnue)\hphantom{\rangle}$
			& {[\SI{e53}{\erg}]}
			& 0.57 &$\pm$ & 0.55
			& 0.92 &$\pm$ & 0.57 & 62
			&& 0.34 &$\pm$ & 0.55
			& 0.93 &$\pm$ & 0.56 & 61\\
			$\langle E(\APnue)\rangle$
			& {[\si{\mega\electronvolt}]}
			& 13.6 &$\pm$ & 19.6
			& 17.5 &$\pm$ & 11.9 & 68
			&& 12.3 &$\pm$ & 19.6
			& 15.8 &$\pm$ & 10.3 & 65\\
			$\alpha(\APnue)\hphantom{\rangle}$ &
			& 2.26 &$\pm$ & 0.87
			& 2.58 &$\pm$ & 0.86 & 33
			&& 2.19 &$\pm$ & 0.87
			& 2.58 &$\pm$ & 0.86 & 33\\
			\cmidrule{3-9}\cmidrule{11-17}
			$\mathcal{E}(\Pnux)\hphantom{\rangle}$
			& {[\SI{e53}{\erg}]}
			& 0.53 &$\pm$ & 0.55
			& 1.03 &$\pm$ & 0.56 & 54
			&& 0.30 &$\pm$ & 0.55
			& 0.93 &$\pm$ & 0.55 & 59\\
			$\langle E(\Pnux)\rangle$
			& {[\si{\mega\electronvolt}]}
			& 12.9 &$\pm$ & 19.6
			& 11.7 &$\pm$ & 5.5 & 47
			&& 12.5 &$\pm$ & 19.6
			& 10.2 &$\pm$ & 5.1 & 50\\
			$\alpha(\Pnux)\hphantom{\rangle}$ &
			& 1.85 &$\pm$ & 0.87
			& 2.63 &$\pm$ & 0.85 & 32
			&& 2.46 &$\pm$ & 0.87
			& 2.59 &$\pm$ & 0.86 & 33\\
			\bottomrule
		\end{tabular}
		\caption{Fluence parameters reconstructed
		by HALO-1kT
		alone, for the two models LS220-s27.0co and
		LS220-z9.6co \protect\cite{Mirizzi:2015eza}.
		For each parameter, we report mean $\mu$,
		standard deviation $\sigma$ and percentage
		accuracy of the accepted points at $3\sigma$
		CL. For each model, the column marked with
		$(\mu^*\pm\sigma^*)$ reports the true value
		$\mu^*$ and the standard deviation
		$\sigma^*$ of a flat distribution in
		the assumed priors \protect\eqref{eq:pris},
		\protect\eqref{eq:prie}, 
		\protect\eqref{eq:pria}.
		}
		\label{tab:Hsol}
\end{table}

\subsection{SK and JUNO}
\label{sec:SKJ}

\begin{figure}[t!]
	\centering
	\subfloat[LS220-s27.0co (\Tnue) SK.]
	{\label{fig:1s14}\includegraphics[width=.32%
	\textwidth]{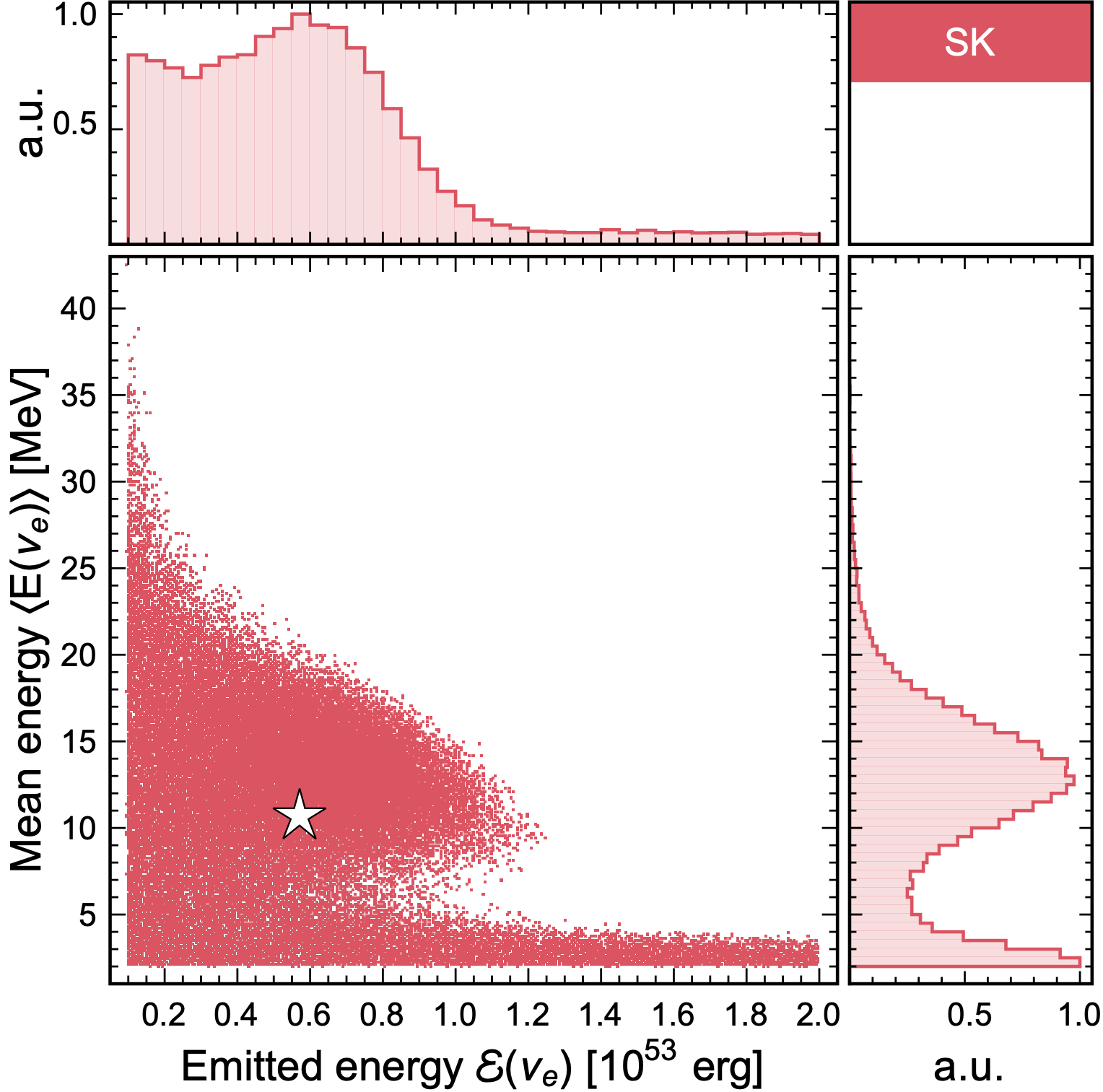}}\hfill
	\subfloat[LS220-s27.0co (\Tnue) JUNO.]
	{\label{fig:1jn14}\includegraphics[width=.32%
	\textwidth]{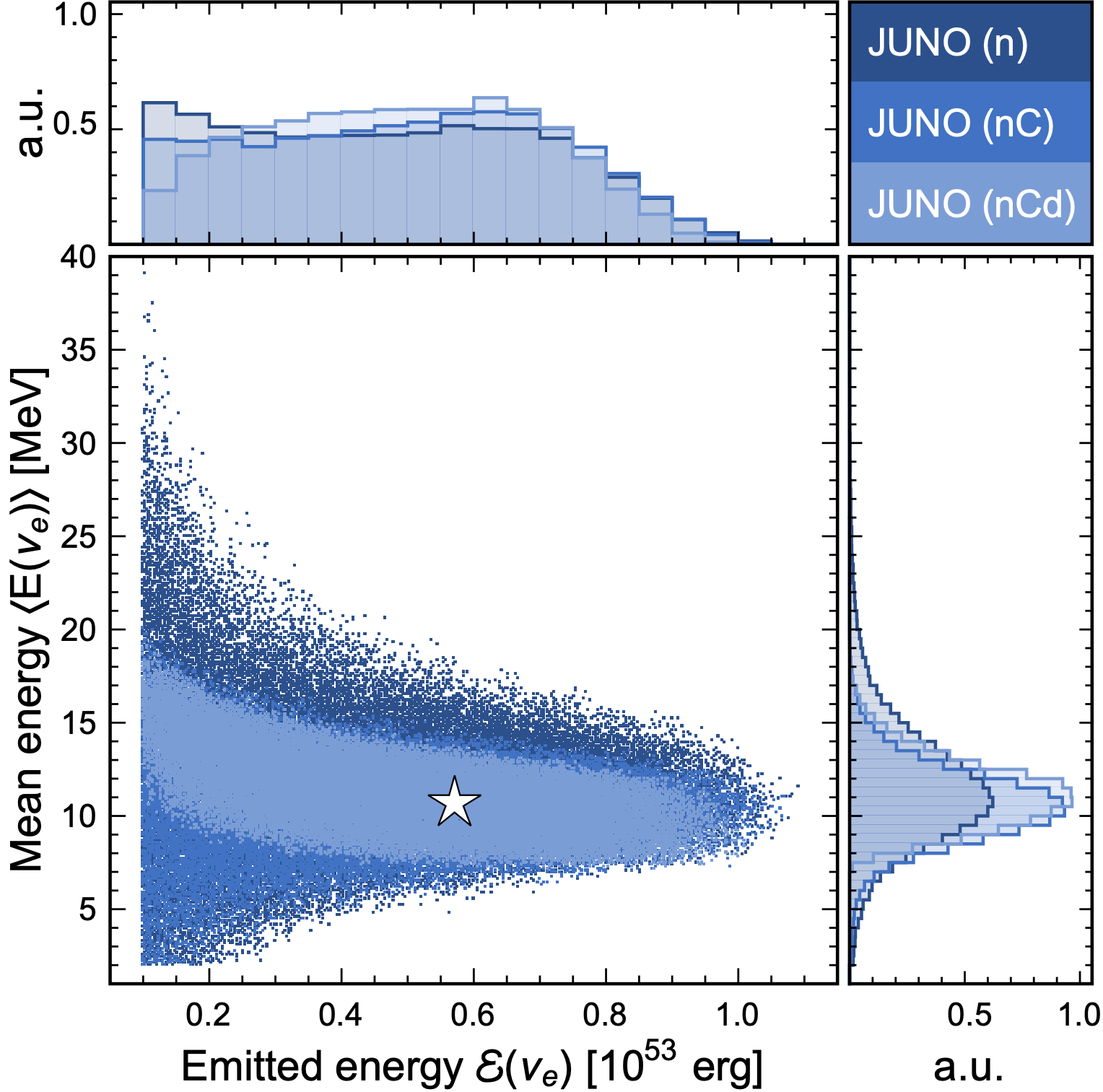}}\hfill
	\subfloat[LS220-s27.0co (\Tnue) JUNO.]
	{\label{fig:1jp14}\includegraphics[width=.32%
	\textwidth]{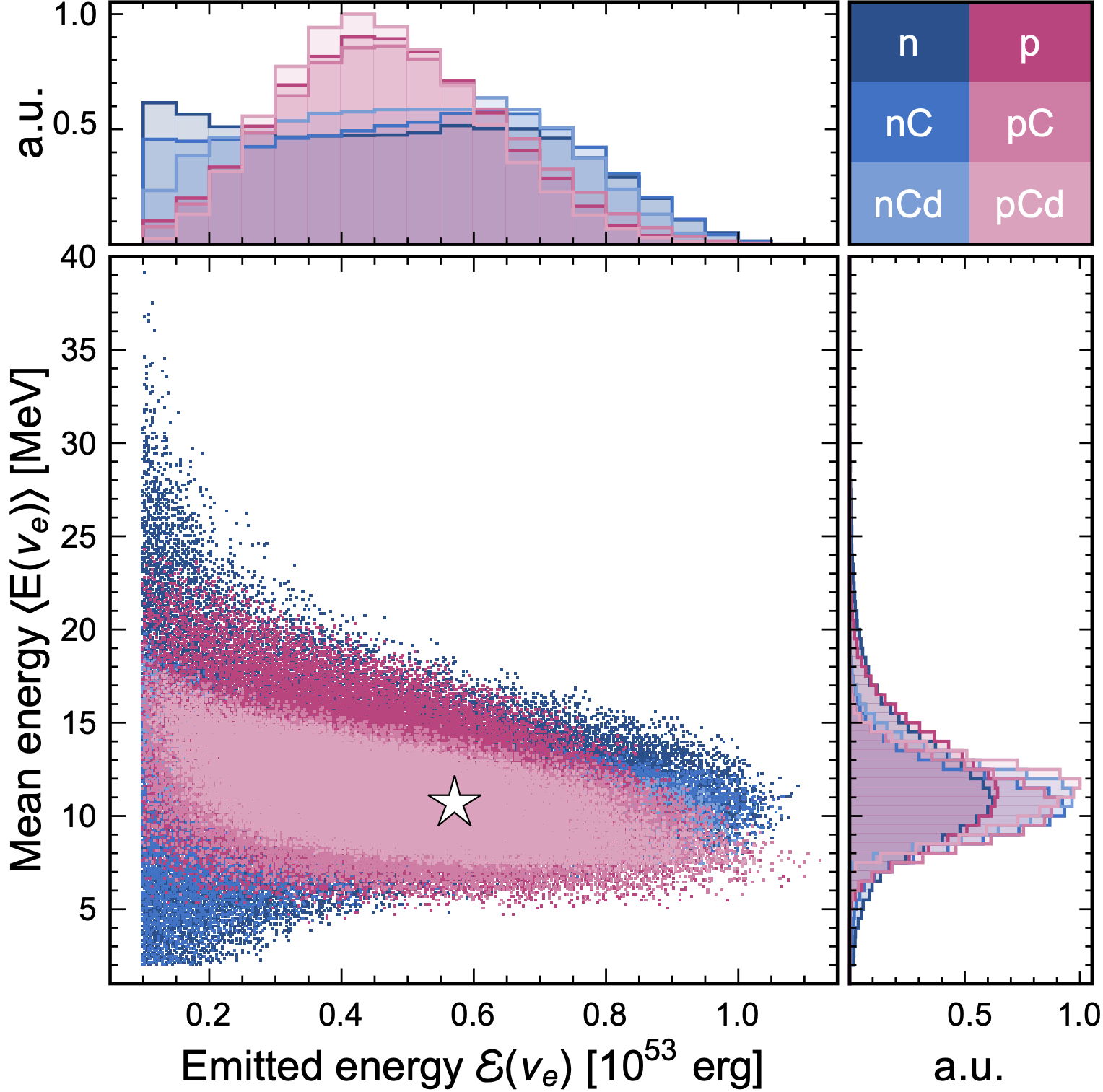}}\\
	\subfloat[LS220-s27.0co, (\Tnux) SK.]
	{\label{fig:1s36}\includegraphics[width=.32%
	\textwidth]{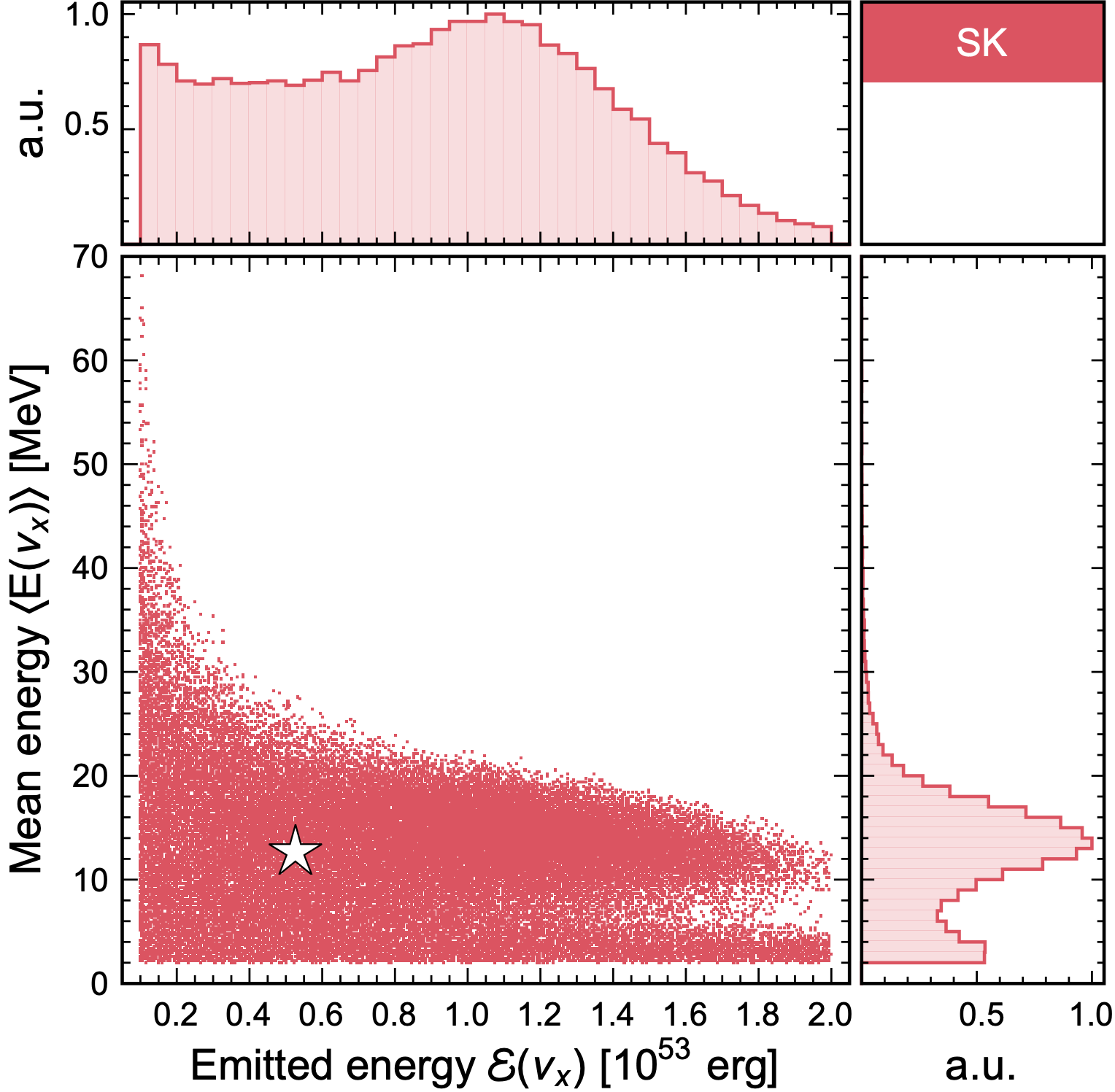}}\hfill
	\subfloat[LS220-s27.0co (\Tnux) JUNO.]
	{\label{fig:1jn36}\includegraphics[width=.32%
	\textwidth]{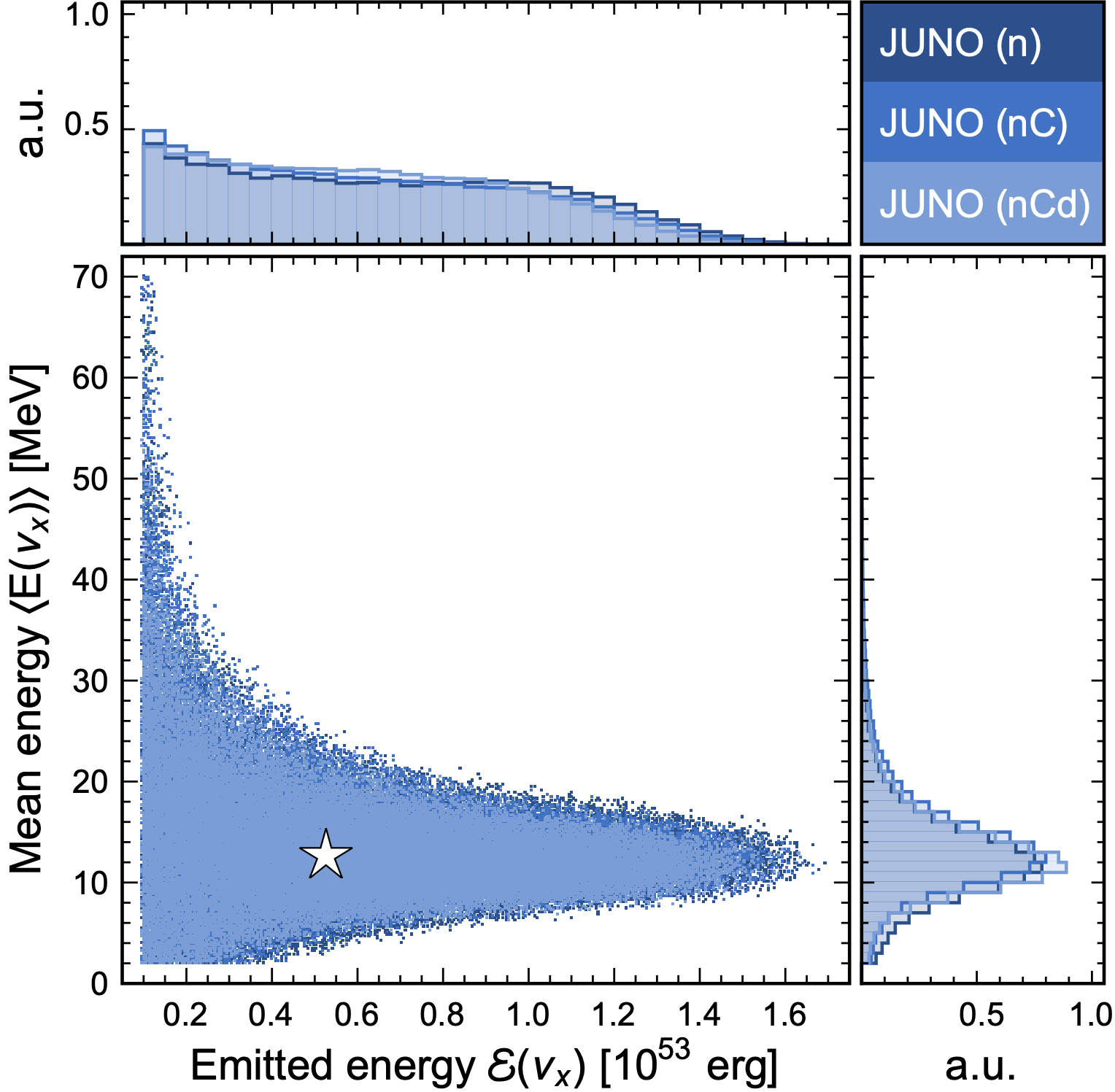}}\hfill
	\subfloat[LS220-s27.0co (\Tnux) JUNO.]
	{\label{fig:1jp36}\includegraphics[width=.32%
	\textwidth]{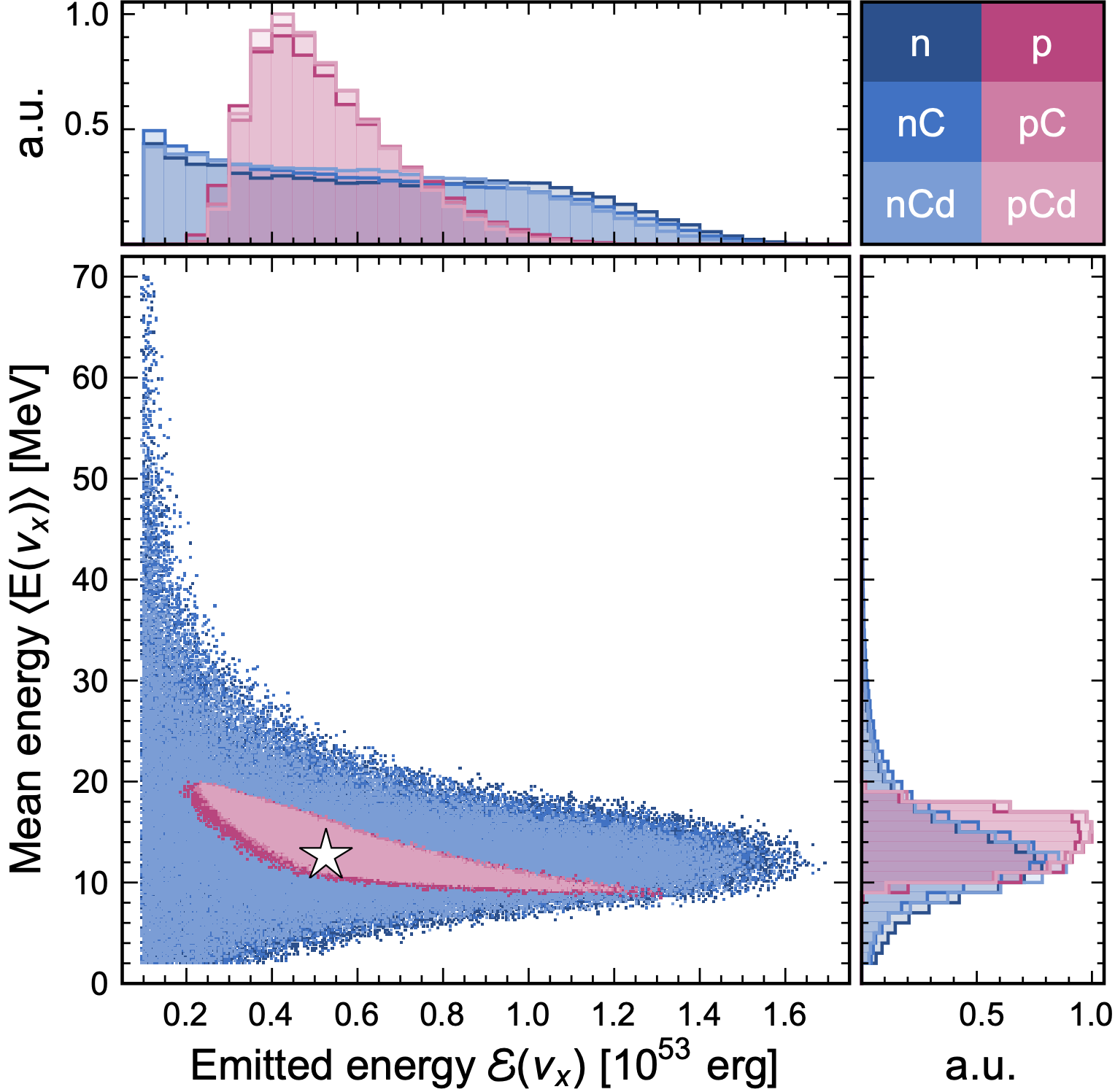}}\\
	\caption{Given the supernova model
	LS220-s27.0co \protect\cite{Mirizzi:2015eza}
	at $D=\SI{10}{\kilo\parsec}$,
	these are the projections onto $\mathcal{E}
	(i)$--$\langle E(i)\rangle$ planes and
	probability distributions of the reconstructed
	parameters for the \Tnue\ (top) and \Tnux\ 
	(bottom) species. The results given by
	Super-Kamiokande (SK) alone are shown in the
	first column in red. In the other two, the blue
	shades represent the results by JUNO when the
	pES channel is not implemented (n/nC/nCd) while
	the pink ones are when it is included
	(p/pC/pCd). For the meaning of the labels see
	Section \protect\ref{sec:JUNO} and Table
	\ref{tab:jcon}. The true values are marked by a
	star.}
	\label{fig:SJsol27}
\end{figure}

\begin{figure}[t!]
	\centering
	\subfloat[LS220-z9.6co (\Tnue) SK.]
	{\label{fig:2s14}\includegraphics[width=.32%
	\textwidth]{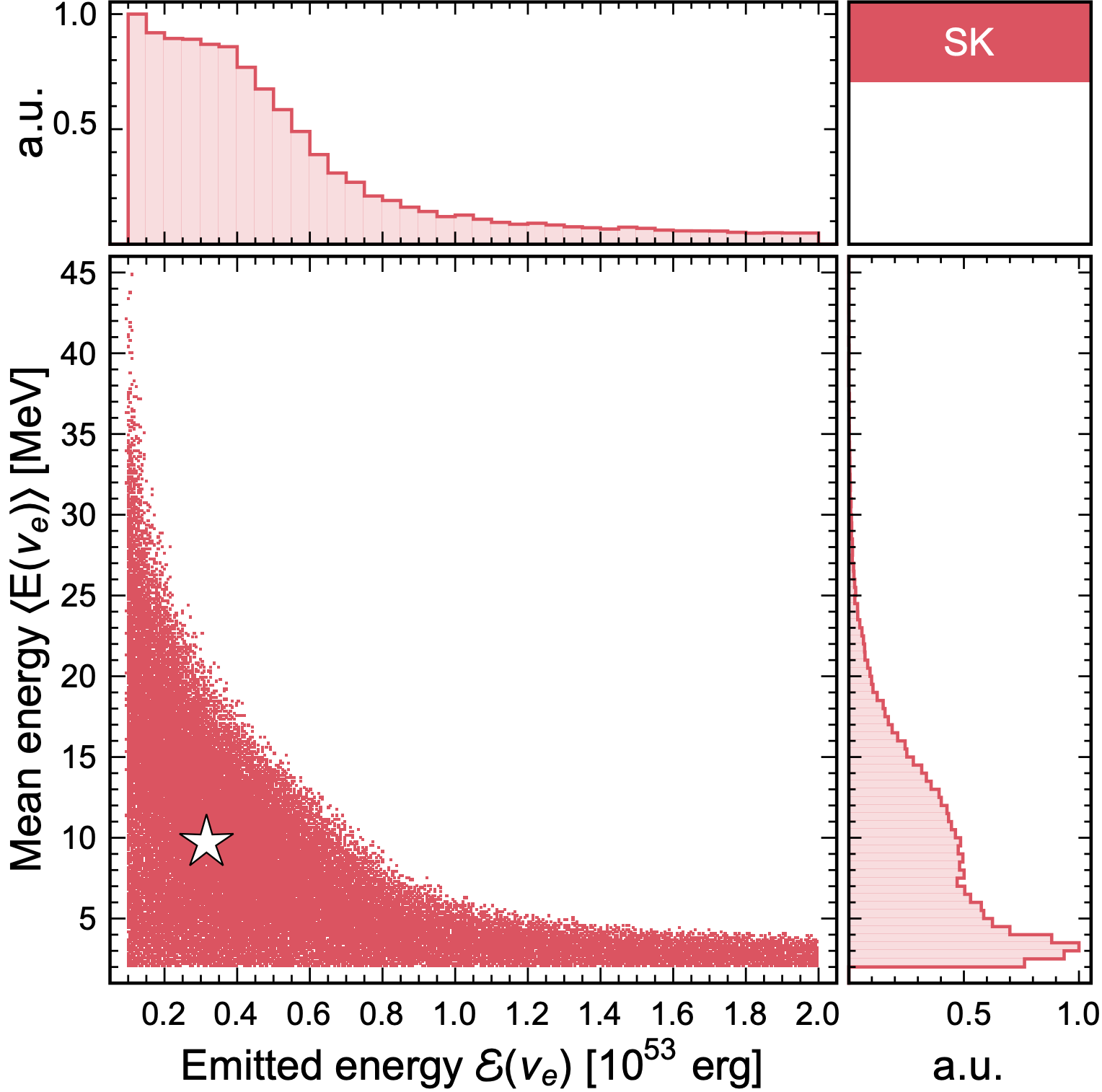}}\hfill
	\subfloat[LS220-z9.6co (\Tnue) JUNO.]
	{\label{fig:2jn14}\includegraphics[width=.32%
	\textwidth]{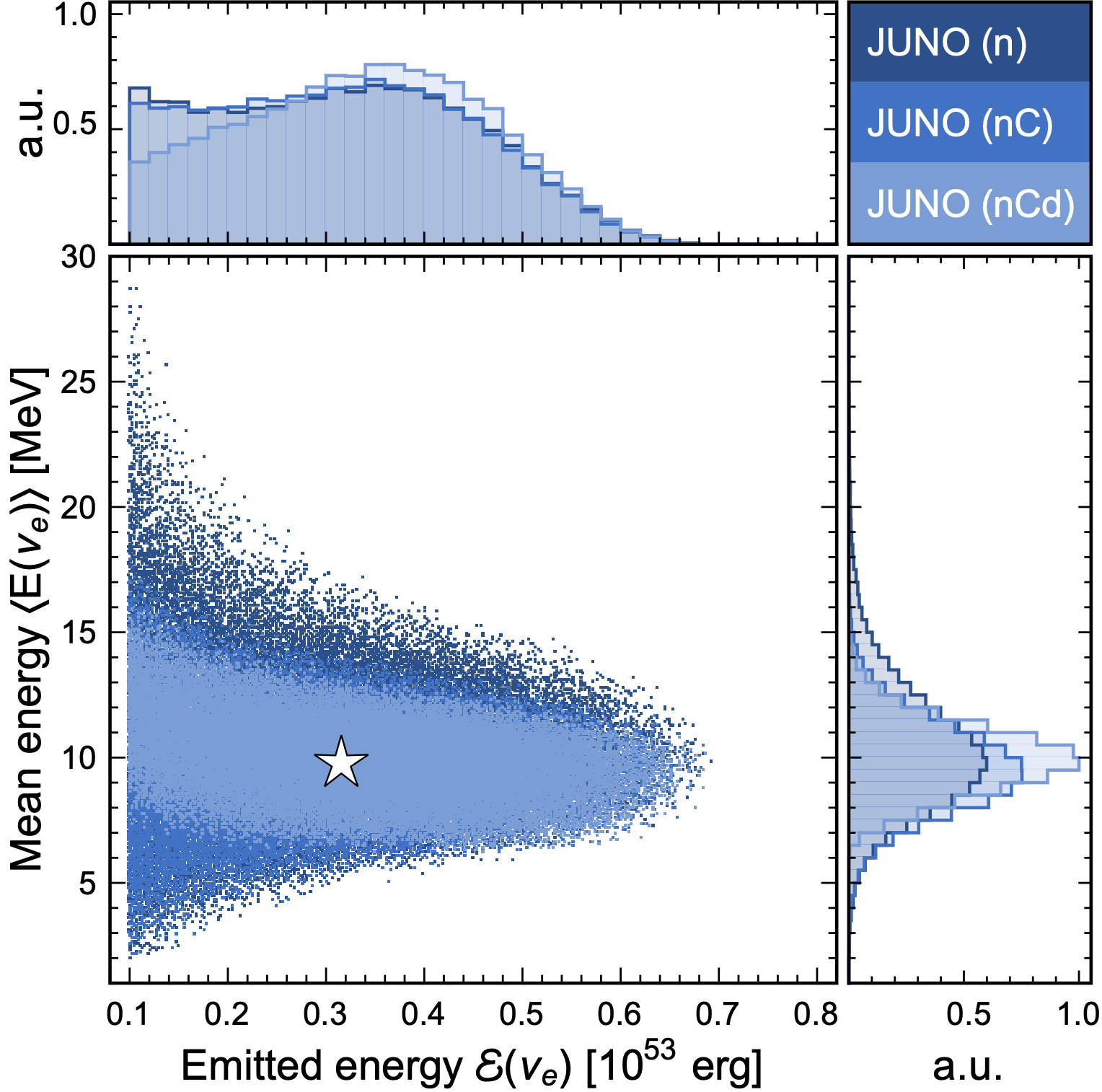}}\hfill
	\subfloat[LS220-z9.6co (\Tnue) JUNO.]
	{\label{fig:2jp14}\includegraphics[width=.32%
	\textwidth]{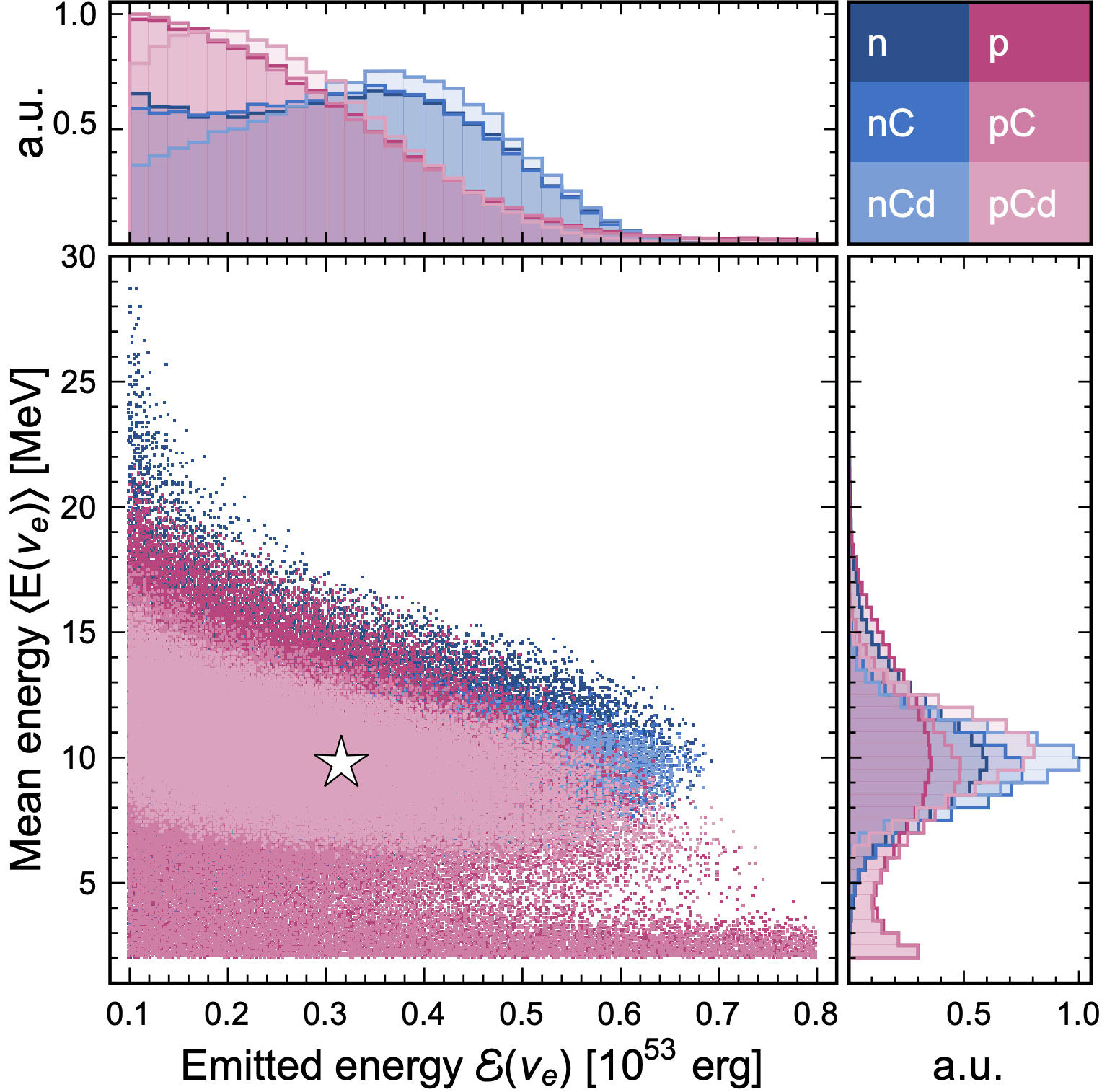}}\\
	\subfloat[LS220-z9.6co (\Tnux) SK.]
	{\label{fig:2s36}\includegraphics[width=.32%
	\textwidth]{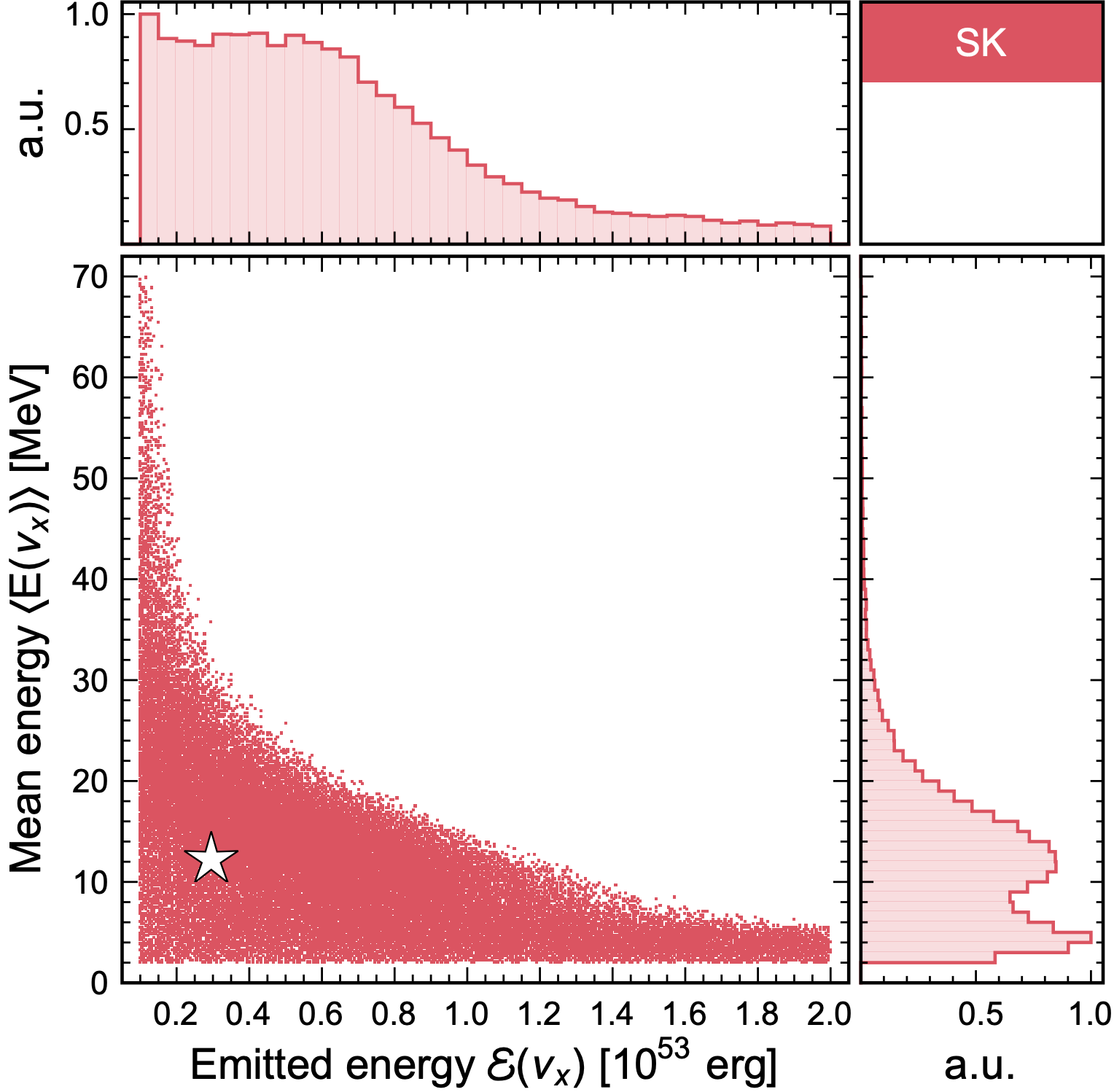}}\hfill
	\subfloat[LS220-z9.6co (\Tnux) JUNO.]
	{\label{fig:2jn36}\includegraphics[width=.32%
	\textwidth]{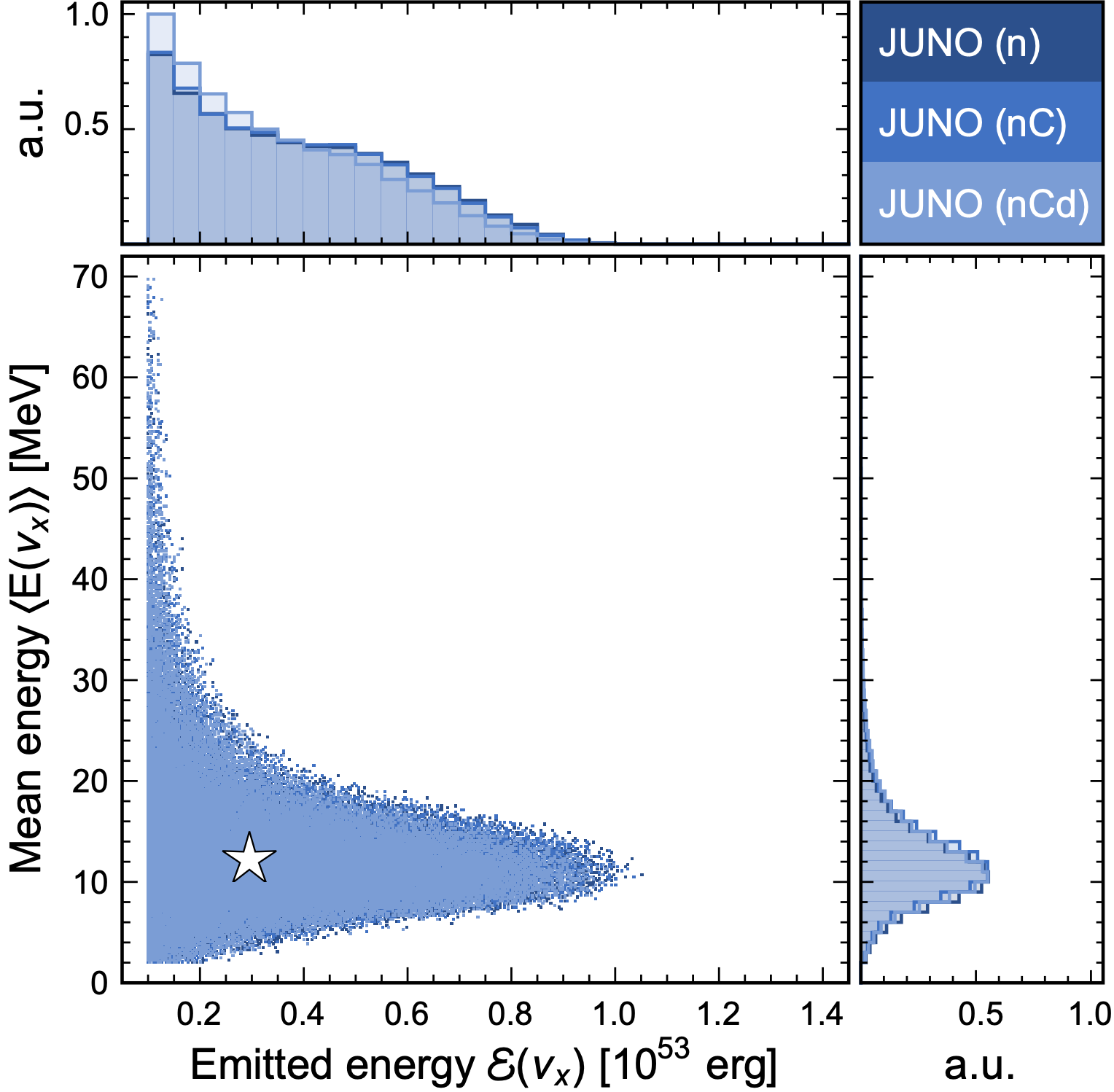}}\hfill
	\subfloat[LS220-z9.6co (\Tnux) JUNO.]
	{\label{fig:2jp36}\includegraphics[width=.32%
	\textwidth]{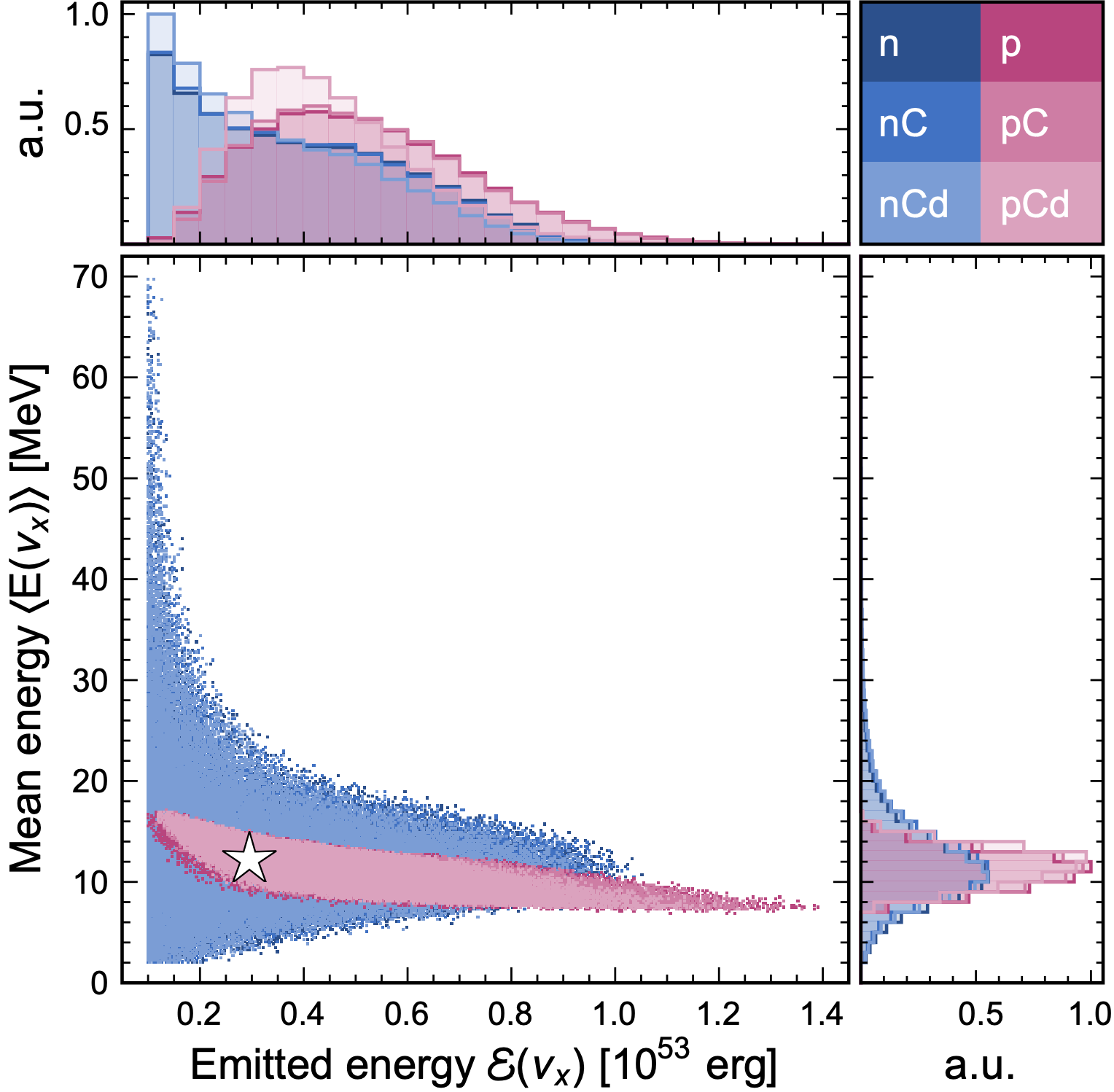}}\\
	\caption{Given the supernova model
	LS220-z9.6co \protect\cite{Mirizzi:2015eza}
	at $D=\SI{10}{\kilo\parsec}$,
	these are the projections onto $\mathcal{E}
	(i)$--$\langle E(i)\rangle$ planes and
	probability distributions of the reconstructed
	parameters for the \Tnue\ (top) and \Tnux\ 
	(bottom) species. The results given by
	Super-Kamiokande (SK) alone are shown in the
	first column in red. In the other two, the blue
	shades represent the results by JUNO when the
	pES channel is not implemented (n/nC/nCd) while
	the pink ones are when it is included
	(p/pC/pCd). For the meaning of the labels see
	Section \protect\ref{sec:JUNO} and Table
	\ref{tab:jcon}. Note that in figure
	\protect\subref{fig:2jp14} the tail of the
	$\mathcal{E}(\Pnue)$ distribution extends up to
	the edge of the prior, namely \SI{2e53}{\erg}.
	The true values are marked by a star.}
	\label{fig:SJsol96}
\end{figure}

Results for the reconstruction of the \Tnue\ and
\Tnux\ parameters from SK and JUNO alone are shown
graphically in Figures \ref{fig:SJsol27} and
\ref{fig:SJsol96}, and quantitatively in Tables
\ref{tab:risex} and \ref{tab:risey}, for models
LS220-s27.0co and LS220-z9.6co respectively. As
already mentioned, the pinching parameters $\alpha(
\Pnue)$ and $\alpha(\Pnux)$ are omitted since their
distributions are always nearly flat within the
prior, no matter what combination of detectors and
channels is considered.

Like HALO, SK does not put a clear constraint
on the $\mathcal{E}(\Pnue)$ and $\mathcal{E}(\Pnux
)$ parameters either, even if it does a slightly
better job than HALO for both models. Indeed, its
distributions present a lower standard deviation
than HALO (i.e., they are more peaked) even if
there  still are huge tails extending to the edge
of the prior, and the reconstructed
distributions somehow flatten at its low-energy
edge. JUNO, on the other hand, gives the best
results on the two emitted energies $\mathcal{
E}_{\uPnue}$, $\mathcal{E}_{\uPnux}$. The
precisions vary with the model and the different
assumptions on the detector, in the raw range $50
\%-30\%$ for \Tnue\ and $60\%-30\%$ for \Tnux.
Moreover, what emerges, in general, is the
importance of the low-energy events that make up
the pES channel. This is understandable, since the
low-energy events are very sensitive to the
normalization of the fluences.

Concerning the mean energies $\langle E(\Pnue)
\rangle$ and $\langle E(\Pnux)\rangle$, the
accuracy in the reconstruction for Super-Kamiokande
is quite poor and comparable with the one of HALO.
On the contrary, JUNO has always the best accuracy,
especially when the CC interactions on \tsup{12}C
are implemented. Depending on the species and the
model, it can reach up to $\sim 20\%$, especially
when more channels are included.

\subsection{The inclusion of HALO}
\label{sec:Hinclus}

\begin{figure}[t!]
	\centering
	\subfloat[SK and HALO.]
	{\label{fig:1sh14}\includegraphics[width=.32%
	\textwidth]{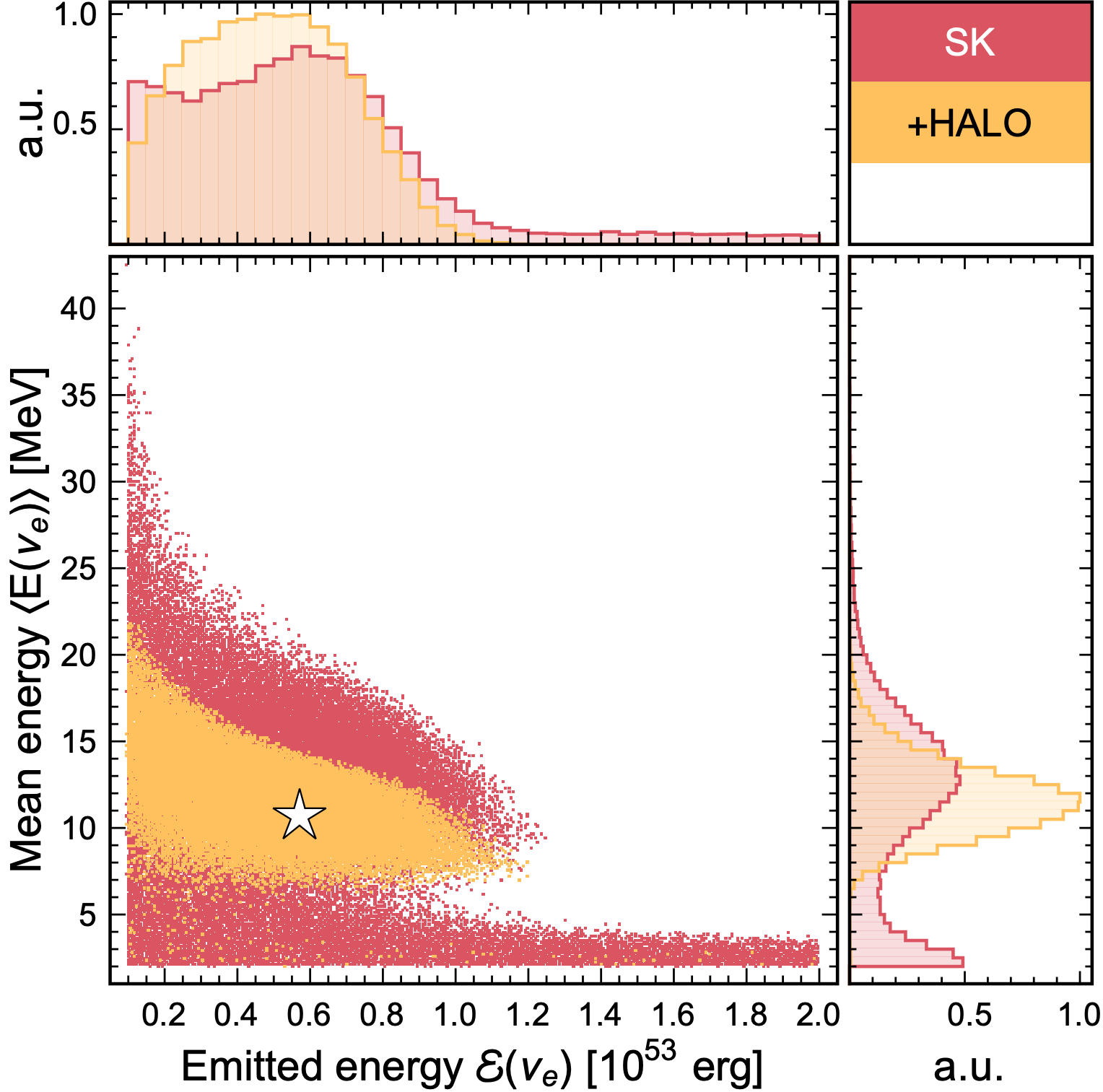}}\hfill
	\subfloat[JUNO (n) and HALO.]
	{\label{fig:1j0h14}\includegraphics[width=.32%
	\textwidth]{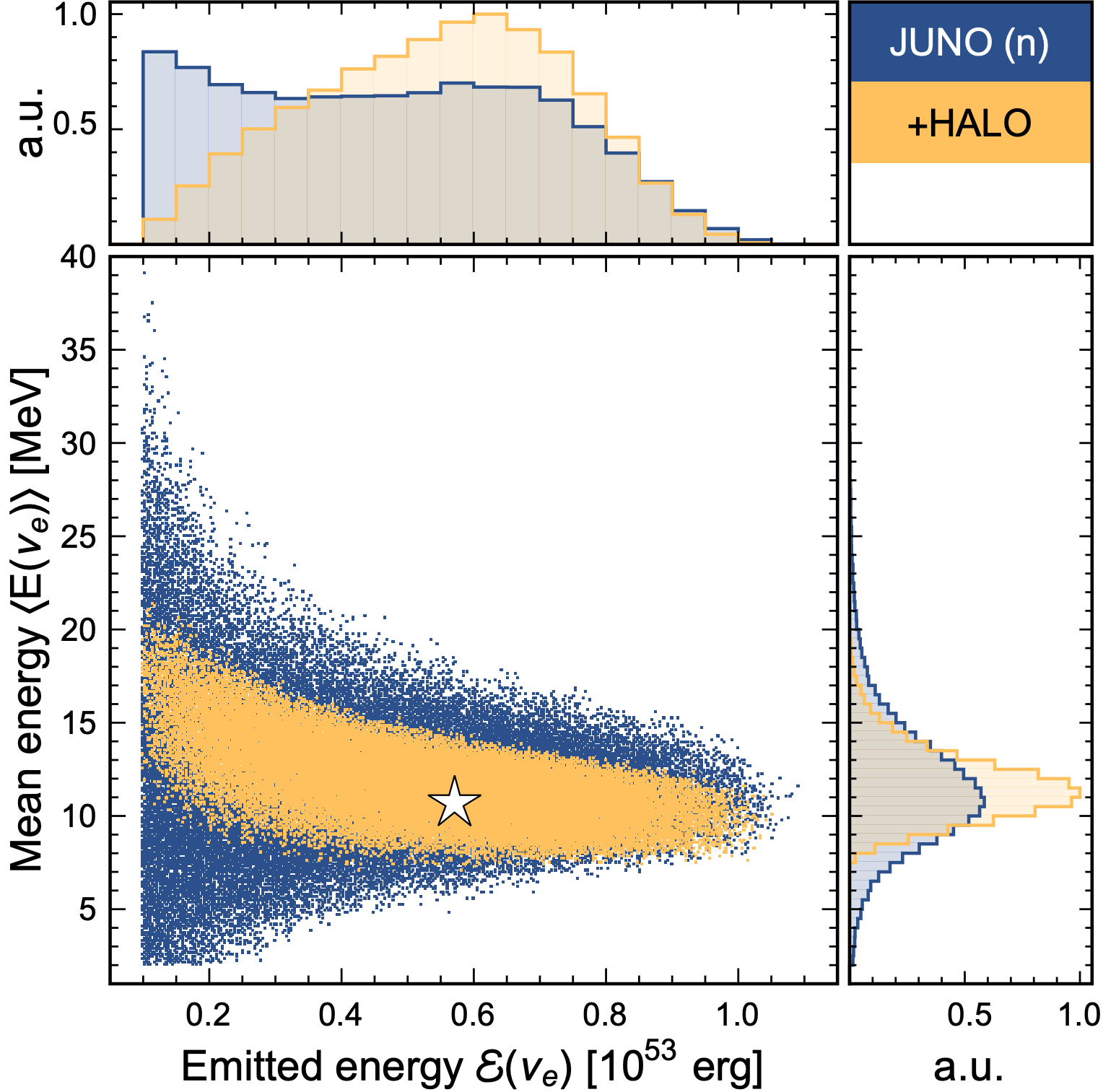}}\hfill
	\subfloat[JUNO (nC) and HALO.]
	{\label{fig:1j1h14}\includegraphics[width=.32%
	\textwidth]{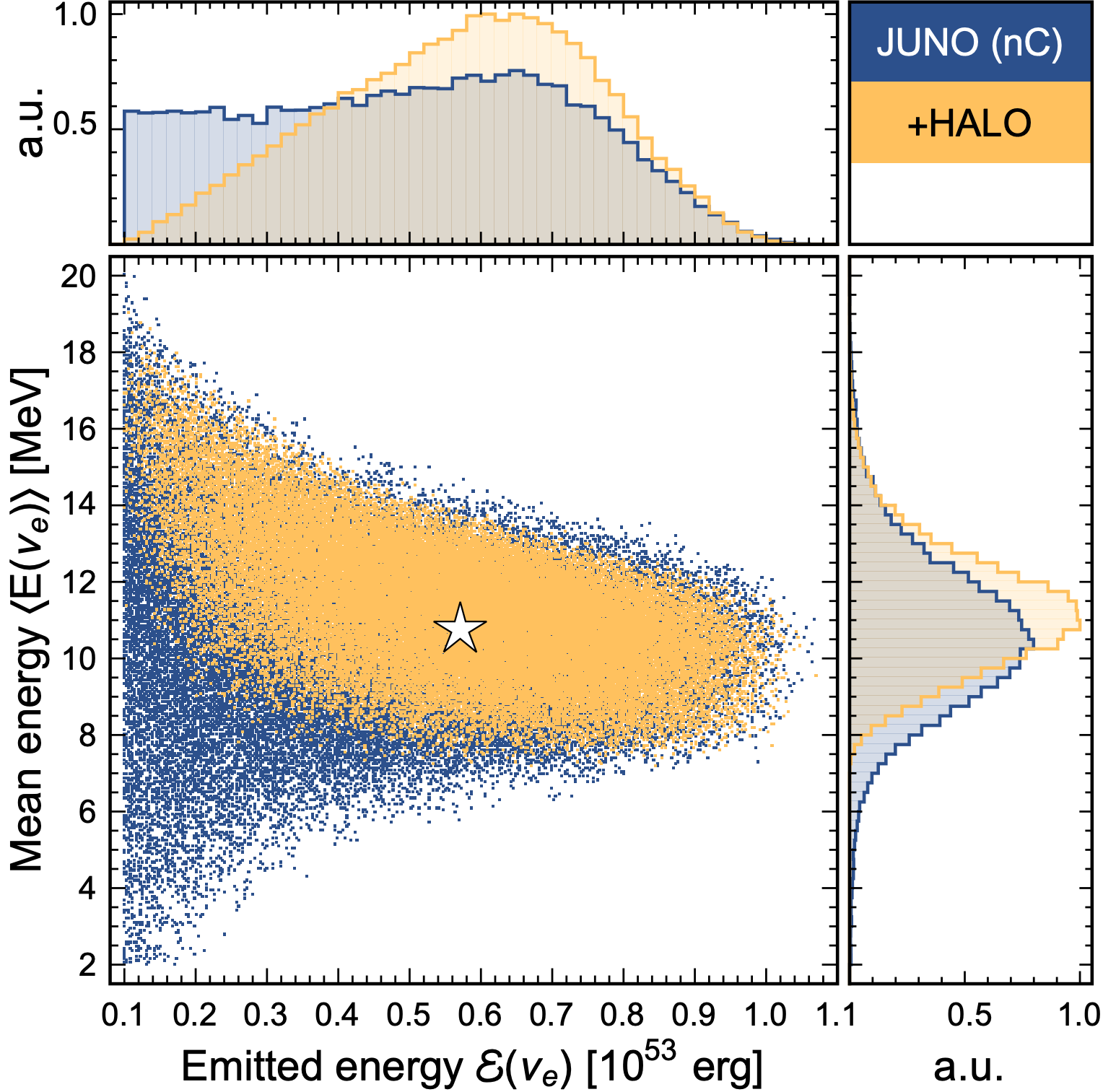}}\\
	\subfloat[JUNO (nCd) and HALO.]
	{\label{fig:1j2h14}\includegraphics[width=.32%
	\textwidth]{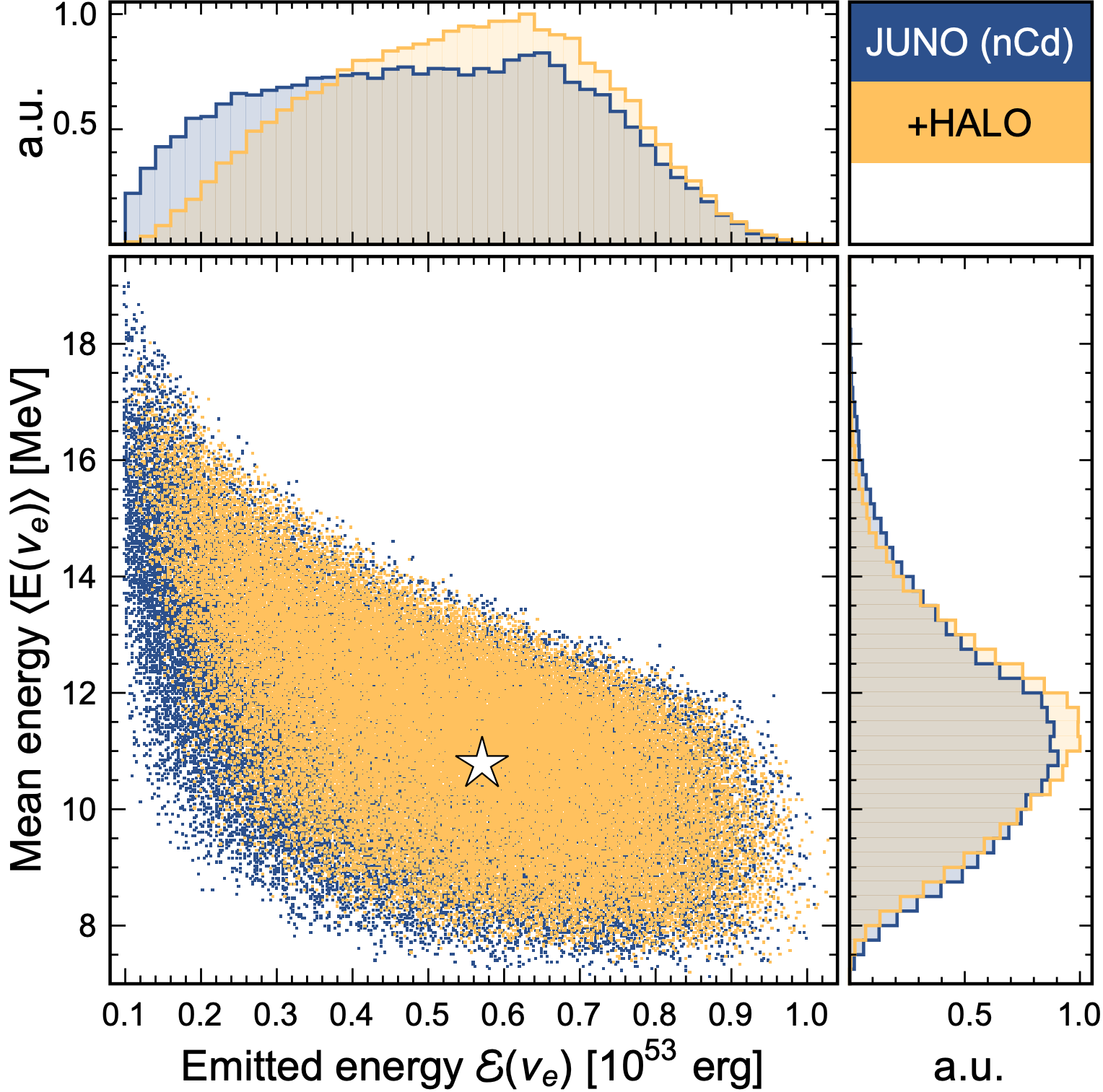}}\hfill
	\subfloat[JUNO (p) and HALO.]
	{\label{fig:1j3h14}\includegraphics[width=.32%
	\textwidth]{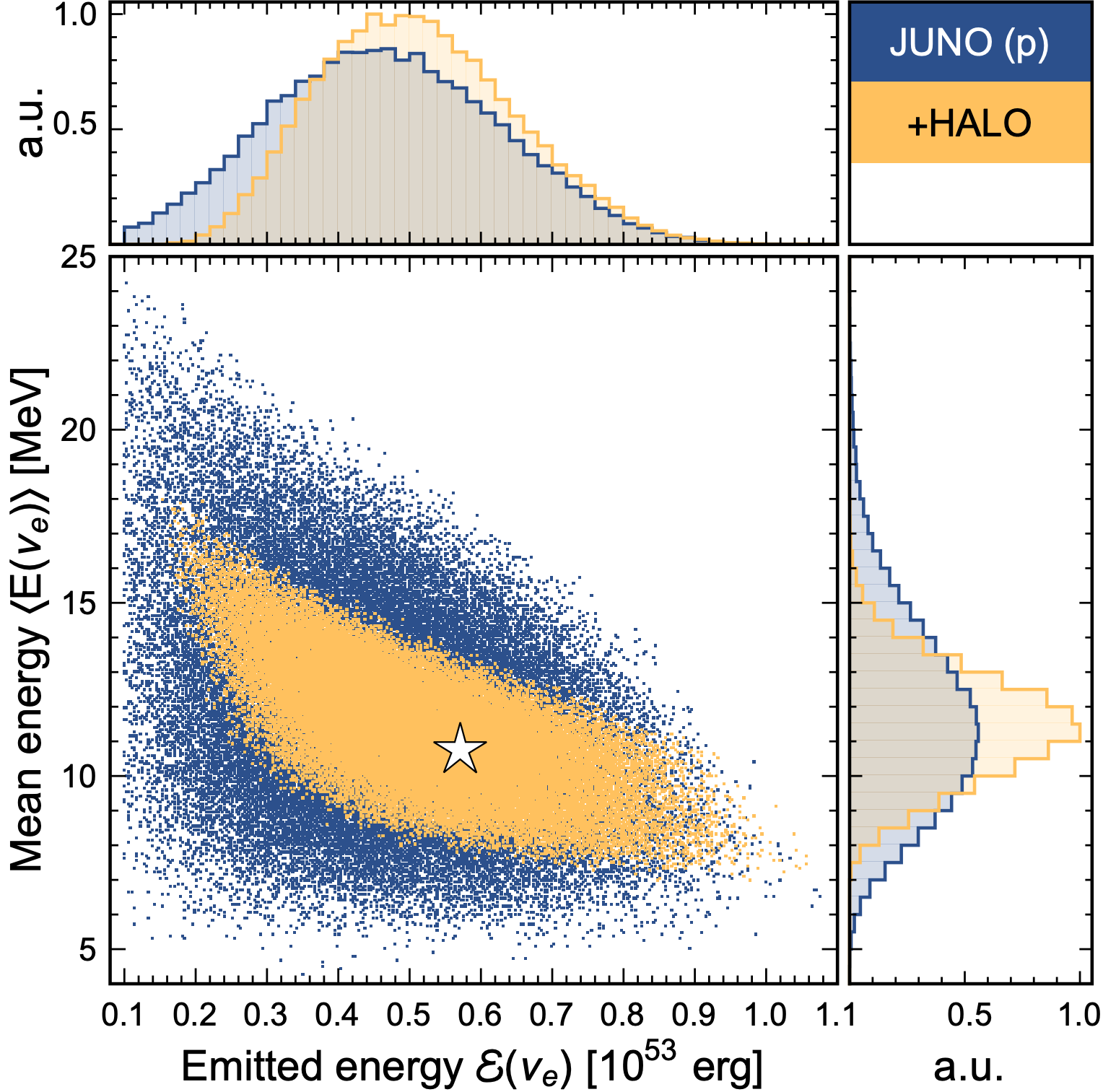}}\hfill
	\subfloat[JUNO (pC) and HALO.]
	{\label{fig:1j4h14}\includegraphics[width=.32%
	\textwidth]{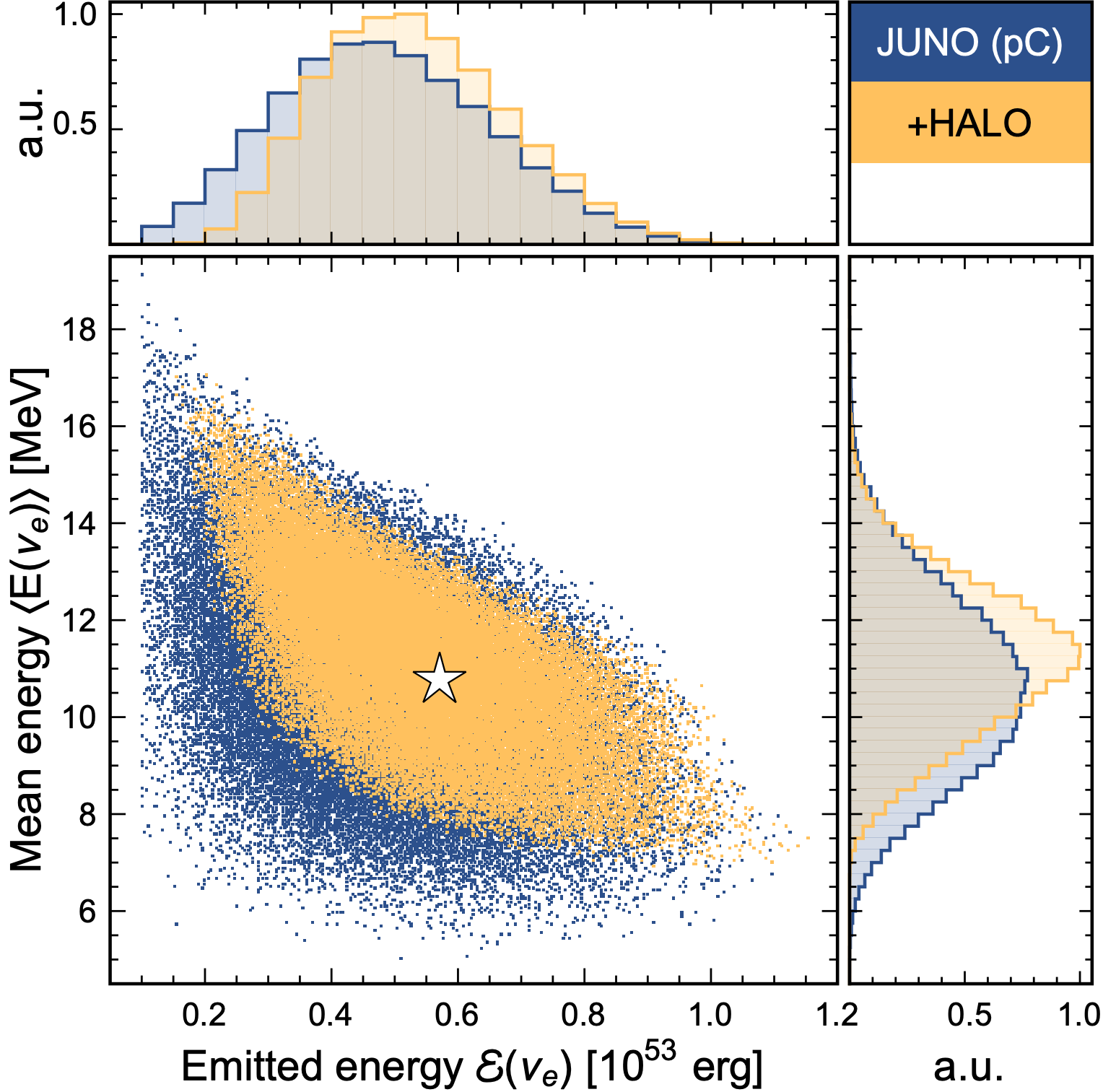}}\\
	\caption{Projections onto $\mathcal{E}(\Pnue)
	$--$\langle E(\Pnue)\rangle$ planes and
	probability distributions, given the supernova
	model LS220-s27.0co \protect\cite{%
	Mirizzi:2015eza}. In all the four panels, the
	orange region shows the results of HALO
	combined with SK \protect\subref{fig:1sh14},
	JUNO in the
	n configuration \protect\subref{fig:1j0h14},
	nC configuration \protect\subref{fig:1j1h14},
	nCd configuration \protect\subref{fig:1j2h14},
	p configuration \protect\subref{fig:1j3h14},
	pC configuration \protect\subref{fig:1j4h14}.
	For the meaning of the JUNO labels see section
	\protect\ref{sec:JUNO} and table
	\ref{tab:jcon}. The true values are marked by
	a star.}
	\label{fig:SJeH27}
\end{figure}

\begin{figure}[t!]
	\centering
	\subfloat[SK and HALO.]
	{\label{fig:2sh14}\includegraphics[width=.32%
	\textwidth]{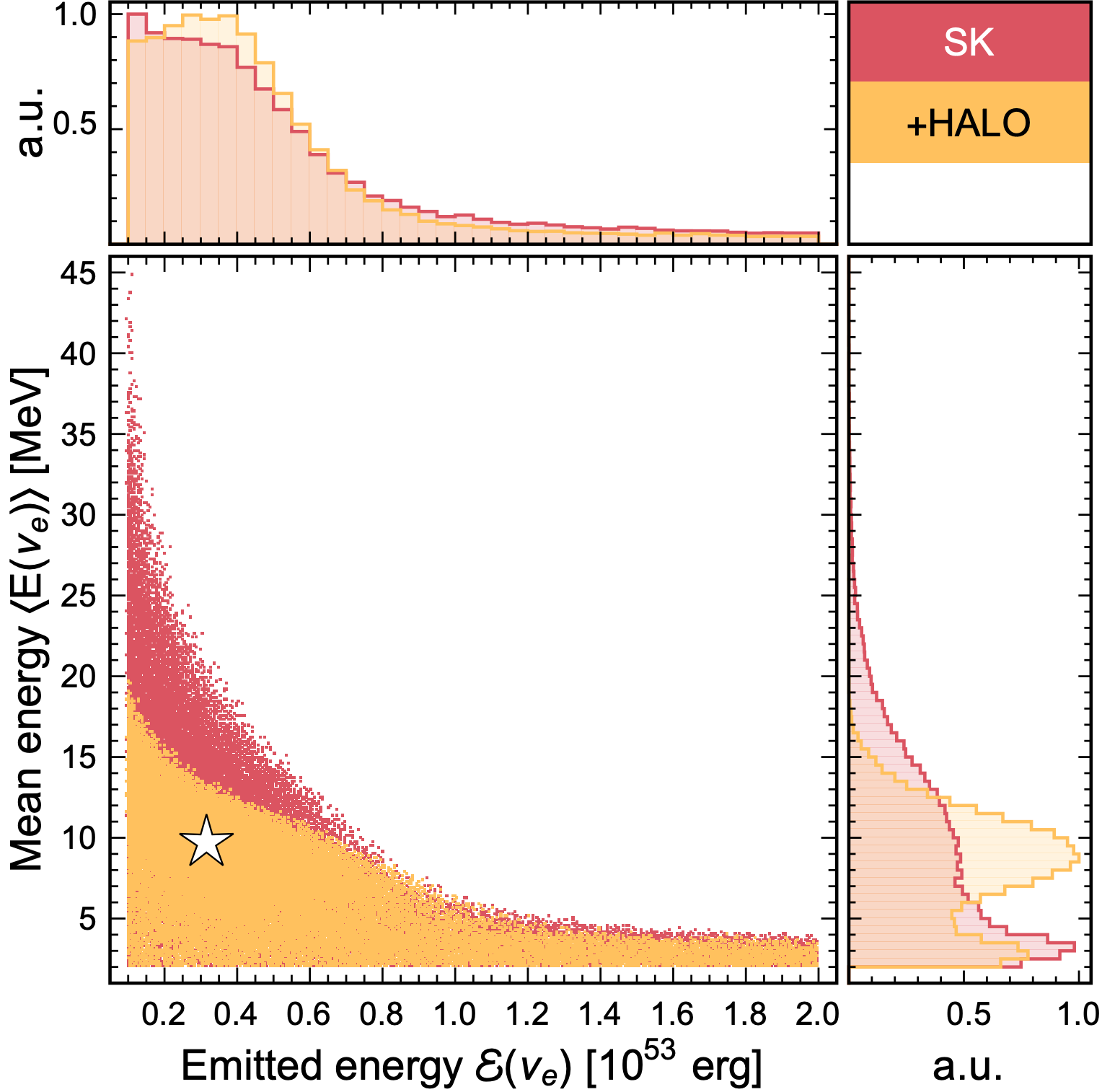}}\hfill
	\subfloat[JUNO (n) and HALO.]
	{\label{fig:2j0h14}\includegraphics[width=.32%
	\textwidth]{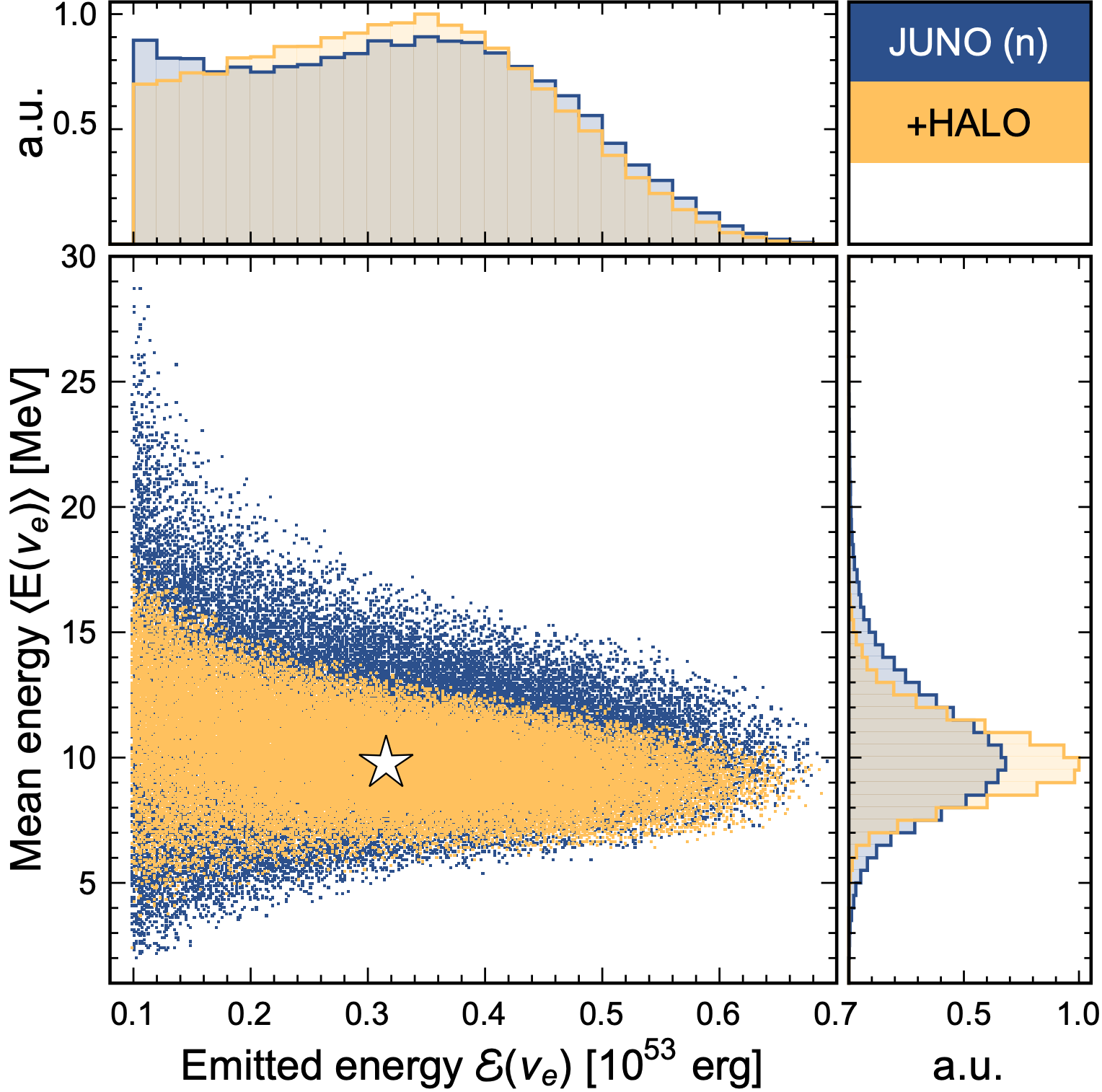}}\hfill
	\subfloat[JUNO (nC) and HALO.]
	{\label{fig:2j1h14}\includegraphics[width=.32%
	\textwidth]{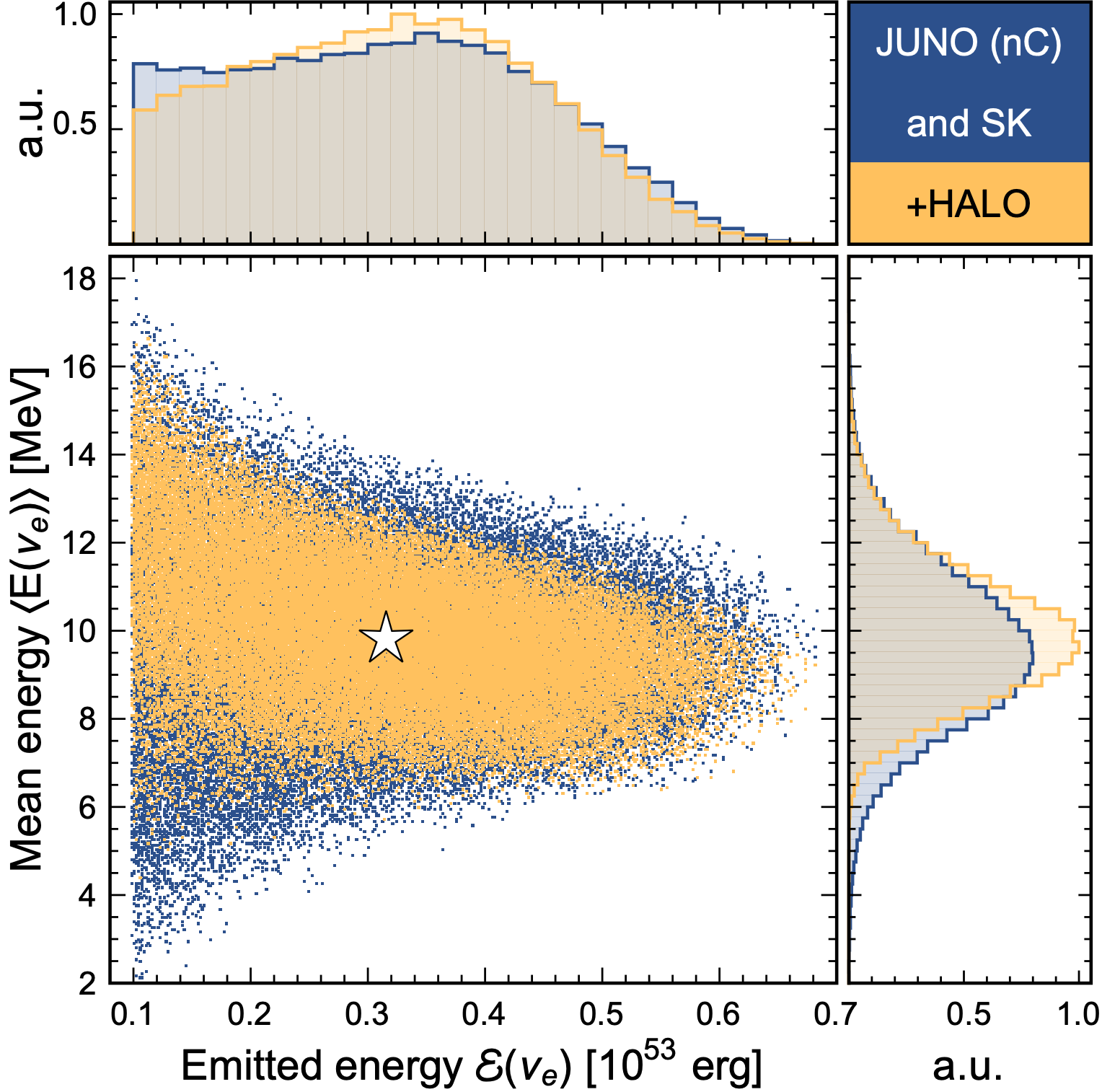}}\\
	\subfloat[JUNO (nCd) and HALO.]
	{\label{fig:2j2h14}\includegraphics[width=.32%
	\textwidth]{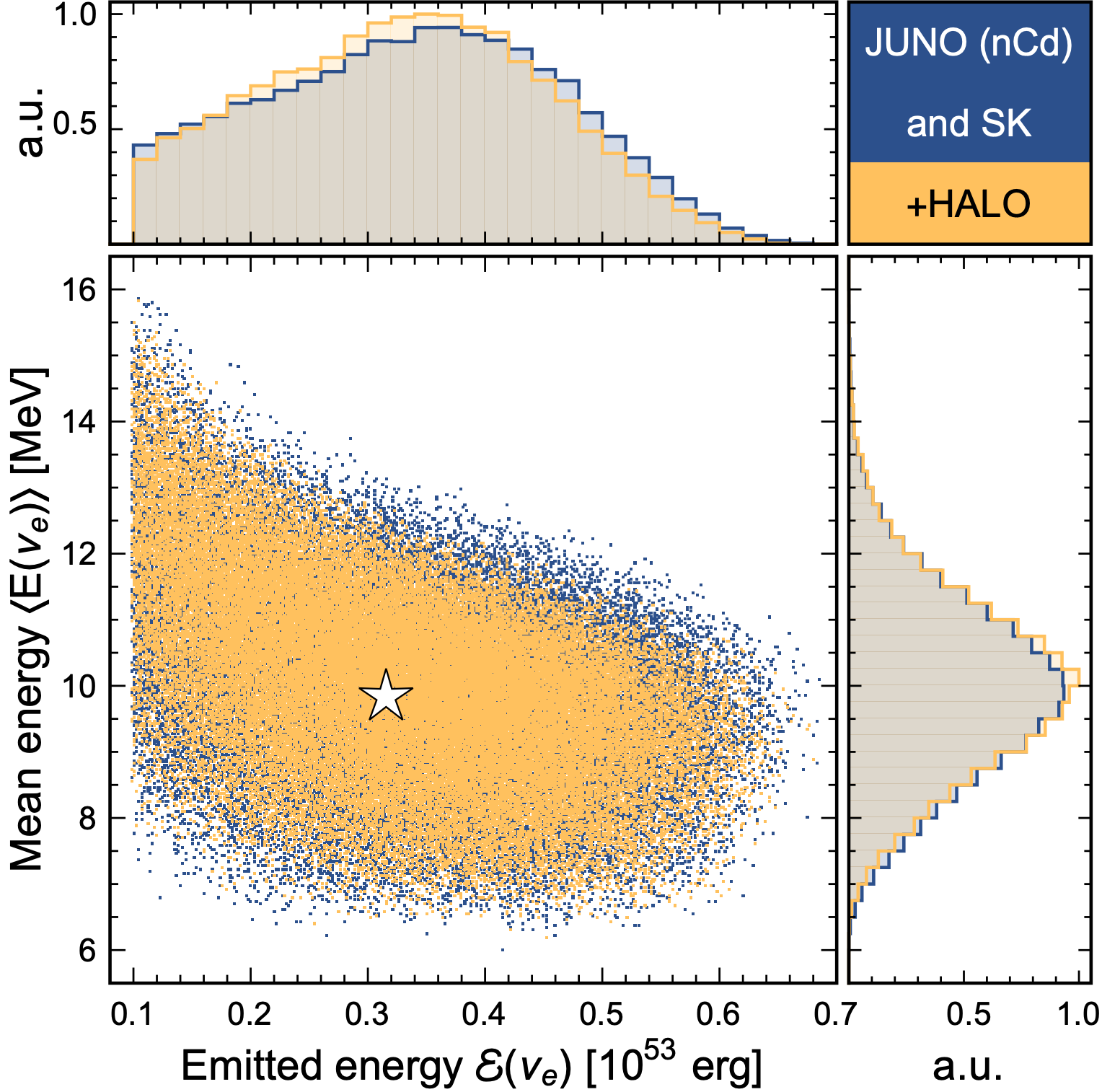}}\hfill
	\subfloat[JUNO (p) and HALO.]
	{\label{fig:2j3h14}\includegraphics[width=.32%
	\textwidth]{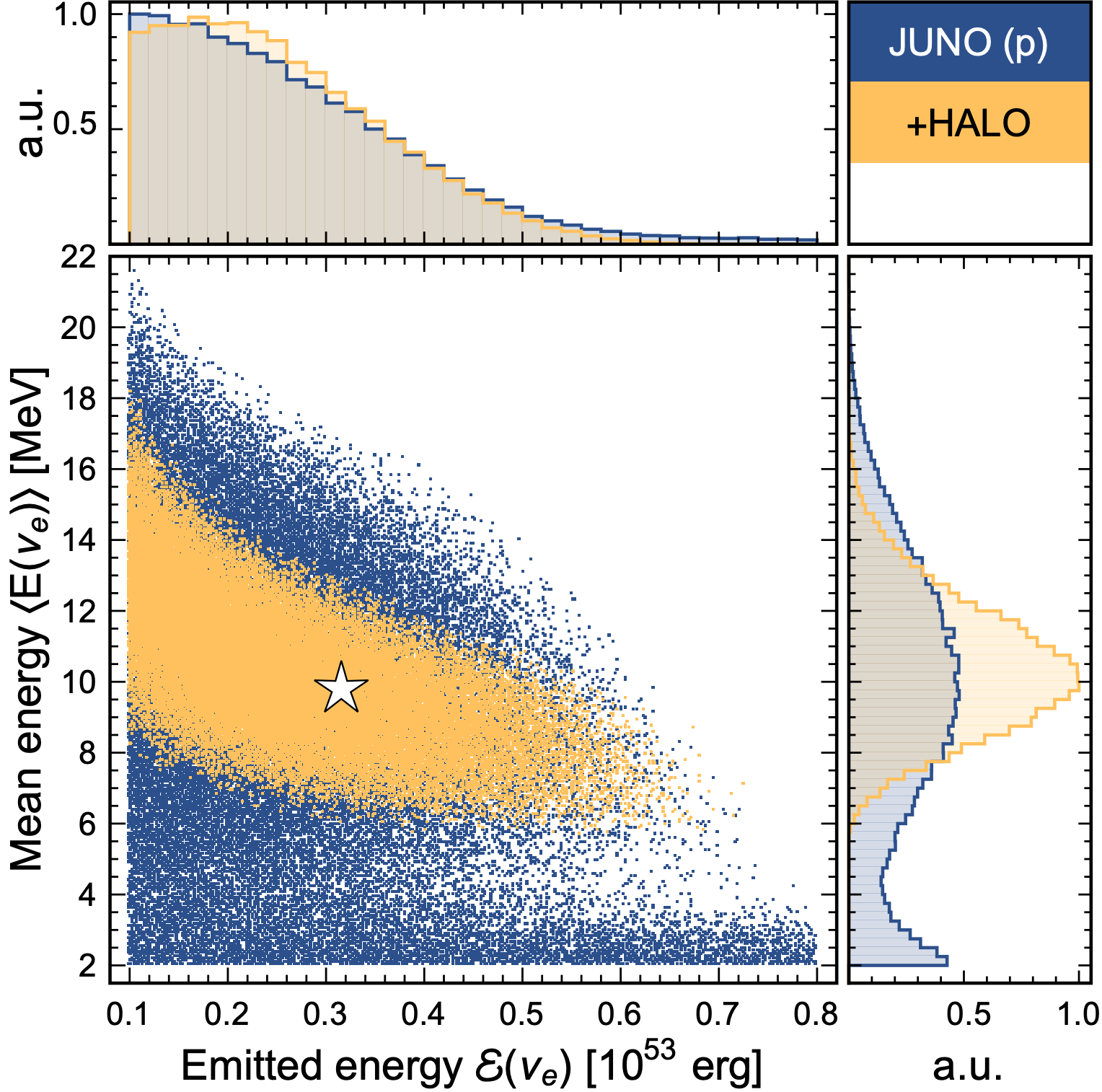}}\hfill
	\subfloat[JUNO (pC) and HALO.]
	{\label{fig:2j4h14}\includegraphics[width=.32%
	\textwidth]{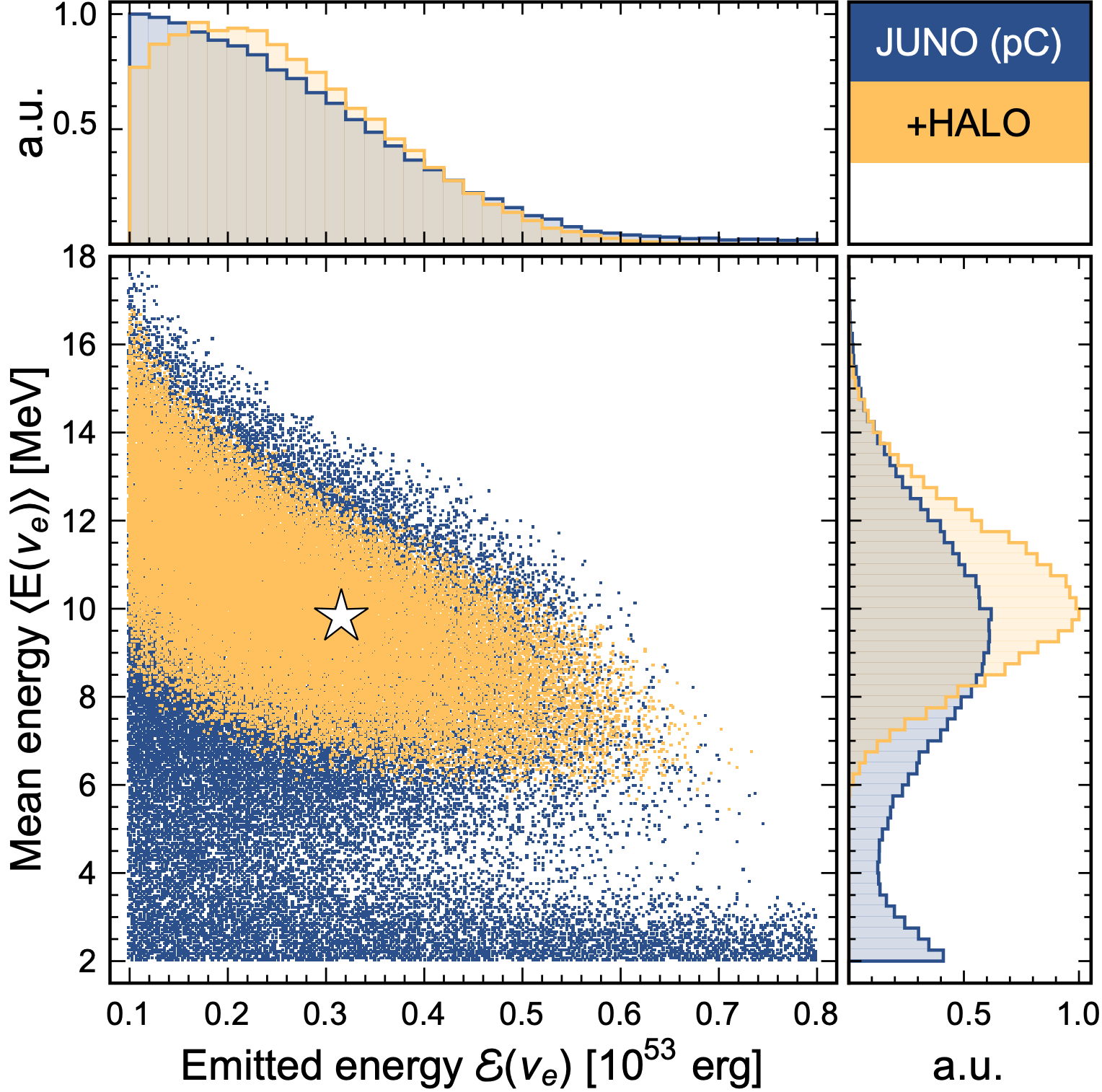}}\\
	\caption{Projections onto $\mathcal{E}(\Pnue)
	$--$\langle E(\Pnue)\rangle$ planes and
	probability distributions, given the supernova
	model LS220-z9.6co \protect\cite{%
	Mirizzi:2015eza}. In all the four panels, the
	orange region shows the results of HALO
	combined with SK \protect\subref{fig:2sh14},
	JUNO in the
	n configuration \protect\subref{fig:2j0h14},
	nC configuration \protect\subref{fig:2j1h14},
	nCd configuration \protect\subref{fig:2j2h14},
	p configuration \protect\subref{fig:2j3h14},
	pC configuration \protect\subref{fig:2j4h14}.
	For the meaning of the JUNO labels see section
	\protect\ref{sec:JUNO} and table
	\ref{tab:jcon}. Note that in figure
	\protect\subref{fig:2j3h14} and
	\protect\subref{fig:2j4h14} the tail of the
	$\mathcal{E}(\Pnue)$ distribution extends up to
	the edge of the prior, namely \SI{2e53}{\erg}.
	The true values are marked by a star.}
	\label{fig:SJeH96}
\end{figure}

The reconstruction of the \Tnue\ species once HALO
is added to SK or JUNO is visually shown in Figures 
\ref{fig:SJeH27} and \ref{fig:SJeH96}, and in
Tables \ref{tab:risex} and \ref{tab:risey}, for
models LS220-s27.0co and LS220-z9.6co respectively.

What clearly emerges from Figures \ref{fig:1sh14}
and \ref{fig:2sh14} is that HALO improves 
the determination of the $\langle E(\Pnue)\rangle$
parameter done by SK. In fact, the precision
improves down to $\approx 20\%$ ($\approx 40\%$)
for LS220-s27.0co (LS220-z9.6co), that is better
than either HALO ($30\%-40\%$) or SK ($40\%-60\%$)
taken alone. This is true for the emitted energy $
\mathcal{E}(\Pnue)$ too, although just for the
LS220-s27.0co model, going from $\approx 60\%$ for
SK and HALO to $\approx 40$ in the combined
analyses. The same applies to \Tnux\ species, 
although to a lesser extent. In general, the
resulting distributions are more peaked and the
uncertainty reduced, as e.g.\ for $\langle E(\Pnux)
\rangle$ in LS220-s27.0co model, going from
$\approx 50\%$ to $\approx 35\%$ accuracy.

The inclusion of HALO data to JUNO gives mixed
results, depending on the number of channels
considered. Concerning the \Tnue\ mean energy, HALO
improves once again the results, especially when
the \tsup{12}C-CC events are not implemented
or not clearly distinguishable from
\tsup{12}C-CC \ATnue\ ones. In this latter case,
the mean energy is reconstructed with an accuracy
around $10\%-20\%$ for both models; that is, better
than either HALO or JUNO alone. As for the emitted
energy $\mathcal{E}(\Pnue)$, combining the two
detectors makes its distribution more peaked in
general, with the maximum of the distribution
shifted towards the true value. The  combination of
these two factors leads to an improvement in the
accuracy of some percent. Moreover, we should note
the inclusion of HALO is able to limit the allowed
region of the parameters; namely, reducing the
low-energy events pushed towards the edge of the
prior and/or the very high-energy ones --- see
e.g.\ Figures \ref{fig:1j0h14}, \ref{fig:2j3h14},
and \ref{fig:2j4h14}.

The inclusion of HALO data to JUNO does little to
the \Tnux\ species, although the $\mathcal{E}(\Pnux
)$ distributions are, also for this species,
shifted. As one can see from Tables \ref{tab:risex}
and \ref{tab:risey}, the reconstructed values
including HALO are lower and thus a little bit
closer to the true values (within statistical
fluctuations). Sometimes, however, neither the JUNO 
or the JUNO+HALO distributions are nicely peaked,
with values piling up towards the low-energy edge
of the prior. This may happen in the analyses that
do not have the pES channel implemented and thus
there is no clear information about the low-energy
events. Concerning the mean energy $\langle E(\Pnux
)\rangle$, the reconstructed  distributions
including HALO are almost identical to the ones
from JUNO alone. When one of the results improves,
that is because the combination gets rid of the
tails in the distribution.

\begin{figure}[t!]
	\centering
	\subfloat[SK+JUNO (n) and HALO]
	{\label{fig:1:0:14}\includegraphics[width=.43%
	\textwidth]{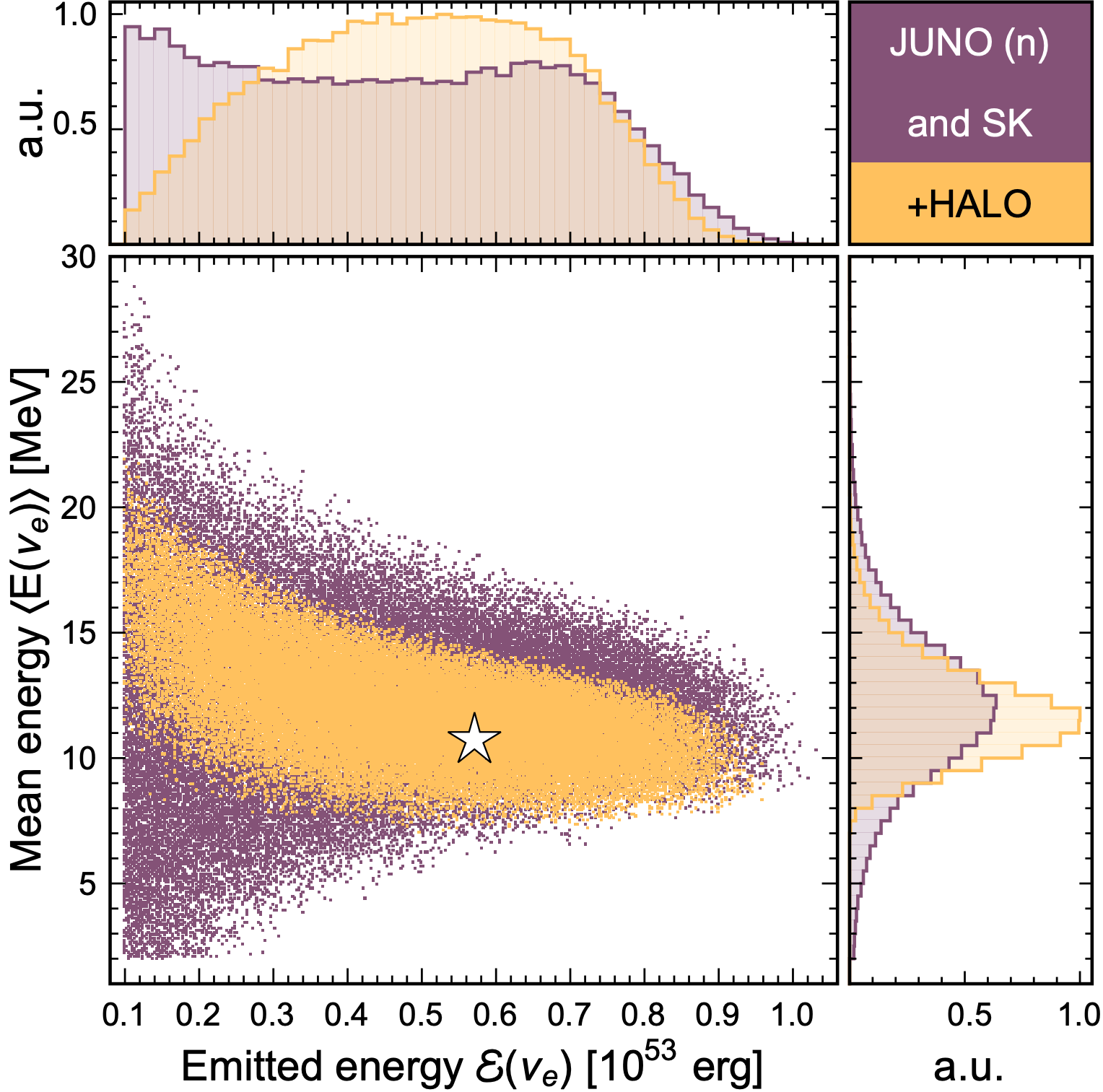}}\hfill
	\subfloat[SK+JUNO (nC) and HALO]
	{\label{fig:1:1:14}\includegraphics[width=.43%
	\textwidth]{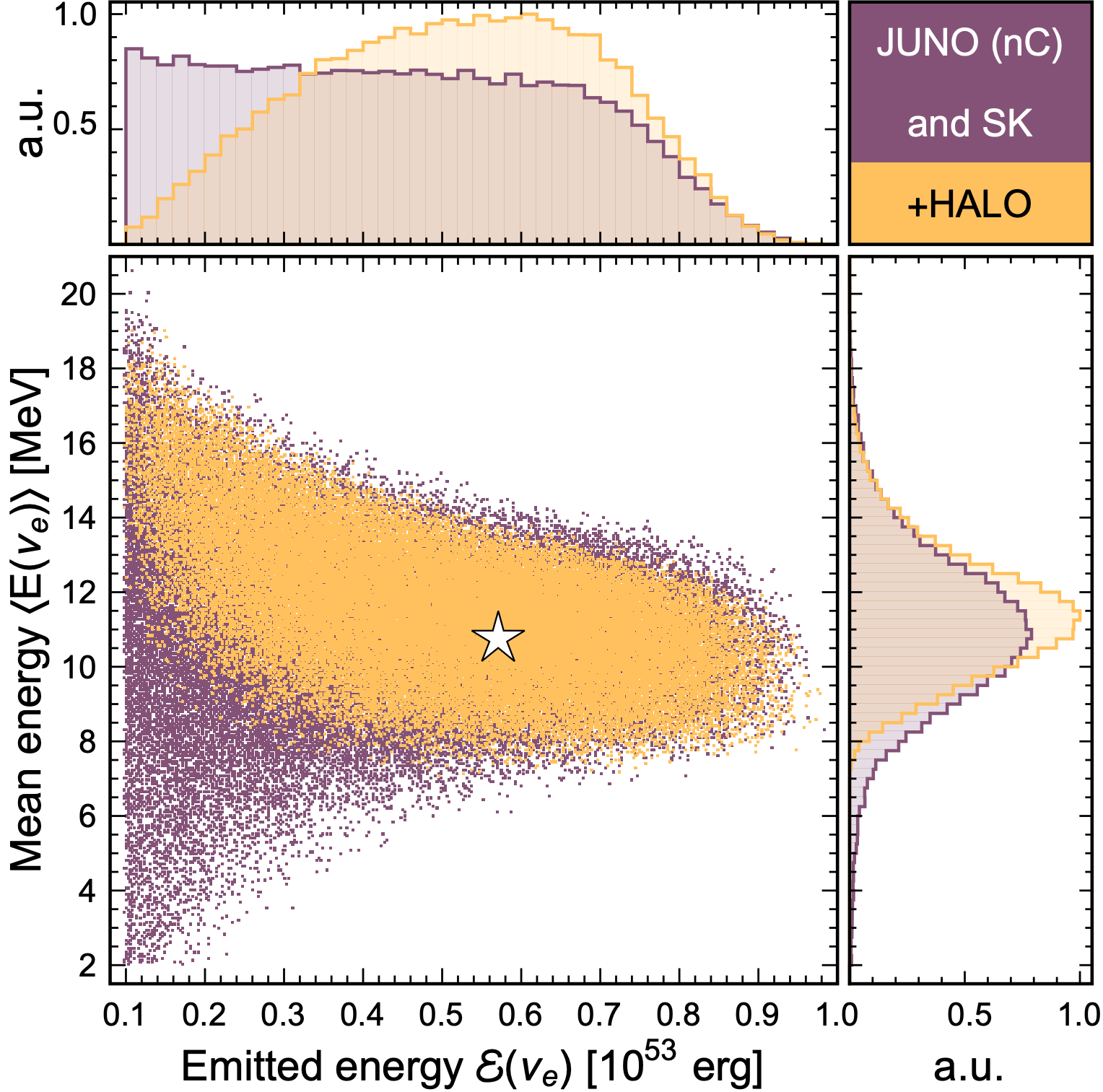}}\\
	\subfloat[SK+JUNO (nCd) and HALO]
	{\label{fig:1:2:14}\includegraphics[width=.43%
	\textwidth]{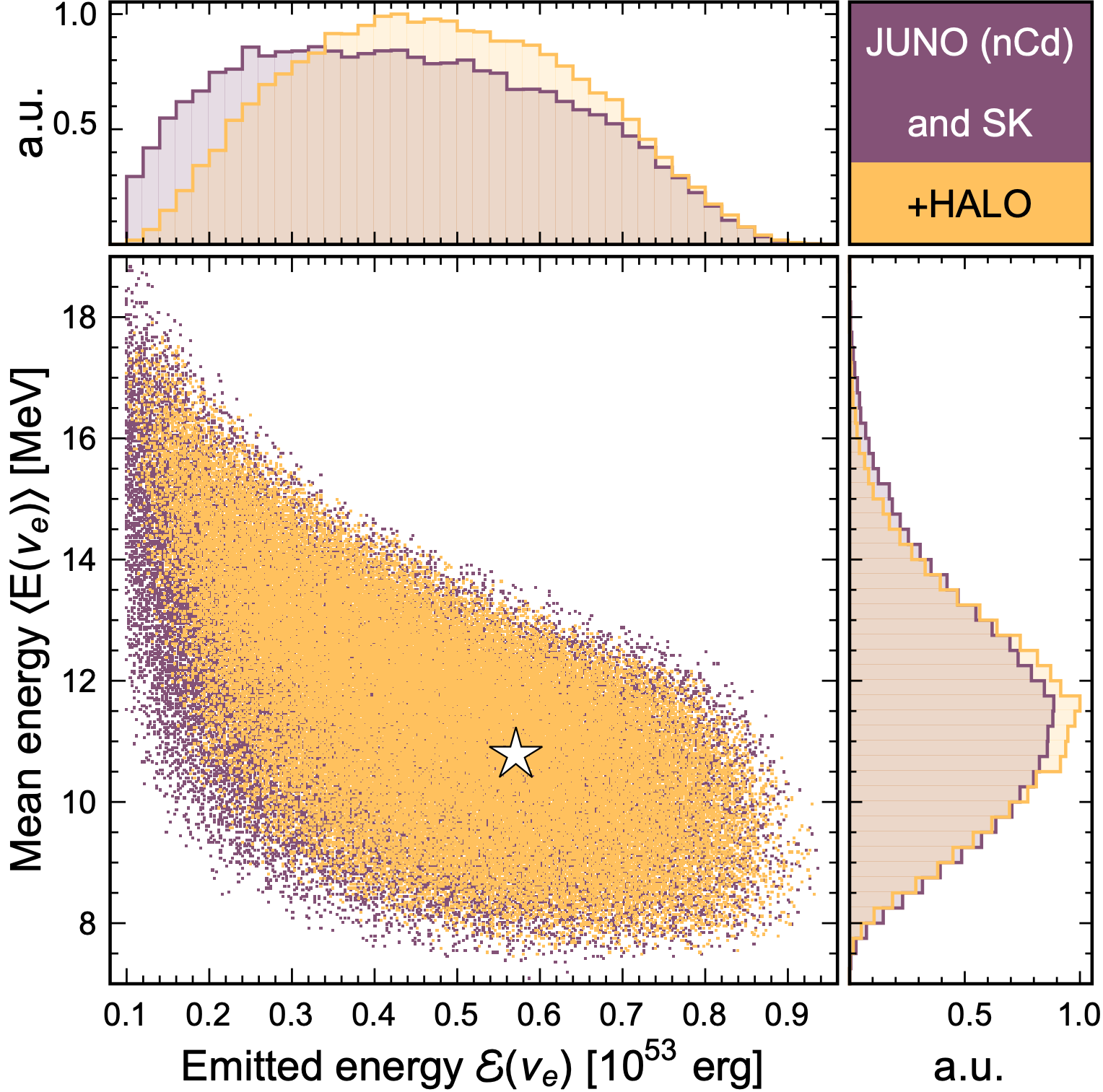}}\hfill
	\subfloat[SK+JUNO (p) and HALO]
	{\label{fig:1:3:14}\includegraphics[width=.43%
	\textwidth]{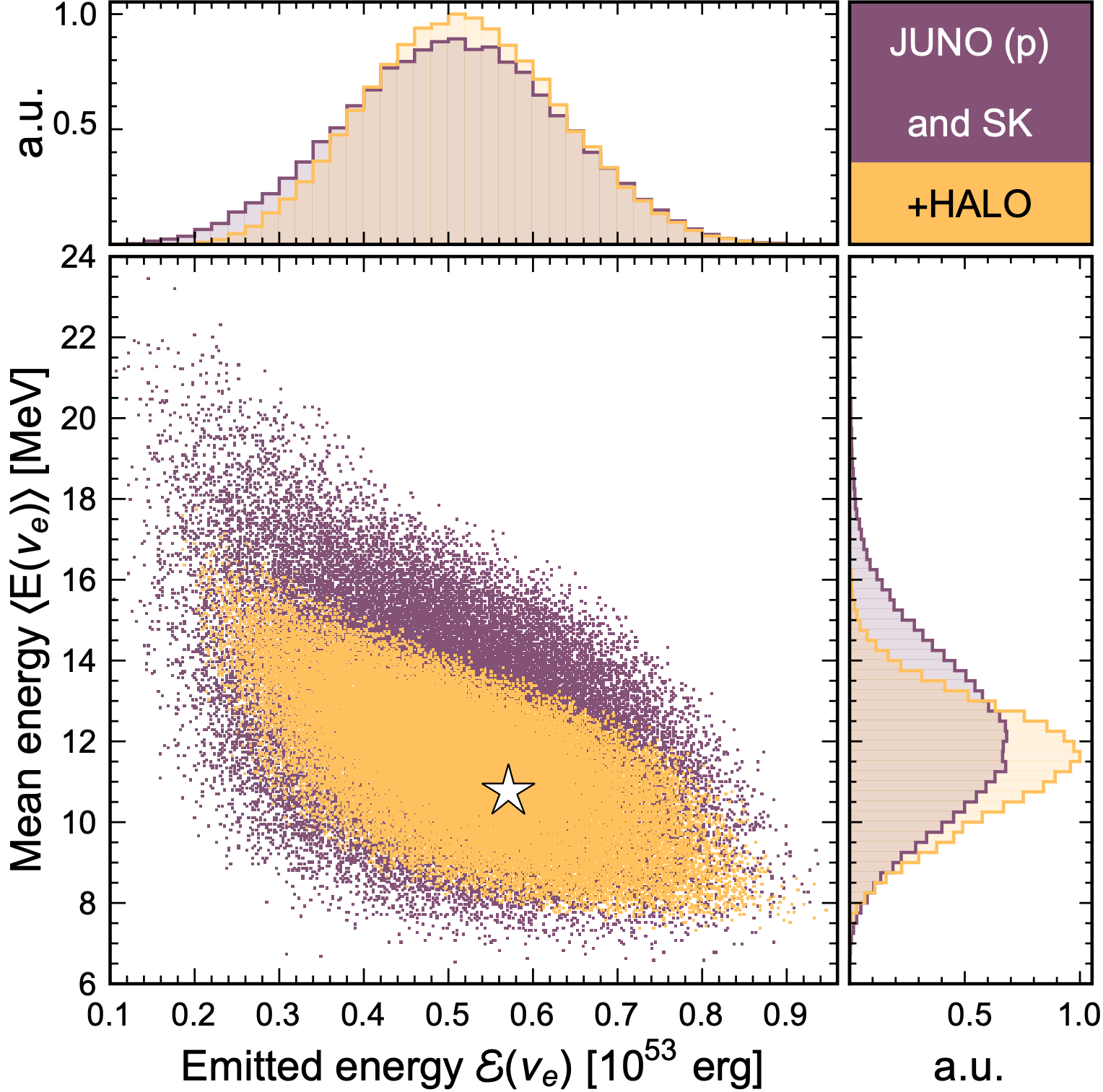}}\\
	\caption{Projections onto $\mathcal{E}(\Pnue)$%
	--$\langle E(\Pnue)\rangle$ planes and
	probability distributions, given the supernova
	model LS220-s27.0co \protect\cite{%
	Mirizzi:2015eza}. In all the four panels, the
	orange regions show the results of HALO
	inclusion to the SK+JUNO results (purple),
	given the n configuration \protect%
	\subref{fig:1:0:14},
	nC configuration \protect\subref{fig:1:1:14},
	nCd configuration \protect\subref{fig:1:2:14},
	p configuration \protect\subref{fig:1:3:14}.
	For the meaning of the JUNO labels see Section
	\protect\ref{sec:JUNO} and Table
	\ref{tab:jcon}. The true values are marked
	by a star.}
	\label{fig:all27}
\end{figure}

\begin{figure}[t!]
	\centering
	\subfloat[SK+JUNO (n) and HALO]
	{\label{fig:2:0:14}\includegraphics[width=.43%
	\textwidth]{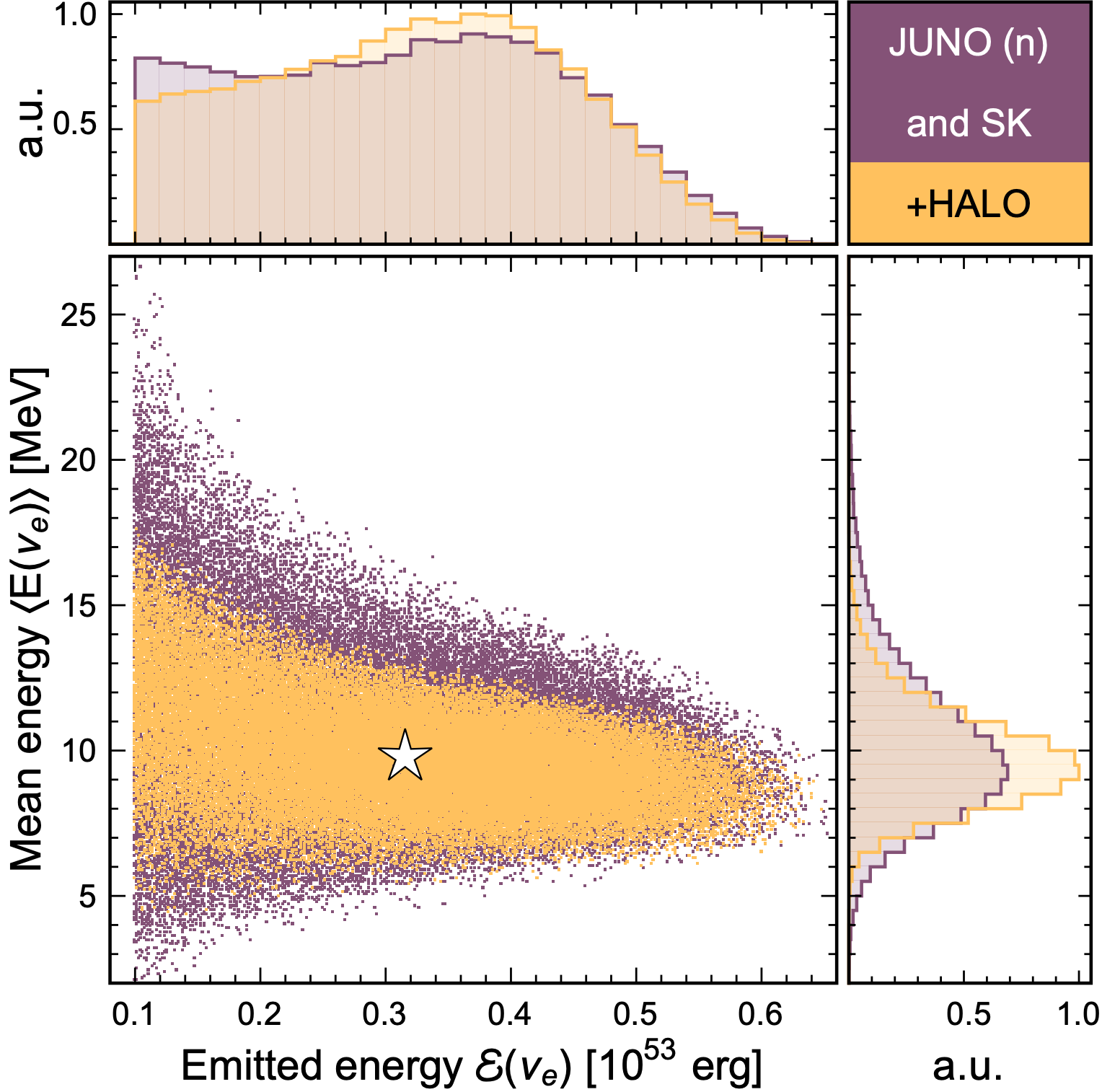}}\hfill
	\subfloat[SK+JUNO (nC) and HALO]
	{\label{fig:2:1:14}\includegraphics[width=.43%
	\textwidth]{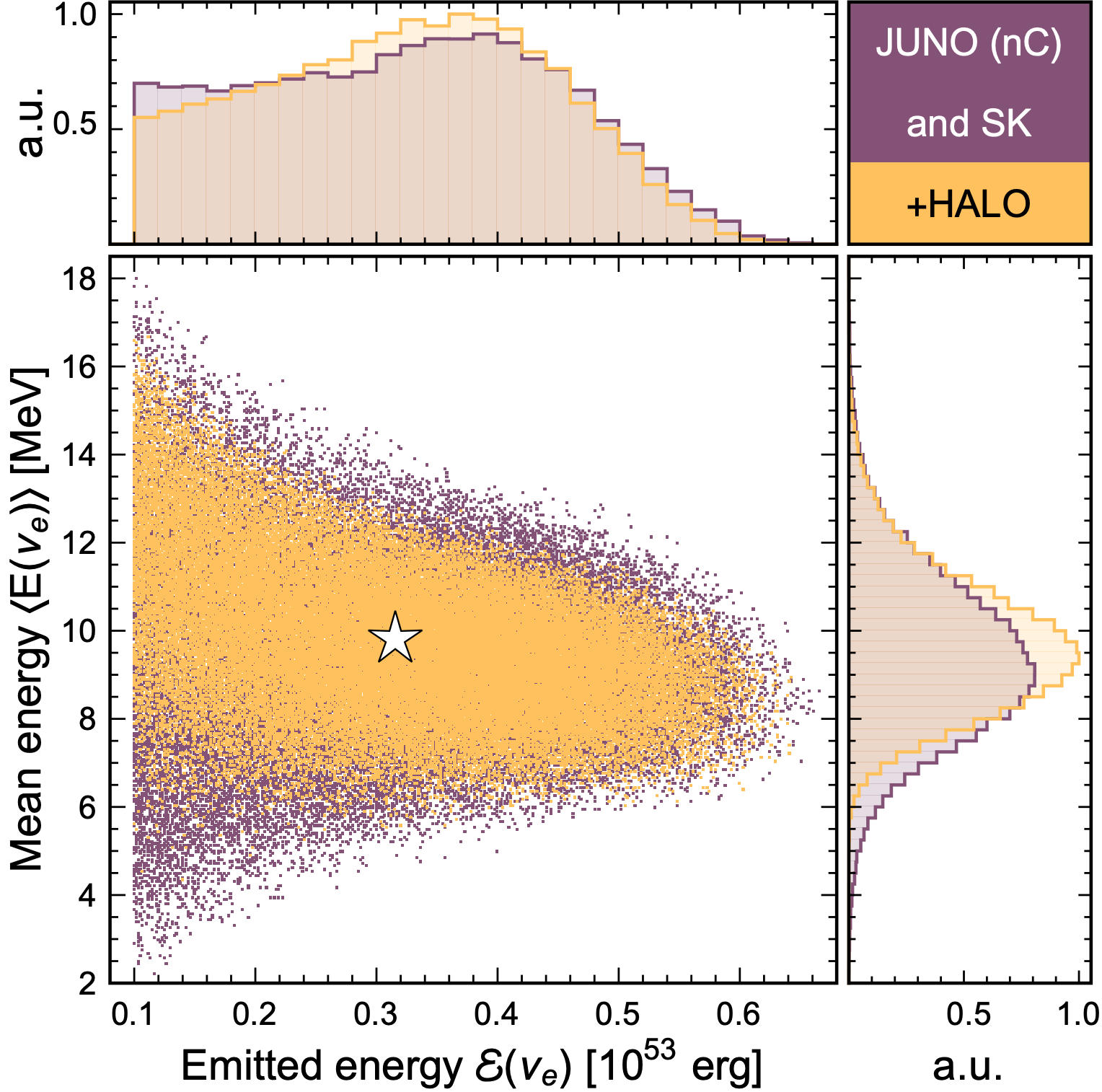}}\\
	\subfloat[SK+JUNO (p) and HALO]
	{\label{fig:2:3:14}\includegraphics[width=.43%
	\textwidth]{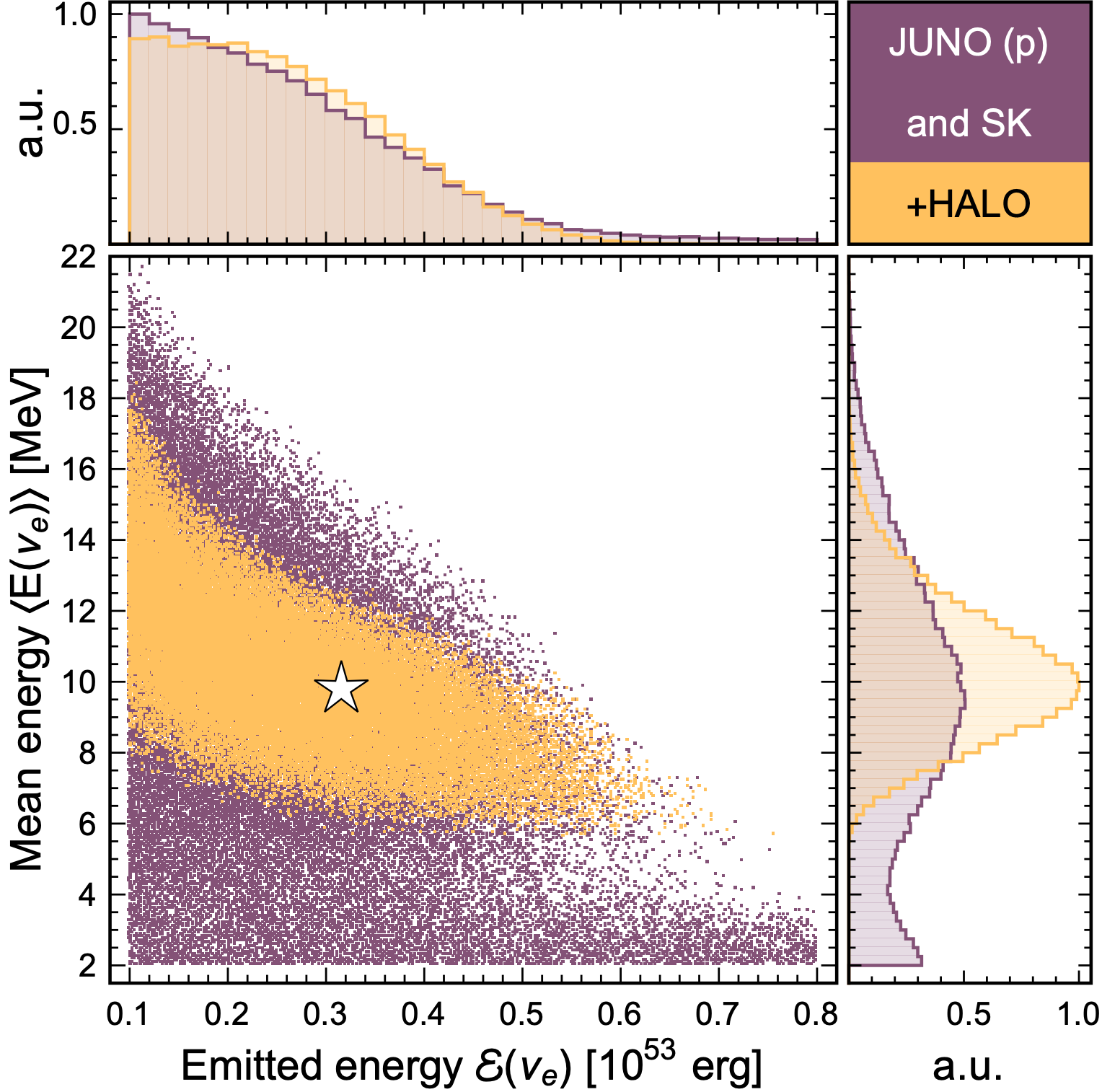}}\hfill
	\subfloat[SK+JUNO (pC) and HALO]
	{\label{fig:2:4:14}\includegraphics[width=.43%
	\textwidth]{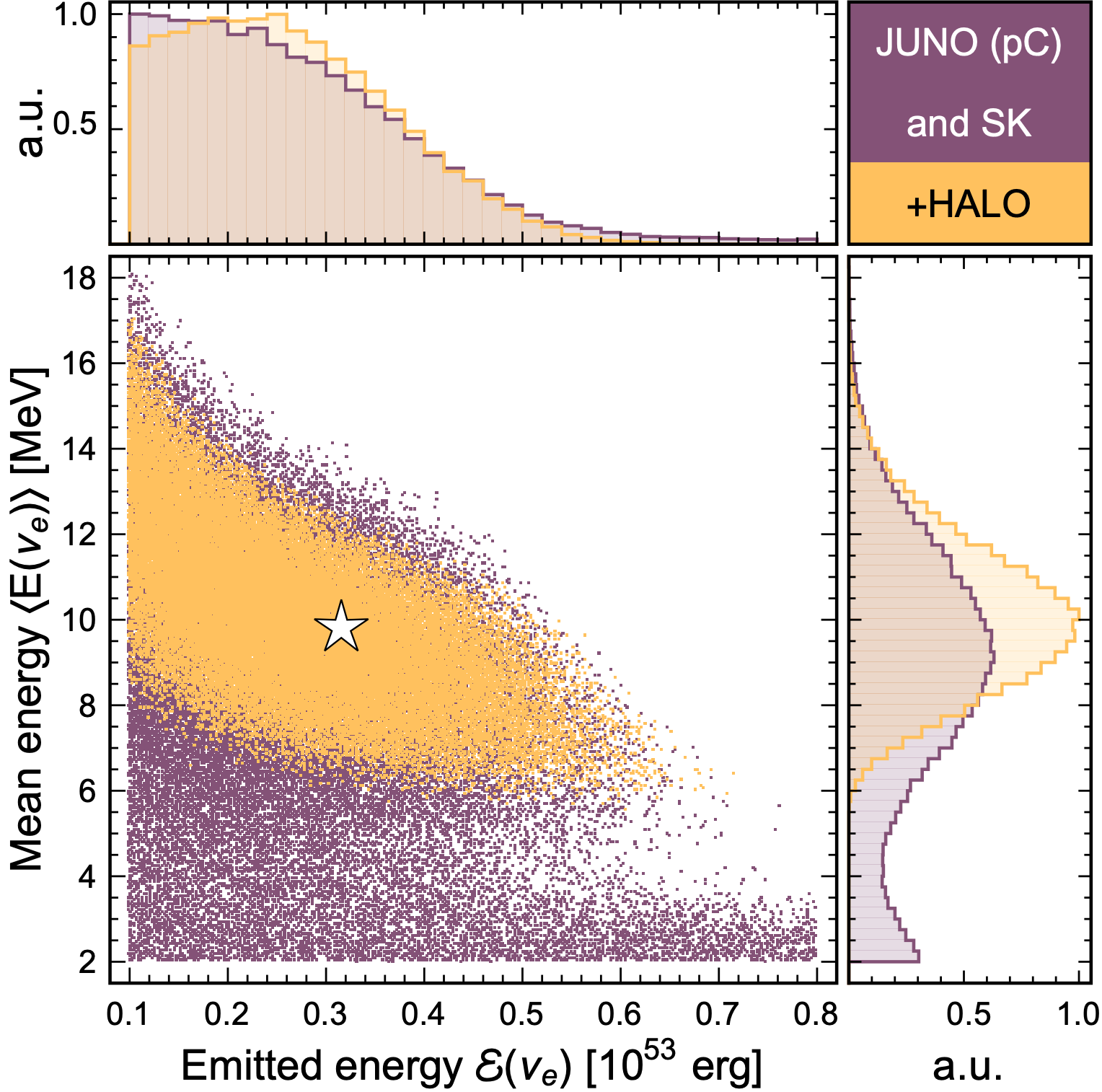}}\\
	\caption{Projections onto $\mathcal{E}(\Pnue)$%
	--$\langle E(\Pnue)\rangle$ planes and
	probability distributions, given the supernova
	model LS220-z9.6co \protect\cite{%
	Mirizzi:2015eza}. In all the four panels, the
	orange regions show the results of HALO
	inclusion to the SK+JUNO results (purple),
	given the n configuration \protect%
	\subref{fig:2:0:14},
	nC configuration \protect\subref{fig:2:1:14},
	p configuration \protect\subref{fig:2:3:14},
	pC configuration \protect\subref{fig:2:4:14}.
	For the meaning of the JUNO labels see Section
	\protect\ref{sec:JUNO} and Table
	\ref{tab:jcon}. Note that in figure
	\protect\subref{fig:2:3:14} and
	\protect\subref{fig:2:4:14} the tail of the
	$\mathcal{E}(\Pnue)$ distribution extends up to
	the edge of the prior, namely \SI{2e53}
	{\erg}.The true values are marked by a star.}
	\label{fig:all96}
\end{figure}

HALO can have an impact not only on SK or JUNO
taken  alone, but also on their combination,
depending on the number of channels implemented in
JUNO. Figures \ref{fig:all27} and \ref{fig:all96}
(for models LS220-s27.0co and LS220-z9.6co
respectively) show the improvement in the 
$\mathcal{E}(\Pnue)$ and $\langle E(\Pnue)\rangle$
reconstructed parameters, once HALO is added to
some of the SK+JUNO configurations.

In general, the inclusion of HALO improves the
distribution of the $\mathcal{E}(\Pnue)$ parameter,
especially when the pES channel and/or
\tsup{12}C-CC interactions are not implemented in
JUNO. The improvement of the accuracy thanks to
HALO is variable but on average it is about $10\%$,
as one can see from Tables \ref{tab:risex} and
\ref{tab:risey}. Moreover, when the
distributions have tails extending to the edge of
the prior (e.g.\ Figures \ref{fig:2:3:14} and
\ref{fig:2:4:14}) the inclusion of HALO helps in
containing the region. Something similar happens
for the \Tnux\ emitted energy, although to
an extent that is sometimes negligible.

Concerning the mean energies, the one of the \Tnue\ 
species has a noticeable improvement if JUNO is
blind to clear \Tnue\ information brought by carbon
events --- i.e.,\ n, p, and even nC, pC
configurations. In those cases, HALO helps to
reduce the uncertainty up to about $50\%$. In
the other cases, the inclusion of HALO makes no
significant difference. As for the \Tnux\ species,
the lead information brings little to no
improvement to the $\langle E(\Pnux)\rangle$
parameter reconstruction.

It is worth noting that, in general, when HALO
improves JUNO results on \Tnue\ species there is no
guarantee that the addition of SK will enhance them
further. On the other hand, when SK improves JUNO
results, the inclusion of HALO always brings
useful information.

\subsection{About the \ATnue\ species}
\label{sec:anue}

Although HALO is almost completely blind to the
\ATnue\ component of the time-integrated flux, it
makes up the most of the signal in SK and JUNO. As
a consequence, the reconstructed parameters of the
\ATnue\ species are the ones presenting the 
strongest constraints. As one can see in Table
\ref{tab:anue}, SK and JUNO give comparable
results. Since the signal is reconstructed through
the IBD channel, there is almost no difference
among the various configurations of JUNO ---
namely, including the handful of events from
\tsup{12}C-CC interactions makes no difference.

\begin{table}
	\centering
		\begin{tabular}{cl%
		r@{\hskip0.15em}c@{\hskip 0.15em}l@{\hskip%
		0.2em}c%
		r@{\hskip0.15em}c@{\hskip 0.15em}l@{\hskip%
		0.2em}c%
		r@{\hskip0.15em}c@{\hskip 0.15em}l@{\hskip%
		0.2em}c}
			\toprule
			& &\multicolumn{3}{c}{$\:
			\mathcal{E}(\APnue)$} &prec.
			&\multicolumn{3}{c}{$
			\hphantom{\langle\rangle}
			\langle E(\APnue)\rangle$} &prec.
			&\multicolumn{3}{c}{$\:
			\alpha(\APnue)$} &prec.\\
			&&\multicolumn{3}{c}{$\:[\SI{e52}
			{\erg}]\vphantom{\frac{1}{2}}$} &\%
			&\multicolumn{3}{c}{$
			\hphantom{\langle\rangle}
			[\si{\mega\electronvolt}]$} &\%
			&&& &\%\\
			\midrule
			\multirow{6}{*}{\rotatebox[origin=c]{90}
			{\parbox[c]{3cm}{\centering 
			LS220-s27.0co}}}
			&$(\mu^*\pm\sigma^*)$
			& {5.68} &$\pm$ & 0.55 &---
			& {13.64} &$\pm$ & 19.6 &---
			& {2.26} &$\pm$ & 0.87 &---\\
			\cmidrule{2-14}
			& SK
			& 5.72 &$\pm$ & 0.17 & 3.0
			& 13.57 &$\pm$ & 0.39 & 2.9
			& 2.26 &$\pm$ & 0.22 & 9.5\\
			& JUNO (n)
			& 5.70 &$\pm$ & 0.16 & 2.8
			& 13.75 &$\pm$ & 0.33 & 2.4
			& 2.36 &$\pm$ & 0.19 & 8.2\\
			& JUNO (p)
			& 5.67 &$\pm$ & 0.15 & 2.6
			& 13.84 &$\pm$ & 0.34 & 2.4
			& 2.41 &$\pm$ & 0.20 & 8.1\\
			& SK+JUNO (n)
			& 5.70 &$\pm$ & 0.11 & 2.0
			& 13.67 &$\pm$ & 0.25 & 1.8
			& 2.31 &$\pm$ & 0.14 & 6.1\\
			& SK+JUNO (p)
			& 5.69 &$\pm$ & 0.11 & 1.9
			& 13.71 &$\pm$ & 0.25 & 1.8
			& 2.33 &$\pm$ & 0.14 & 6.0\\
			\midrule
			\multirow{6}{*}{\rotatebox[origin=c]{90}
			{\parbox[c]{3cm}{\centering 
			LS220-z9.6co}}}
			&$(\mu^*\pm\sigma^*)$
			& {3.38} &$\pm$ & 0.55 &---
			& {12.26} &$\pm$ & 19.6 &---
			& {2.19} &$\pm$ & 0.87 &---\\
			\cmidrule{2-14}
			& SK
			& 3.44 &$\pm$ & 0.15 & 4.3
			& 11.93 &$\pm$ & 0.53 & 4.5
			& 2.06 &$\pm$ & 0.30 & 14\\
			& JUNO (n)
			& 3.32 &$\pm$ & 0.12 & 3.7
			& 12.33 &$\pm$ & 0.41 & 3.3
			& 2.21 &$\pm$ & 0.25 & 11\\
			& JUNO (p)
			& 3.36 &$\pm$ & 0.13 & 4.0
			& 12.14 &$\pm$ & 0.49 & 4.1
			& 2.10 &$\pm$ & 0.28 & 13\\
			& SK+JUNO (n)
			& 3.37 &$\pm$ & 0.09 & 2.7
			& 12.17 &$\pm$ & 0.31 & 2.6
			& 2.15 &$\pm$ & 0.18 & 8.5\\
			& SK+JUNO (p)
			& 3.40 &$\pm$ & 0.10 & 2.9
			& 12.04 &$\pm$ & 0.36 & 3.0
			& 2.07 &$\pm$ & 0.20 & 9.7\\
			\bottomrule
		\end{tabular}
		\caption{Reconstructed parameters of the 
		\ATnue\ fluence, from the supernova models
		LS220-s27.0co and LS220-z9.6co \protect
		\cite{Mirizzi:2015eza}, in Super-Kamiokande
		(SK), JUNO and a combination of the two.
		The analyses involving JUNO have been
		performed with (p) and without (n) the pES
		scattering --- see Section \protect
		\ref{sec:JUNO} and Table \ref{tab:jcon}.
		The first rows of each block, marked with
		$(\mu^*\pm\sigma^*)$, report the true value
		$\mu^*$ and the standard deviation
		$\sigma^*$ of a flat distribution in the
		assumed priors \protect\eqref{eq:pris},
		\protect\eqref{eq:prie}, 
		\protect\eqref{eq:pria}.
		}
		\label{tab:anue}
\end{table}

Combining the two detectors gives the best results,
with the precision depending on the considered
model. Indeed, for the $\mathcal{E}(\APnue)$
parameter the overall uncertainties are almost the
same, but the true value is different. Therefore,
the $\mathcal{E}(\APnue)$ precision for the SK+JUNO
analysis goes from $\approx 2\%$ (LS220-s27.0co) to 
$\approx 3\%$ (LS220-z9.6co), while the one
for $\langle E(\APnue)\rangle$ goes from $\approx
2\%$ (LS220-s27.0co) to $\approx 3\%$
(LS220-z9.6co). Finally, the $\alpha(\APnue)$
parameter is the only pinching parameter that gets
constrained, with an accuracy of around $10\%$.

\subsection{About the total emitted energy}
\label{sec:etot}

For a given analysis, once a point $P$ satisfies
the likelihood condition and thus is accepted in
the set, it can be used to reconstruct \emph{a
posteriori} the total emitted energy $\ETOT$, as
\begin{equation}
	\label{eq:etot}
	\mathcal{E}_{\text{tot}}|_P = \mathcal{E}
	(\Pnue)|_P + \mathcal{E}(\APnue)|_P +
	4\mathcal{E}(\Pnux)|_P.
\end{equation}
The reconstructed results, for both models, are
summarized in Table \ref{tab:etot}.

The first thing that clearly emerges is that HALO
is not able to put any constraint on $\ETOT$.
Indeed, the mean and standard deviation of the
reconstructed distributions practically coincide
with mean (\SI{6.3e53}{\erg}) and standard
deviation (\SI{2.3e53}{\erg}) of the prior
distribution; namely, the one given by eq.\ 
\eqref{eq:etot} when all the $\mathcal{E}(\nu_i)$
vary randomly inside their prior \eqref{eq:pris}.
We note that, in this latter case, the prior
accuracy is $37\%$.

Super-Kamkokande does a little better, although
remaining unsatisfactory. The overall precision is
no different from the prior one, but the
reconstructed value does get shifted from the prior
one towards the true one while the overall error
decrease. Surprisingly, the inclusion of HALO seems
to improve this trend, in general. This is
especially evident when considering the combination
with JUNO, the detector leading the $\ETOT$
reconstruction. Alone, JUNO has an accuracy ranging
from $\approx 30\%$ to $\approx 20\%$, depending on
the model and the channels implemented. When
combined with HALO, the value is always
reconstructed closer to the true one and the
uncertainty reduced by about $10\%$. Remarkably,
adding HALO to JUNO gives results that are
comparably or slightly better than the ones from
JUNO and SK. Similar to the discussion in Section
\ref{sec:Hinclus}, this means that HALO still has
some (sometimes minor) impact when added to the
combination of JUNO+SK, while the one of SK to the
combination HALO+JUNO is sometimes null.

\begin{table}
	\centering
		\begin{tabular}{l%
		r@{\hskip0.15em}c@{\hskip 0.15em}l@{\hskip%
		0.2em}c%
		r@{\hskip0.15em}c@{\hskip 0.15em}l@{\hskip%
		0.2em}c}
			\toprule
			&\multicolumn{4}{c}{LS220-s27.0co}
			&\multicolumn{4}{c}{LS220-z9.6co}\\
			\cmidrule{2-9}
			&\multicolumn{3}{c}{$\:
			\mathcal{E}_{\mathrm{tot}}$} &prec.
			&\multicolumn{3}{c}{$\:
			\mathcal{E}_{\mathrm{tot}}$} &prec.\\
			&\multicolumn{3}{c}{$\:[\SI{e53}
			{\erg}]\vphantom{\frac{1}{2}}$} &\%
			&\multicolumn{3}{c}{$\:[\SI{e53}
			{\erg}]\vphantom{\frac{1}{2}}$} &\%\\
			\midrule
			$(\mu^*\pm\sigma^*)$
			& {3.24} &$\pm$ & 2.33 &---
			& {1.83} &$\pm$ & 2.33 &---\\
			\midrule
			\lk HALO
			&\lk 5.98 &$\qm\pm$ &\lk 2.34 &\lk 39
			&\lk 5.56 &$\qm\pm$ &\lk 2.33 &\lk 41\\
			\midrule
			SK
			& 4.71 &$\pm$ & 1.79 & 38
			& 3.50 &$\pm$ & 1.64 & 46\\
			\multicolumn{1}{|r|}{\lk + HALO}
			&\lk 4.02 &$\qm\pm$ &\lk 1.41 &\lk 35
			&\lk 3.41 &$\qm\pm$ &\lk 1.57 &\lk 46\\
			\midrule
			JUNO (n)
			& 3.72 &$\pm$ & 1.28 & 34
			& 2.17 &$\pm$ & 0.70 & 32\\
			\multicolumn{1}{|r|}{\lk + HALO}
			&\lk 3.31 &$\qm\pm$ &\lk 1.10 &\lk 33
			&\lk 2.17 &$\qm\pm$ &\lk 0.70 &\lk 32\\
			\cdashlinelr{1-9}
			JUNO (nC)
			& 3.56 &$\pm$ & 1.24 & 34
			& 2.15 &$\pm$ & 0.69 & 32\\
			\multicolumn{1}{|r|}{\lk + HALO}
			&\lk 3.19 &$\qm\pm$ &\lk 1.05 &\lk 32
			&\lk 1.82 &$\qm\pm$ &\lk 0.68 &\lk 37\\
			\cdashlinelr{1-9}
			JUNO (nCd)
			& 3.52 &$\pm$ & 1.16 & 33
			& 2.04 &$\pm$ & 0.65 & 31\\
			\multicolumn{1}{|r|}{\lk + HALO}
			&\lk 3.33 &$\qm\pm$ &\lk 1.07 &\lk 32
			&\lk 1.75 &$\qm\pm$ &\lk 0.65 &\lk 37\\
			\cdashlinelr{1-9}
			JUNO (p)
			& 3.16 &$\pm$ & 0.61 & 19
			& 2.72 &$\pm$ & 0.87 & 32\\
			\multicolumn{1}{|r|}{\lk + HALO}
			&\lk 3.03 &$\qm\pm$ &\lk 0.50 &\lk 16
			&\lk 2.41 &$\qm\pm$ &\lk 0.62 &\lk 25\\
			\cdashlinelr{1-9}
			JUNO (pC)
			& 3.18 &$\pm$ & 0.58 & 18
			& 2.71 &$\pm$ & 0.85 & 31\\
			\multicolumn{1}{|r|}{\lk + HALO}
			&\lk 3.06 &$\qm\pm$ &\lk 0.52 &\lk 16
			&\lk 2.40 &$\qm\pm$ &\lk 0.61 &\lk 25\\
			\cdashlinelr{1-9}
			JUNO (pCd)
			& 3.12 &$\pm$ & 0.55 & 17
			& 2.37 &$\pm$ & 0.61 & 25\\
			\multicolumn{1}{|r|}{\lk + HALO}
			&\lk 3.05 &$\qm\pm$ &\lk 0.51 &\lk 16
			&\lk 2.37 &$\qm\pm$ &\lk 0.60 &\lk 25\\
			\midrule
			SK+JUNO (n)
			& 3.82 &$\pm$ & 1.27 & 33
			& 2.19 &$\pm$ & 0.69 & 31\\
			\multicolumn{1}{|r|}{\lk + HALO}
			&\lk 3.56 &$\qm\pm$ &\lk 1.09 &\lk 30
			&\lk 2.14 &$\qm\pm$ &\lk 0.65 &\lk 30\\
			\cdashlinelr{1-9}
			SK+JUNO (nC)
			& 3.87 &$\pm$ & 1.20 & 31
			& 2.15 &$\pm$ & 0.68 & 31\\
			\multicolumn{1}{|r|}{\lk + HALO}
			&\lk 3.48 &$\qm\pm$ &\lk 1.05 &\lk 30
			&\lk 2.13 &$\qm\pm$ &\lk 0.65 &\lk 30\\
			\cdashlinelr{1-9}
			SK+JUNO (nCd)
			& 3.88 &$\pm$ & 1.07 & 27
			& 1.99 &$\pm$ & 0.62 & 31\\
			\multicolumn{1}{|r|}{\lk + HALO}
			&\lk 3.67 &$\qm\pm$ &\lk 0.99 &\lk 26
			&\lk 2.04 &$\qm\pm$ &\lk 0.62 &\lk 30\\
			\cdashlinelr{1-9}
			SK+JUNO (p)
			& 3.13 &$\pm$ & 0.60 & 19
			& 2.87 &$\pm$ & 0.90 & 31\\
			\multicolumn{1}{|r|}{\lk + HALO}
			&\lk 3.08 &$\qm\pm$ &\lk 0.50 &\lk 16
			&\lk 2.48 &$\qm\pm$ &\lk 0.62 &\lk 25\\
			\cdashlinelr{1-9}
			SK+JUNO (pC)
			& 3.15 &$\pm$ & 0.55 & 17
			& 2.80 &$\pm$ & 0.88 & 31\\
			\multicolumn{1}{|r|}{\lk + HALO}
			&\lk 3.08 &$\qm\pm$ &\lk 0.51 &\lk 16
			&\lk 2.46 &$\qm\pm$ &\lk 0.62 &\lk 25\\
			\cdashlinelr{1-9}
			SK+JUNO (pCd)
			& 3.18 &$\pm$ & 0.55 & 17
			& 2.43 &$\pm$ & 0.61 & 25\\
			\multicolumn{1}{|r|}{\lk + HALO}
			&\lk 3.11 &$\qm\pm$ &\lk 0.51 &\lk 16
			&\lk 2.40 &$\qm\pm$ &\lk 0.59 &\lk 24\\
			\bottomrule
		\end{tabular}
		\caption{Reconstructed total energy
		$\mathcal{E}_{\mathrm{tot}}$ \eqref{eq:etot} 
		given different combination of detectors and
		supernova models: LS220-s27.0co and
		LS220-z9.6co \protect\cite{Mirizzi:2015eza}.
		Each row is doubled to show the impact of
		including HALO-1kT to the analysis.
		The first
		row, marked with $(\mu^*\pm\sigma^*)$,
		reports the true value $\mu^*$ and the
		standard deviation $\sigma^*$ of the prior
		distribution; i.e.\ the one given by the
		$\mathcal{E}(\nu_i)$ in eq.\ 
		\eqref{eq:etot} varying randomly inside
		their prior \eqref{eq:pris}. Each analysis
		involving JUNO is performed six times,
		corresponding to the six assumptions on its
		configuration summarized in table \protect%
		\ref{tab:jcon}, namely: pES implemented (p),
		$\Pnue/\APnue$ \tsup{12}C-CC reactions
		included (C), $\Pnue/\APnue$ \tsup{12}C-CC
		reactions included and distinguished (d).}
		\label{tab:etot}
\end{table}

\section{Conclusions}
\label{sec:conclu}

In this paper we have assessed the impact that a
lead-based detector, such as HALO-1kT, can
have if combined with high-statistics supernova
detector such as Super-Kamiokande and JUNO. We
performed a full-parameter likelihood analysis on 
the time-integrated fluxes (fluences), taking into
account two supernova models without implementing
any oscillation mechanism. This justifies the
quantitative differences one might find with our
previous works
\cite{GalloRosso:2017hbp,GalloRosso:2017mdz}.

Different combinations of detectors and channels
were taken into account. What emerges from the
results are, on one hand, that HALO-1kT alone is
not able to give a strong constraint on the
fluences. On the other hand, the orthogonal source
of information it provides is precious when
combined with other detectors. In that case, the
combined results are in general better than the
results of the detectors taken singularly,
especially if they do not have a \Tnue\ sensitive
channel. HALO can also give a (minor) contribution
in constraining the total emitted energy $\ETOT$,
although its impact strongly depends on the
supernova model and the channels included in
the analyses.

The analysis procedure followed in the present
work can be extended in many ways. First of all, a
realistic detector response can be implemented.
Moreover, realistic uncertainties can be taken
into account both on the neutrino-lead cross
sections and on the neutrino-carbon ones. Finally,
the number and kind of supernova models should be
extended, together with the ones concerning the
flavor transformation in dense environments.

The next supernova detection will surely be a
once-in-a-lifetime event for many physical branches
and it is eagerly awaited by the scientific
community. While neutrino physics still justifies
the construction of bigger and bigger detectors,
HALO-1kT is a small, economical, and stable
detector able to provide supplementary information
and, hopefully, to act as an important piece of the
supernova puzzle.

\section*{Acknowledgements}

The author would like to thank C.J.\ Virtue,
F.\ Vissani, and M.C.\ Volpe for their invaluable
support, help, and comments. Without them, the
realization of this work would have never been
possible.

\newpage

\printbibliography

\begin{table}
	\centering
		\resizebox{\textwidth}{!}{
		\begin{tabular}{l%
		r@{\hskip0.15em}c@{\hskip 0.15em}l@{\hskip%
		0.2em}c%
		r@{\hskip0.15em}c@{\hskip 0.15em}l@{\hskip%
		0.2em}c%
		r@{\hskip0.15em}c@{\hskip 0.15em}l@{\hskip%
		0.2em}c%
		r@{\hskip0.15em}c@{\hskip 0.15em}l@{\hskip%
		0.2em}c}
			\toprule
			\multicolumn{1}{c}{\multirow{2}{*}{%
			LS220-s27.0co}}
			&\multicolumn{3}{c}{$\:
			\mathcal{E}(\Pnue)$} &prec.
			&\multicolumn{3}{c}{$
			\hphantom{\langle\rangle}
			\langle E(\Pnue)\rangle$} &prec.
			&\multicolumn{3}{c}{$\:
			\mathcal{E}(\Pnux)$} &prec.
			&\multicolumn{3}{c}{$
			\hphantom{\langle\rangle}
			\langle E(\Pnux)\rangle$} &prec.\\
			&\multicolumn{3}{c}{$\:[\SI{e53}
			{\erg}]\vphantom{\frac{1}{2}}$} &\%
			&\multicolumn{3}{c}{$
			\hphantom{\langle\rangle}
			[\si{\mega\electronvolt}]$} &\%
			&\multicolumn{3}{c}{$\:[\SI{e53}
			{\erg}]\vphantom{\frac{1}{2}}$} &\%
			&\multicolumn{3}{c}{$
			\hphantom{\langle\rangle}
			[\si{\mega\electronvolt}]$} &\%\\
			\midrule
			$(\mu^*\pm\sigma^*)$
			& {0.57} &$\pm$ & 0.55 &---
			& {10.8} &$\pm$ & 19.6 &---
			& {0.53} &$\pm$ & 0.55 &---
			& {12.9} &$\pm$ & 19.6 &---\\
			\midrule
			\lk HALO-1kT
			&\lk 0.93 &$\qm\pm$ &\lk 0.54 &\lk 58
			&\lk 9.3 &$\qm\pm$ &\lk 3.2 &\lk 34
			&\lk 1.03 &$\qm\pm$ &\lk 0.56 &\lk 54
			&\lk 11.7 &$\qm\pm$ &\lk 5.5 &\lk 47\\
			\midrule
			Super-K
			& 0.59 &$\pm$ & 0.35 & 59
			& 11.3 &$\pm$ & 5.3 & 47
			& 0.89 &$\pm$ & 0.45 & 51
			& 12.6 &$\pm$ & 6.0 & 47\\
			\multicolumn{1}{r}{\lk + HALO}
			&\lk 0.49 &$\qm\pm$ &\lk 0.21 &\lk 43
			&\lk 11.7 &$\qm\pm$ &\lk 2.1 &\lk 18
			&\lk 0.74 &$\qm\pm$ &\lk 0.38 &\lk 52
			&\lk 12.8 &$\qm\pm$ &\lk 4.5 &\lk 35\\
			\midrule
			JUNO (n)
			& 0.47 &$\pm$ & 0.23 & 49
			& 11.7 &$\pm$ & 9.6 & 31
			& 0.67 &$\pm$ & 0.37 & 56
			& 12.9 &$\pm$ & 5.5 & 42\\
			\multicolumn{1}{r}{\lk + HALO}
			&\lk 0.55 &$\qm\pm$ &\lk 0.19 &\lk 35
			&\lk 11.7 &$\qm\pm$ &\lk 1.8 &\lk 16
			&\lk 0.55 &$\qm\pm$ &\lk 0.32 &\lk 58
			&\lk 12.1 &$\qm\pm$ &\lk 4.2 &\lk 34\\
			\cdashlinelr{1-17}
			JUNO (nC)
			& 0.50 &$\pm$ & 0.22 & 45
			& 10.6 &$\pm$ & 2.1 & 20
			& 0.62 &$\pm$ & 0.36 & 57
			& 14.2 &$\pm$ & 5.9 & 42\\
			\multicolumn{1}{r}{\lk + HALO}
			&\lk 0.57 &$\qm\pm$ &\lk 0.18 &\lk 32
			&\lk 11.3 &$\qm\pm$ &\lk 1.5 &\lk 14
			&\lk 0.51 &$\qm\pm$ &\lk 0.30 &\lk 58
			&\lk 12.9 &$\qm\pm$ &\lk 4.3 &\lk 34\\
			\cdashlinelr{1-17}
			JUNO (nCd)
			& 0.49 &$\pm$ & 0.20 & 41
			& 11.3 &$\pm$ & 1.8 & 16
			& 0.61 &$\pm$ & 0.33 & 54
			& 13.2 &$\pm$ & 4.8 & 36\\
			\multicolumn{1}{r}{\lk + HALO}
			&\lk 0.54 &$\qm\pm$ &\lk 0.18 &\lk 33
			&\lk 11.3 &$\qm\pm$ &\lk 1.6 &\lk 14
			&\lk 0.55 &$\qm\pm$ &\lk 0.31 &\lk 55
			&\lk 12.8 &$\qm\pm$ &\lk 3.7 &\lk 29\\
			\cdashlinelr{1-17}
			JUNO (p)
			& 0.47 &$\pm$ & 0.16 & 34
			& 11.8 &$\pm$ & 2.7 & 23
			& 0.53 &$\pm$ & 0.17 & 33
			& 13.9 &$\pm$ & 2.3 & 17\\
			\multicolumn{1}{r}{\lk + HALO}
			&\lk 0.51 &$\qm\pm$ &\lk 0.13 &\lk 26
			&\lk 11.5 &$\qm\pm$ &\lk 1.5 &\lk 13
			&\lk 0.49 &$\qm\pm$ &\lk 0.14 &\lk 28
			&\lk 14.5 &$\qm\pm$ &\lk 2.2 &\lk 15\\
			\cdashlinelr{1-17}
			JUNO (pC)
			& 0.48 &$\pm$ & 0.17 & 34
			& 10.6 &$\pm$ & 1.9 & 17
			& 0.53 &$\pm$ & 0.16 & 30
			& 14.2 &$\pm$ & 2.3 & 16\\
			\multicolumn{1}{r}{\lk + HALO}
			&\lk 0.53 &$\qm\pm$ &\lk 0.15 &\lk 28
			&\lk 11.2 &$\qm\pm$ &\lk 1.5 &\lk 13
			&\lk 0.49 &$\qm\pm$ &\lk 0.14 &\lk 29
			&\lk 14.5 &$\qm\pm$ &\lk 2.2 &\lk 15\\
			\cdashlinelr{1-17}
			JUNO (pCd)
			& 0.46 &$\pm$ & 0.15 & 32
			& 11.2 &$\pm$ & 1.6 & 15
			& 0.52 &$\pm$ & 0.15 & 29
			& 14.2 &$\pm$ & 2.3 & 16\\
			\multicolumn{1}{r}{\lk + HALO}
			&\lk 0.51 &$\qm\pm$ &\lk 0.13 &\lk 26
			&\lk 11.2 &$\qm\pm$ &\lk 1.5 &\lk 14
			&\lk 0.50 &$\qm\pm$ &\lk 0.14 &\lk 28
			&\lk 14.5 &$\qm\pm$ &\lk 2.2 &\lk 15\\
			\midrule
			SK+JUNO (n)
			& 0.46 &$\pm$ & 0.22 & 48
			& 12.0 &$\pm$ & 3.2 & 27
			& 0.70 &$\pm$ & 0.37 & 53
			& 12.5 &$\pm$ & 4.1 & 33\\
			\multicolumn{1}{r}{\lk + HALO}
			&\lk 0.50 &$\qm\pm$ &\lk 0.18 &\lk 37
			&\lk 11.9 &$\qm\pm$ &\lk 1.9 &\lk 16
			&\lk 0.63 &$\qm\pm$ &\lk 0.31 &\lk 50
			&\lk 13.0 &$\qm\pm$ &\lk 3.4 &\lk 26\\
			\cdashlinelr{1-17}
			SK+JUNO (nC)
			& 0.44 &$\pm$ & 0.21 & 47
			& 10.9 &$\pm$ & 2.2 & 20
			& 0.71 &$\pm$ & 0.35 & 49
			& 14.2 &$\pm$ & 3.6 & 25\\
			\multicolumn{1}{r}{\lk + HALO}
			&\lk 0.52 &$\qm\pm$ &\lk 0.18 &\lk 35
			&\lk 11.5 &$\qm\pm$ &\lk 1.6 &\lk 14
			&\lk 0.60 &$\qm\pm$ &\lk 0.30 &\lk 50
			&\lk 13.8 &$\qm\pm$ &\lk 3.4 &\lk 24\\
			\cdashlinelr{1-17}
			SK+JUNO (nCd)
			& 0.44 &$\pm$ & 0.18 & 42
			& 11.6 &$\pm$ & 1.8 & 16
			& 0.72 &$\pm$ & 0.31 & 43
			& 14.0 &$\pm$ & 3.0 & 21\\
			\multicolumn{1}{r}{\lk + HALO}
			&\lk 0.48 &$\qm\pm$ &\lk 0.16 &\lk 34
			&\lk 11.5 &$\qm\pm$ &\lk 1.7 &\lk 14
			&\lk 0.66 &$\qm\pm$ &\lk 0.28 &\lk 43
			&\lk 13.8 &$\qm\pm$ &\lk 2.7 &\lk 19\\
			\cdashlinelr{1-17}
			SK+JUNO (p)
			& 0.51 &$\pm$ & 0.13 & 25
			& 12.3 &$\pm$ & 2.1 & 17
			& 0.52 &$\pm$ & 0.17 & 33
			& 14.0 &$\pm$ & 2.4 & 17\\
			\multicolumn{1}{r}{\lk + HALO}
			&\lk 0.52 &$\qm\pm$ &\lk 0.11 &\lk 22
			&\lk 11.5 &$\qm\pm$ &\lk 1.4 &\lk 12
			&\lk 0.50 &$\qm\pm$ &\lk 0.14 &\lk 28
			&\lk 14.4 &$\qm\pm$ &\lk 2.2 &\lk 16\\
			\cdashlinelr{1-17}
			SK+JUNO (pC)
			& 0.51 &$\pm$ & 0.13 & 24
			& 11.2 &$\pm$ & 1.6 & 14
			& 0.52 &$\pm$ & 0.16 & 30
			& 14.3 &$\pm$ & 2.3 & 16\\
			\multicolumn{1}{r}{\lk + HALO}
			&\lk 0.53 &$\qm\pm$ &\lk 0.12 &\lk 22
			&\lk 11.3 &$\qm\pm$ &\lk 1.4 &\lk 12
			&\lk 0.50 &$\qm\pm$ &\lk 0.14 &\lk 28
			&\lk 14.5 &$\qm\pm$ &\lk 2.2 &\lk 15\\
			\cdashlinelr{1-17}
			SK+JUNO (pCd)
			& 0.50 &$\pm$ & 0.12 & 25
			& 11.2 &$\pm$ & 1.5 & 13
			& 0.53 &$\pm$ & 0.16 & 30
			& 14.1 &$\pm$ & 2.3 & 16\\
			\multicolumn{1}{r}{\lk + HALO}
			&\lk 0.52 &$\qm\pm$ &\lk 0.11 &\lk 22
			&\lk 11.2 &$\qm\pm$ &\lk 1.4 &\lk 13
			&\lk 0.51 &$\qm\pm$ &\lk 0.14 &\lk 28
			&\lk 14.4 &$\qm\pm$ &\lk 2.2 &\lk 15\\
			\bottomrule
		\end{tabular}
		}
		\caption{Reconstructed parameters for the
		time-integrated fluxes from the supernova
		model LS220-s27.0co \protect%
		\cite{Mirizzi:2015eza}, given different
		combination of detectors. Each row is 
		doubled to show the impact of including
		HALO-1kT
		to the analysis. The first row, marked with
		$(\mu^*\pm\sigma^*)$, reports the true value
		$\mu^*$ and the standard deviation
		$\sigma^*$ of a flat distribution in the
		assumed priors \protect\eqref{eq:pris},
		\protect\eqref{eq:prie}, 
		\protect\eqref{eq:pria}. Each analysis
		involving JUNO is performed six times,
		corresponding to the six assumptions on its
		configuration summarized in Table \protect%
		\ref{tab:jcon}, namely: pES implemented (p),
		$\Pnue/\APnue$ \tsup{12}C-CC reactions
		included (C), $\Pnue/\APnue$ \tsup{12}C-CC
		reactions included and distinguished (d).
		}
		\label{tab:risex}
\end{table}
\begin{table}
	\centering
		\resizebox{\textwidth}{!}{
		\begin{tabular}{l%
		r@{\hskip0.15em}c@{\hskip 0.15em}l@{\hskip%
		0.2em}c%
		r@{\hskip0.15em}c@{\hskip 0.15em}l@{\hskip%
		0.2em}c%
		r@{\hskip0.15em}c@{\hskip 0.15em}l@{\hskip%
		0.2em}c%
		r@{\hskip0.15em}c@{\hskip 0.15em}l@{\hskip%
		0.2em}c}
			\toprule
			\multicolumn{1}{c}{\multirow{2}{*}{%
			LS220-z9.6co}}
			&\multicolumn{3}{c}{$\:
			\mathcal{E}(\Pnue)$} &prec.
			&\multicolumn{3}{c}{$
			\hphantom{\langle\rangle}
			\langle E(\Pnue)\rangle$} &prec.
			&\multicolumn{3}{c}{$\:
			\mathcal{E}(\Pnux)$} &prec.
			&\multicolumn{3}{c}{$
			\hphantom{\langle\rangle}
			\langle E(\Pnux)\rangle$} &prec.\\
			&\multicolumn{3}{c}{$\:[\SI{e53}
			{\erg}]\vphantom{\frac{1}{2}}$} &\%
			&\multicolumn{3}{c}{$
			\hphantom{\langle\rangle}
			[\si{\mega\electronvolt}]$} &\%
			&\multicolumn{3}{c}{$\:[\SI{e53}
			{\erg}]\vphantom{\frac{1}{2}}$} &\%
			&\multicolumn{3}{c}{$
			\hphantom{\langle\rangle}
			[\si{\mega\electronvolt}]$} &\%\\
			\midrule
			$(\mu^*\pm\sigma^*)$
			& {0.32} &$\pm$ & 0.55 &---
			& {9.9} &$\pm$ & 19.6 &---
			& {0.30} &$\pm$ & 0.55 &---
			& {12.5} &$\pm$ & 19.6 &---\\
			\midrule
			\lk HALO-1kT
			&\lk 0.91 &$\qm\pm$ &\lk 0.56 &\lk 61
			&\lk 6.4 &$\qm\pm$ &\lk 2.7 &\lk 42
			&\lk 0.93 &$\qm\pm$ &\lk 0.55 &\lk 59
			&\lk 10.2 &$\qm\pm$ &\lk 5.1 &\lk 49\\	
			\midrule
			Super-K
			& 0.53 &$\pm$ & 0.41 & 77
			& 9.2 &$\pm$ & 5.7 & 62
			& 0.66 &$\pm$ & 0.42 & 64
			& 12.4 &$\pm$ & 7.9 & 64\\
			\multicolumn{1}{|r|}{\lk + HALO}
			&\lk 0.49 &$\qm\pm$ &\lk 0.36 &\lk 73
			&\lk 8.0 &$\qm\pm$ &\lk 3.3 &\lk 41
			&\lk 0.64 &$\qm\pm$ &\lk 0.39 &\lk 60
			&\lk 11.3 &$\qm\pm$ &\lk 5.3 &\lk 46\\
			\midrule
			JUNO (n)
			& 0.32 &$\pm$ & 0.13 & 41
			& 10.4 &$\pm$ & 2.6 & 25
			& 0.38 &$\pm$ & 0.20 & 53
			& 12.2 &$\pm$ & 5.1 & 42\\
			\multicolumn{1}{|r|}{\lk + HALO}
			&\lk 0.32 &$\qm\pm$ &\lk 0.12 &\lk 38
			&\lk 10.0 &$\qm\pm$ &\lk 1.6 &\lk 16
			&\lk 0.38 &$\qm\pm$ &\lk 0.20 &\lk 51
			&\lk 11.4 &$\qm\pm$ &\lk 3.4 &\lk 30\\
			\cdashlinelr{1-17}
			JUNO (nC)
			& 0.32 &$\pm$ & 0.13 & 40
			& 9.6 &$\pm$ & 1.8 & 19
			& 0.38 &$\pm$ & 0.20 & 53
			& 12.8 &$\pm$ & 5.3 & 42\\
			\multicolumn{1}{|r|}{\lk + HALO}
			&\lk 0.37 &$\qm\pm$ &\lk 0.12 &\lk 32
			&\lk 9.8 &$\qm\pm$ &\lk 1.5 &\lk 15
			&\lk 0.28 &$\qm\pm$ &\lk 0.19 &\lk 68
			&\lk 11.5 &$\qm\pm$ &\lk 3.3 &\lk 28\\
			\cdashlinelr{1-17}
			JUNO (nCd)
			& 0.34 &$\pm$ & 0.12 & 36
			& 10.0 &$\pm$ & 1.4 & 14
			& 0.34 &$\pm$ & 0.19 & 54
			& 12.8 &$\pm$ & 5.7 & 44\\
			\multicolumn{1}{|r|}{\lk + HALO}
			&\lk 0.38 &$\qm\pm$ &\lk 0.12 &\lk 30
			&\lk 9.8 &$\qm\pm$ &\lk 1.3 &\lk 13
			&\lk 0.26 &$\qm\pm$ &\lk 0.18 &\lk 71
			&\lk 11.3 &$\qm\pm$ &\lk 3.3 &\lk 28\\
			\cdashlinelr{1-17}
			JUNO (p)
			& 0.31 &$\pm$ & 0.25 & 80
			& 9.2 &$\pm$ & 4.0 & 44
			& 0.52 &$\pm$ & 0.20 & 39
			& 11.2 &$\pm$ & 1.7 & 16\\
			\multicolumn{1}{|r|}{\lk + HALO}
			&\lk 0.26 &$\qm\pm$ &\lk 0.11 &\lk 41
			&\lk 10.5 &$\qm\pm$ &\lk 1.8 &\lk 17
			&\lk 0.46 &$\qm\pm$ &\lk 0.17 &\lk 36
			&\lk 11.5 &$\qm\pm$ &\lk 1.8 &\lk 15\\
			\cdashlinelr{1-17}
			JUNO (pC)
			& 0.30 &$\pm$ & 0.23 & 77
			& 8.4 &$\pm$ & 3.3 & 39
			& 0.52 &$\pm$ & 0.20 & 38
			& 11.2 &$\pm$ & 1.8 & 16\\
			\multicolumn{1}{|r|}{\lk + HALO}
			&\lk 0.26 &$\qm\pm$ &\lk 0.11 &\lk 41
			&\lk 10.3 &$\qm\pm$ &\lk 1.7 &\lk 16
			&\lk 0.45 &$\qm\pm$ &\lk 0.16 &\lk 36
			&\lk 11.6 &$\qm\pm$ &\lk 1.8 &\lk 15\\
			\cdashlinelr{1-17}
			JUNO (pCd)
			& 0.26 &$\pm$ & 0.11 & 41
			& 10.3 &$\pm$ & 1.6 & 16
			& 0.44 &$\pm$ & 0.16 & 37
			& 11.6 &$\pm$ & 1.8 & 15\\
			\multicolumn{1}{|r|}{\lk + HALO}
			&\lk 0.27 &$\qm\pm$ &\lk 0.10 &\lk 39
			&\lk 10.3 &$\qm\pm$ &\lk 1.6 &\lk 15
			&\lk 0.44 &$\qm\pm$ &\lk 0.16 &\lk 36
			&\lk 11.6 &$\qm\pm$ &\lk 1.8 &\lk 15\\
			\midrule
			SK+JUNO (n)
			& 0.32 &$\pm$ & 0.12 & 39
			& 10.2 &$\pm$ & 2.7 & 26
			& 0.39 &$\pm$ & 0.20 & 52
			& 12.5 &$\pm$ & 4.8 & 38\\
			\multicolumn{1}{|r|}{\lk + HALO}
			&\lk 0.32 &$\qm\pm$ &\lk 0.12 &\lk 36
			&\lk 9.8 &$\qm\pm$ &\lk 1.7 &\lk 16
			&\lk 0.37 &$\qm\pm$ &\lk 0.19 &\lk 50
			&\lk 12.3 &$\qm\pm$ &\lk 3.6 &\lk 29\\
			\cdashlinelr{1-17}
			SK+JUNO (nC)
			& 0.32 &$\pm$ & 0.12 & 39
			& 9.4 &$\pm$ & 1.9 & 20
			& 0.37 &$\pm$ & 0.20 & 53
			& 13.5 &$\pm$ & 5.0 & 37\\
			\multicolumn{1}{|r|}{\lk + HALO}
			&\lk 0.32 &$\qm\pm$ &\lk 0.12 &\lk 36
			&\lk 9.7 &$\qm\pm$ &\lk 1.5 &\lk 15
			&\lk 0.37 &$\qm\pm$ &\lk 0.19 &\lk 50
			&\lk 12.3 &$\qm\pm$ &\lk 3.5 &\lk 28\\
			\cdashlinelr{1-17}
			SK+JUNO (nCd)
			& 0.35 &$\pm$ & 0.12 & 34
			& 9.8 &$\pm$ & 1.4 & 14
			& 0.33 &$\pm$ & 0.18 & 55
			& 13.7 &$\pm$ & 5.1 & 38\\
			\multicolumn{1}{|r|}{\lk + HALO}
			&\lk 0.34 &$\qm\pm$ &\lk 0.11 &\lk 33
			&\lk 9.8 &$\qm\pm$ &\lk 1.4 &\lk 14
			&\lk 0.34 &$\qm\pm$ &\lk 0.18 &\lk 52
			&\lk 12.1 &$\qm\pm$ &\lk 3.5 &\lk 28\\
			\cdashlinelr{1-17}
			SK+JUNO (p)
			& 0.31 &$\pm$ & 0.26 & 82
			& 9.1 &$\pm$ & 4.0 & 44
			& 0.55 &$\pm$ & 0.21 & 38
			& 10.9 &$\pm$ & 1.8 & 16\\
			\multicolumn{1}{|r|}{\lk + HALO}
			&\lk 0.26 &$\qm\pm$ &\lk 0.11 &\lk 41
			&\lk 10.3 &$\qm\pm$ &\lk 1.9 &\lk 18
			&\lk 0.47 &$\qm\pm$ &\lk 0.17 &\lk 36
			&\lk 11.3 &$\qm\pm$ &\lk 1.8 &\lk 16\\
			\cdashlinelr{1-17}
			SK+JUNO (pC)
			& 0.31 &$\pm$ & 0.24 & 77
			& 8.4 &$\pm$ & 3.2 & 38
			& 0.54 &$\pm$ & 0.21 & 38
			& 11.0 &$\pm$ & 1.79 & 16\\
			\multicolumn{1}{|r|}{\lk + HALO}
			&\lk 0.26 &$\qm\pm$ &\lk 0.10 &\lk 39
			&\lk 10.2 &$\qm\pm$ &\lk 1.7 &\lk 17
			&\lk 0.46 &$\qm\pm$ &\lk 0.17 &\lk 36
			&\lk 11.4 &$\qm\pm$ &\lk 1.8 &\lk 16\\
			\cdashlinelr{1-17}
			SK+JUNO (pCd)
			& 0.26 &$\pm$ & 0.10 & 39
			& 10.1 &$\pm$ & 1.7 & 17
			& 0.46 &$\pm$ & 0.17 & 37
			& 11.4 &$\pm$ & 1.8 & 16\\
			\multicolumn{1}{|r|}{\lk + HALO}
			&\lk 0.27 &$\qm\pm$ &\lk 0.10 &\lk 37
			&\lk 10.1 &$\qm\pm$ &\lk 1.6 &\lk 15
			&\lk 0.45 &$\qm\pm$ &\lk 0.16 &\lk 36
			&\lk 11.5 &$\qm\pm$ &\lk 1.8 &\lk 15\\
			\bottomrule
		\end{tabular}
		}
		\caption{Reconstructed parameters for the
		time-integrated fluxes from the supernova
		model LS220-z9.6co \protect%
		\cite{Mirizzi:2015eza}, given different
		combination of detectors. Each row is
		doubled to show the impact of including
		HALO-1kT
		to the analysis. The first row, marked with
		$(\mu^*\pm\sigma^*)$, reports the true value
		$\mu^*$ and the standard deviation
		$\sigma^*$ of a flat distribution in the
		assumed priors \protect\eqref{eq:pris},
		\protect\eqref{eq:prie}, 
		\protect\eqref{eq:pria}. Each analysis
		involving JUNO is performed six times,
		corresponding to the six assumptions on its
		configuration summarized in Table \protect%
		\ref{tab:jcon}, namely: pES implemented (p),
		$\Pnue/\APnue$ \tsup{12}C-CC reactions
		included (C), $\Pnue/\APnue$ \tsup{12}C-CC
		reactions included and distinguished (d).
		}
		\label{tab:risey}
\end{table}

\end{document}